\documentclass[journal=jctcce,manuscript=article]{achemso}

\usepackage{amsmath}
\usepackage{bm}
\usepackage{color}      % Needed for correction macros
\usepackage{graphicx}
\usepackage{multicol}
\usepackage{multirow}
\usepackage{xpatch}     % To change format of citenum

\newcommand{\bra}[1]{\ensuremath{\langle \: #1 \: |}}
\newcommand{\ket}[1]{\ensuremath{| \: #1 \: \rangle}}

\newcommand{\ExV}[3]{\ensuremath{\langle \: #1 \mid #2 \mid #3 \: \rangle}}

\title{Electron Dynamics with the Time-Dependent Density Matrix Renormalization Group}
\author{Alberto Baiardi}
\affiliation{ETH Z\"{u}rich, Laboratorium f\"{u}r Physikalische Chemie, Vladimir-Prelog-Weg 2, 8093 Z\"{u}rich, Switzerland.}
\email{alberto.baiardi@phys.chem.ethz.ch}
\date{\today}

\newcommand{\quotes}[1]{``#1''}

\abbreviations{}

\begin{document}

\begin{tocentry}
\includegraphics[width=\textwidth]{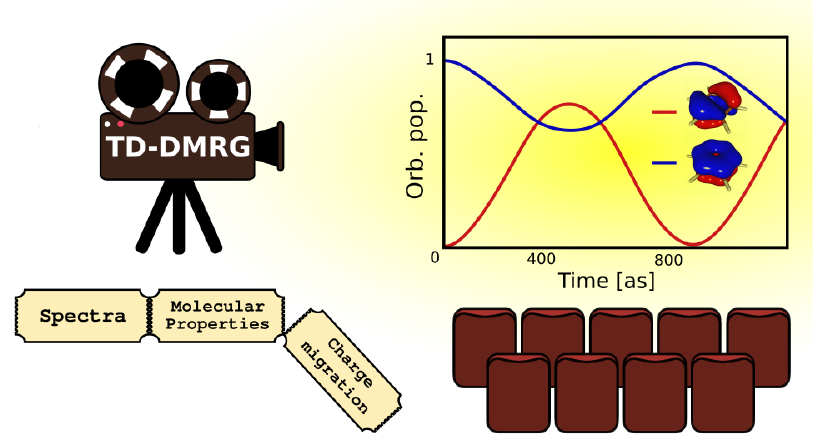}
\end{tocentry}

\begin{abstract}
\noindent In this work, we simulate the electron dynamics in molecular systems with the Time-Dependent Density Matrix Renormalization Group (TD-DMRG) algorithm.
We leverage the generality of the so-called tangent-space TD-DMRG formulation and design a computational framework in which the dynamics is driven by the exact non-relativistic electronic Hamiltonian.
We show that, by parametrizing the wave function as a matrix product state, we can accurately simulate the dynamics of systems including up to 20 electrons and 32 orbitals.
We apply the TD-DMRG algorithm to three problems that are hardly targeted by time-independent methods: the calculation of molecular (hyper)polarizabilities, the simulation of electronic absorption spectra, and the study of ultrafast ionization dynamics.
\end{abstract}

\section{Introduction}
\label{sec:intro}

Ultrafast spectroscopic techniques based on attosecond light pulses\cite{Corkum2007_AttosecondScience,Vrakking2007_AttosecondMolecularDynamics,Krausz2009_AttosecondPhysicsReview,Keller2012_AttosecondScience,Martin2019_Attosecond-Review} can probe electronic motions in molecular systems under strong non-equilibrium conditions.\cite{Worner2015_Iodoacetylene,Hutten2019_Kripton-Attosecond,Martinez2019_Br2-Attosecond}
The interpretation of the resulting experimental data calls for new computational methods to simulate electron dynamics by properly including quantum effects.
The design of so-called \quotes{real-time electronic-structure methods}\cite{Li2020_RTES-Review} has focused so far on solving the electronic time-dependent (TD) Schr\"{o}dinger equation based on wave function parametrization borrowed from time-independent (TI) quantum chemical algorithms.
However, the relative accuracy of TI-based wave function parameterizations changes drastically when extended to the time domain.
In fact, a molecule with a single-reference ground-state wave function may become strongly correlated when driven out of equilibrium.
Therefore, the design of systematically improvable wave function-based TD algorithms is the key to reliable electron dynamics simulations.
Real-time time-dependent density functional theory (RT-TD-DFT) is the only algorithm that can target molecules with several dozens of atoms.\cite{VanVoorhis2006_PredictorCorrector,Lopata2011_RealTimeTDDFT,Repisky2015_RealTime4c,Isborn2016_ElectronDynamics-TDDFT,Li2018_RealTime-Review}
However, exchange-correlation functionals that are parametrized on TI problems do not reproduce fundamental effects, such as Rabi oscillations in few-particles systems.\cite{Isborn2014_RabiOscillations}
Wave function-based methods, such as the TD formulations of the Coupled Cluster,\cite{Pigg2012_TDCC-ImaginaryTime,Kvaal2012_TDCC,Nascimento2016_TDCC,DePrince2017_CoreCC-RealTime,DePrince2019_EOM-TD-Spectroscopy,Li2019_Relativistic-RTCC,BondoPedersen2020_Stability-TDCC,Balbi2020-RT_CC} Configuration Interaction (TD-CI),\cite{Mazziotti2010_TDCI,Li2018_RealTime-GUGA_CI,Peng2018_TDCI} and Complete Active Space Self-Consistent Field\cite{Miranda2011_TD-CASSCF,Sato2013_TD-CAS,Madsen2014_TD-RASSCF,Sato2015_TD-RAS,Chan2018_TD-CASSCF-Surfaces,Sato2019_TDCASSCF-HHG} methods do not suffer from this limitation.
However, their range of application is limited by their high computational cost.

\noindent In the present work, we show that large-scale TD-CI simulations are feasible based on the time-dependent density matrix renormalization group (TD-DMRG) method.\cite{Paeckel2019_Review}
In its standard time-independent formulation,\cite{White1992_DMRGBasis,White1993_DMRGBasis} DMRG is an optimization algorithm for wave functions expressed as matrix product states (MPSs).\cite{McCulloch2007_FromMPStoDMRG}
Applications to quantum-chemical problems have demonstrated that full CI (or complete active space CI) molecular wave functions for up to 100 orbitals can be encoded as compact MPSs\cite{Chan2008_Review,Zgid2009_Review,Chan2011_Review,Wouters2013_Review,Keller2014,Kurashige2014_Review,Olivares2015_DMRGInPractice,Szalay2015_Review,Yanai2015,Knecht2016_Chimia,Baiardi2020_Review} and can, therefore, be optimized efficiently with DMRG.
However, the application of TD-DMRG to electronic quantum dynamics is much less explored.\cite{Frahm2019_TD-DMRG_Ultrafast}
The area law\cite{Hastings2007_AreaLaw} guarantees that the ground state wave function of short-ranged Hamiltonians can be represented as a compact MPS, but it does not apply to time-dependent simulations.
In fact, the wave function entanglement increases under non-equilibrium conditions\cite{Montangero2006_EntanglementIncrease,Cirac2008_EntanglementScaling,Zwolak2020_KramersCrossover,Zwolak2020_Transport-EntanglementBarrier} and such an effect, known as entanglement barrier, is not captured by monodimensional tensor network factorizations, such as the MPS.
Nevertheless, the area law does not apply also to time-independent quantum chemical problems,\cite{Baiardi2020_Review} but DMRG remains in practice more efficient than other full CI algorithms.
In the present work we tailor TD-DMRG to quantum chemical problems to assess how the entanglement barrier affects, in practice, the accuracy of electron dynamics simulations.
This is the first step towards the definition of tensor network methods providing the best compromise between cost and accuracy for molecular quantum dynamics.
Among the various TD-DMRG variants,\cite{Vidal2004_TEBD,Feiguin2005_Adaptive-TDDMRG,Haegeman2011_TDDMRG-MPSMPO,Zaletel2015,Ronca2017_TDDMRG-Targeting} the approach presented here relies on the tangent-space formulation\cite{Lubich2014_TimeIntegrationTT,Haegeman2016_MPO-TDDMRG} for two reasons.
First, it leads to very robust and numerically stable propagation algorithms.
Moreover, it relies on the so-called matrix product state/matrix product operator (MPS/MPO) DMRG formalism and can be, therefore, combined with our general DMRG framework that supports vibrational,\cite{Baiardi2017_VDMRG,Muolo2020_NEAP-DMRG} vibronic,\cite{Baiardi2019_TDDMRG} and electronic\cite{Keller2015_MPS-MPO-SQHamiltonian,Keller2016_SpinAdapted} quantum chemical Hamiltonians.

\noindent We extend the framework that we designed for vibrational and vibronic quantum dynamics\cite{Baiardi2019_TDDMRG} to the nonrelativistic Coulomb Hamiltonian, including spin symmetry\cite{Zgid2008_DMRGSpinAdaptation,Wouters2012_SpinAdapted,Sharma2015_GeneralNonAbelian,Keller2016_SpinAdapted} and supporting time-dependent perturbations.
We benchmark the TD-DMRG accuracy on three problems that are standard targets of real-time electronic-structure algorithms.
First, we simulate the ionization dynamics of benzene for active spaces including up to 26 orbitals.
Then, we show how electronic absorption spectra can be efficiently calculated with TD-DMRG.
Finally, we calculate high-order molecular properties based on a time-dependent finite-difference approach originally designed for TD-CI.\cite{Saalfrank2007_Properties-RealTime,Ding2013_Polarizability-RealTimeDFT,Li2018_RealTime-GUGA_CI}

\section{Electronic TD-DMRG theory}
\label{sec:theory}

\subsection{MPS/MPO-based DMRG}

DMRG encodes a full CI (or complete active space CI) wave function of an $L$-orbital system,

\begin{equation}
\ket{\Psi_\text{FCI}}
= \sum_{\sigma_1,\ldots,\sigma_L} C_{\sigma_1,\ldots,\sigma_L} \ket{\sigma_1 \cdots \sigma_L} \, ,
\label{eq:FullCI}
\end{equation}
as a matrix product state (MPS)

\begin{equation}
\ket{\Psi_\text{MPS}} = \sum_{\sigma_1 \cdots \sigma_L} \sum_{a_1,\ldots,a_{L-1}}^{m}
M_{1,a_1}^{\sigma_1} M_{a_1,a_2}^{\sigma_2} \cdots M_{a_{L-1},1}^{\sigma_L} 
\ket{\sigma_1 \cdots \sigma_L} \, .
\label{eq:MPS_Definition}
\end{equation}

\noindent Eq.~(\ref{eq:MPS_Definition}) expresses the CI tensor $C_{\sigma_1,\ldots,\sigma_L}$ as a product of $L$ 3-dimensional tensors $M_{a_{i-1},a_i}^{\sigma_i}$, where the $\sigma_i$ index labels the allowed occupations of orbital $i$, and the indexes $a_{i-1}$ and $a_i$ have maximum dimension $m$, usually referred to as \quotes{bond dimension}.
Any full CI wave function is represented exactly by an MPS with a bond dimension that grows exponentially with $L$, but the area law\cite{Hastings2007_AreaLaw} ensures that the ground state of short-ranged Hamiltonians can be encoded with much more compact MPSs. 
It has been shown\cite{Chan2008_Review,Zgid2009_Review,Chan2011_Review,Wouters2013_Review,Keller2014,Kurashige2014_Review,Olivares2015_DMRGInPractice,Szalay2015_Review,Yanai2015,Knecht2016_Chimia,Baiardi2020_Review} that the electronic ground state of molecular systems can often be encoded as a compact MPS, even though the area law prerequisites are not met.
Time-independent DMRG (TI-DMRG) optimizes the tensors $M_{a_{i-1},a_i}^{\sigma_i}$ based on the variational principle.
To do so, the non-relativistic electronic Hamiltonian $\mathcal{H}_\text{ele}$ that reads, in second-quantization, as 

\begin{equation}
\mathcal{H}_\text{ele} = \sum_{pq} h_{pq} a_p^\dagger a_q 
+ \frac{1}{2} \sum_{pqrs} \langle pq | rs \rangle a_p^\dagger a_q^\dagger a_s a_r \,
\label{eq:SQEleHams}
\end{equation}
where $h_{pq}$ and $\langle pq | rs \rangle$ are one- and two-electron integrals, respectively, is encoded as a matrix product operator (MPO)\cite{McCulloch2007_FromMPStoDMRG}

\begin{equation}
\mathcal{H}_\text{ele} = \sum_{\bm{\sigma},\bm{\sigma}'} 
\sum_{b_1=1}^{b_1^\text{max}} \cdots \sum_{b_{L-1}=1}^{b_{L-1}^\text{max}}
H_{1b_1}^{\sigma_1,\sigma_1'} \cdots H_{b_{L-1}1}^{\sigma_L,\sigma_L'}
\ket{\bm{\sigma}} \bra{\bm{\sigma}'}
\label{eq:MPO_Definition}
\end{equation}
as proposed in Ref.~\citenum{Keller2015_MPS-MPO-SQHamiltonian}. 
The energy functional $E[\ket{\Psi_\text{MPS}}]$ obtained by combining Eqs.~(\ref{eq:MPS_Definition}) and (\ref{eq:MPO_Definition}) is minimized iteratively one tensor at the time, starting from the first orbital (or \quotes{site}, in DMRG language). 
This minimization is equivalent to the alternating least squares algorithm\cite{Oseledets2012_ALS} and leads, for site $i$, to the following eigenvalue problem

\begin{equation}
\sum_{b_i,b_{i+1}} \sum_{\sigma_i,\sigma_i'} \sum_{a_i',a_{i+1}'}
L_{a_i,a_i'}^{b_i} W_{b_i,b_{i+1}}^{\sigma_i,\sigma_i'} 
R_{a_{i+1},a_{i+1}'}^{b_{i+1}} M_{a_i',a_{i+1}'}^{\sigma_i'} = E M_{a_i,a_{i+1}}^{\sigma_i} \, ,
\label{eq:EigenvalueProblem}
\end{equation}
where $\bm{L}$ and $\bm{R}$ are so-called boundaries that collect the partial contraction between the MPO and MPS for sites $[1,i-1]$ and $[i+1,L]$, respectively.\cite{Schollwoeck2011_Review-DMRG}

\noindent The MPS representation given in Eq.~(\ref{eq:MPS_Definition}) is not unique, \textit{i.e.} there exist gauge transformations that change $\bm{M}^{\sigma_i}$ without changing the underlying wave function $\ket{\Psi_\text{MPS}}$.
A sufficient condition to fix this gauge freedom is to left-normalize all tensors up to site $i$, such that\cite{Holtz2012_ManifoldTT}

\begin{equation}
\sum_{a_{i-1},\sigma_i} M_{a_{i-1},a_i}^{\sigma_i} M_{a_{i-1},a_i'}^{\sigma_i} 
= \delta_{a_i,a_i'} \, ,
\label{eq:LeftNormalization}
\end{equation}
and to right-normalize the remaining tensors, so that

\begin{equation}
\sum_{a_i,\sigma_i} M_{a_{i-1},a_i}^{\sigma_i} M_{a_{i-1}',a_i}^{\sigma_i} 
= \delta_{a_{i-1},a_{i-1}'} \, .
\label{eq:RightNormalized}
\end{equation}

\noindent It is useful to introduce the left- and right-renormalized bases for site $i$ ($\ket{a_{i}^{(l)}}$ and $\ket{a_{i}^{(r)}}$, respectively) defined recursively as follows:

\begin{equation}
\begin{aligned}
\ket{a_{i}^{(l)}} &= \sum_{a_{i-1}} \sum_{\sigma_i}
M_{a_{i-1},a_i}^{\sigma_i} \ket{a_{i-1}^{(l)} \sigma_i} \\
\ket{a_{i}^{(r)}} &= \sum_{a_{i+1}} \sum_{\sigma_{i+1}}
M_{a_i,a_{i+1}}^{\sigma_{i+1}} \ket{a_{i+1}^{(r)} \sigma_{i+1}} \, . \\
\end{aligned}
\label{eq:RenormalizedBases}
\end{equation}

\noindent $\ket{a_i^{(l)}}$ and $\ket{a_{i}^{(r)}}$ constitute an effective $m$-dimensional subspace of the $\ket{\sigma_1 \cdots \sigma_i}$ and $\ket{\sigma_{i+1} \cdots \sigma_L}$ bases, respectively.
In terms of Eq.~(\ref{eq:RenormalizedBases}), Eq.~(\ref{eq:EigenvalueProblem}) is obtained by projecting the full CI eigenvalue problem in the $\ket{a_{i-1}^{(l)} \sigma_i a_i^{(r)}}$ basis.

\subsection{Tangent-space TD-DMRG}

The tangent-space TD-DMRG theory\cite{Haegeman2016_MPO-TDDMRG} derives the equation of motion governing the propagation of an MPS from the Dirac-Frenkel variational principle.\cite{VanLeuven1988_EquivalenceTDPrinciple} 
The best MPS representation $\ket{\Psi_\text{MPS}(t)}$ of the exact full CI wave function $\ket{\Psi_\text{FCI}(t)}$ is obtained minimizing, at each time $t$, the functional $F(t)$ defined as follows (in Hartree atomic units)

\begin{equation}
F(t) = \left\| \mathcal{H} \ket{\Psi_\text{MPS}(t)} 
- \mathrm{i} \frac{\partial \ket{\Psi_\text{MPS}(t)}}{\partial t} \right\|^2 \, .
\label{eq:DiracFunctional}
\end{equation}

\noindent $\mathcal{H} \ket{\Psi_\text{MPS}(t)}$ is represented by an MPS with bond dimension larger than that of $\ket{\Psi_\text{MPS}(t)}$.\cite{Schollwoeck2011_Review-DMRG} 
Therefore, an exact MPS representation of the time-evolving wave function, corresponding to $F(t)=0$, would require a continuous increase of $m$ with time, an effect known as entanglement barrier.\cite{Cirac2008_EntanglementScaling,Osborne2008_EntanglementBarrier,Goto2019_LongTime-TDDMRG,Legeza2019_OrbitalOptimization-TD,Zwolak2020_KramersCrossover,Zwolak2020_Transport-EntanglementBarrier}
To avoid that, TD-DMRG minimizes $F(t)$ to obtain the best MPS representation of the wave function for a given bond dimension $m$.
The optimal MPS is obtained by solving of the differential equation\cite{Lubich2014_TimeIntegrationTT} 

\begin{equation}
\mathrm{i} \frac{\partial \ket{\Psi_\text{MPS}(t)}}{\partial t} 
= \mathcal{P}_{\ket{\Psi_\text{MPS}(t)}} \mathcal{H} \ket{\Psi_\text{MPS}(t)} \, ,
\label{eq:ProjectedTDSchrodinger}
\end{equation}
where $\mathcal{P}_{\ket{\Psi_\text{MPS}(t)}}$ is the so-called tangent space projector to the manifold $\mathcal{T}_{m}$ of MPSs with bond dimension $m$. 
In practice, all MPSs with a given bond dimension form a subspace of the full CI space, and $\mathcal{T}_{m}$ is the linear approximation of this space in the vicinity of $\ket{\Psi_\text{MPS}(t)}$. 
$\mathcal{P}_{\ket{\Psi_\text{MPS}(t)}}$ is the projection operator onto $\mathcal{T}_{m}$.
The closed-form for $\mathcal{P}_{\ket{\Psi_\text{MPS}(t)}}$ reads\cite{Lubich2014_TimeIntegrationTT,Haegeman2016_MPO-TDDMRG}

\begin{equation}
\begin{aligned}
\mathcal{P}_{\ket{\Psi_\text{MPS}(t)}} 
&= \sum_{i=1}^L \sum_{a_{i-1}^{(l)} \sigma_i a_i^{(r)}} 
\ket{a_{i-1}^{(l)} \sigma_i a_i^{(r)}} 
\bra{a_{i-1}^{(l)} \sigma_i a_i^{(r)}}
- \sum_{i=1}^{L-1} \sum_{a_i^{(l)}, a_i^{(r)}}
\ket{a_i^{(r)} a_i^{(r)}} 
\bra{a_i^{(l)} a_i^{(r)}} \\
&= \sum_{i=1}^L \mathcal{P}_i^{(1)} - \sum_{i=1}^{L-1} \mathcal{P}_i^{(2)} \, .
\end{aligned}
\label{eq:TangentSpace}
\end{equation}
With these approximations, Eq.~(\ref{eq:ProjectedTDSchrodinger}) becomes

\begin{equation}
\mathrm{i} \frac{\partial \ket{\Psi_\text{MPS}(t)}}{\partial t} =
\left( \sum_{i=1}^L \mathcal{P}_i^{(1)} - \sum_{i=1}^{L-1} \mathcal{P}_i^{(2)} \right)
\mathcal{H} \ket{\Psi_\text{MPS}(t)} \, .
\label{eq:ProjectedTDSchrodinger2}
\end{equation}

\noindent The solution of Eq.~\ref{eq:ProjectedTDSchrodinger2} can be approximated based on the Trotter factorization, so that the tensors are propagated sequentially as:

\begin{equation}
\bm{M}^{\sigma_i}(t+\Delta t) = e^{- \mathrm{i} \bm{H}_i^{(1)} \Delta t } \bm{M}^{\sigma_i}(t) \, ,
\label{eq:ForwardPropagation}
\end{equation}
where $\bm{H}_i^{(1)}$ is the representation of $\mathcal{H}$ in the $\ket{a_{i-1}^{(l)} \sigma_i a_i^{(r)}}$ renormalized basis obtained with the MPS canonized on site $i$.
Additionally, terms involving the $\mathcal{P}_i^{(2)}$ operator determine the time-evolution of the zero-site tensor $\bm{N}$ for site $i$, defined as $\bm{S}\bm{V}$ where

\begin{equation}
M_{a_{i-1}a_i}^{\sigma_i}
= \sum_{a_i'} U_{a_{i-1}a_i'}^{\sigma_i} S_{a_i',a_i'} V_{a_i',a_i} \, .
\label{eq:SVD}
\end{equation}

\noindent The parallel of Eq.~(\ref{eq:ForwardPropagation}) for the zero-site tensor reads:

\begin{equation}
\bm{N}(t + \Delta t) = e^{\mathrm{i} \bm{H}_i^{(2)} \Delta t } \bm{N}(t) \, .
\label{eq:BackwardPropagation}
\end{equation}

\noindent where $\bm{H}_i^{(2)}$ is the representation of $\mathcal{H}$ in the $\ket{a_i^{(l)} a_i^{(r)}}$ basis.
Eq.~(\ref{eq:BackwardPropagation}) propagates the zero-site tensor backward in time.
Since the $\ket{a_i^{(l)}}$ basis functions are already included in the $\ket{a_{i-1}^{(l)} \sigma_i}$ basis (see Eq.~(\ref{eq:RenormalizedBases})), this back-propagation step ensures that the wave function projection onto the $\ket{a_i^{(l)}}$ basis is not propagated twice.
We solve Eqs.~(\ref{eq:ForwardPropagation}) and (\ref{eq:BackwardPropagation}) by approximating the exponential operator based on the Lanczos algorithm,\cite{Saad1992_MatrixExponential,Lubich1997_MatrixExponential,VanDerEshof2006_MatrixExponentialPreconditioner} so that both the wave function energy and norm are conserved.\cite{Haegeman2016_MPO-TDDMRG}

\noindent The TD-DMRG algorithm outlined above can be extended to propagate simultaneously two consecutive sites, as is done in two-site TI-DMRG.
In this case, $\mathcal{P}_i^{(1)}$ is the projector for the $\ket{a_{i-1}^{(l)} \sigma_i \sigma_{i+1} a_{i+1}^{(r)}}$ basis, and Eq.~(\ref{eq:ForwardPropagation}) determines the time evolution of the two-site tensor $\bm{T}^{\sigma_i,\sigma_{i+1}}$, defined as

\begin{equation}
T_{a_{i-1},a_{i+1}}^{\sigma_i,\sigma_{i+1}} 
= \sum_{a_i} M_{a_{i-1},a_i}^{\sigma_i} M_{a_i,a_{i+1}}^{\sigma_{i+1}} \, .
\label{eq:TSTensor}
\end{equation}

\noindent Moreover, $\mathcal{P}_i^{(2)}$ becomes the projection onto the $\ket{a_{i-1}^{(l)} \sigma_i a_i^{(r)}}$ basis.
We will refer in the following to this two-site variant as TD-DMRG(TS), and to the single-site formulation as TD-DMRG(SS).

\noindent Tangent-space TD-DMRG approximates TD full CI in three respects.
First, the wave function is encoded as an MPS with a fixed bond dimension $m$. 
We will refer to the related error as the \quotes{truncation error}. 
This approximation is the same as for TI-DMRG and can be controlled monitoring the convergence of the target observables with the bond dimension $m$.
The second approximation is the Trotter factorization of the time-evolution operator.
We will refer to the corresponding error as \quotes{Trotterization error}.
The impact of this approximation can be quantified by considering that the Trotter factorization represents the first-order term of the Baker-Campbell-Hausdorff formula for an exponential operator. 
Therefore, the error will be proportional to the first correction, \textit{i.e.} the second-order term.
Once applied onto an MPS $\ket{\Psi_\text{MPS}}$, such terms would include, for instance, the following contribution:

\begin{equation}
\begin{aligned}
\left[ \mathcal{P}_i^{(1)} \mathcal{H}, \mathcal{P}_i^{(2)} \mathcal{H} \right]
\ket{\Psi_\text{MPS}}& \\
= \sum_{a_{i-1}^{(l)},\sigma_{i},a_i^{(r)}} \sum_{a_i^{(l)},\bar{a}_i^{(r)}}
& \left(
\ket{a_{i-1}^{(l)} \sigma_i a_i^{(r)}}
\bra{a_{i-1}^{(l)} \sigma_i a_i^{(r)}} \mathcal{H}
\ket{a_i^{(l)} \bar{a}_i^{(r)}} \bra{a_i^{(l)} \bar{a}_i^{(r)}} 
\mathcal{H} \ket{\Psi_\text{MPS}} \right. \\
& \left.
- \ket{a_i^{(l)} \bar{a}_i^{(r)}} \bra{a_i^{(l)} \bar{a}_i^{(r)}}
\mathcal{H}
\ket{a_{i-1}^{(l)} \sigma_i a_i^{(r)}}
\bra{a_{i-1}^{(l)} \sigma_i a_i^{(r)}} \mathcal{H} \ket{\Psi_\text{MPS}}
\right)
\end{aligned}
\label{eq:TrotterError}
\end{equation}

\noindent In the first term of the right-hand side of Eq.~(\ref{eq:TrotterError}), $\mathcal{H}$ is applied on the MPS, then projected onto the $\ket{a_{i-1}^{(l)} \sigma_i a_i^{(r)}}$ basis, $\mathcal{H}$ is applied a second time and, lastly, the wave function is projected onto the $\ket{a_i^{(l)} a_i^{(r)}}$ basis. 
The second term applies the two projectors in the reversed order 
The representation of $\mathcal{H} \ket{\Psi_\text{MPS}}$ in the $\ket{a_i^{(l)}}$ basis will be less accurate than that in the $\ket{a_{i-1}^{(l)} \sigma_i}$ basis and, therefore, the two terms of Eq.~(\ref{eq:TrotterError}) will differ.
In the limit of a converged MPS, the contribution of the basis functions included in $\ket{a_{i-1}^{(l)} \sigma_i}$ and not in $\ket{a_i^{(l)}}$ is negligible 
and, therefore, the error will decrease with $m$, as pointed out in Ref.~\citenum{Paeckel2019_Review}.
A third approximation is the solution of Eqs.~(\ref{eq:ForwardPropagation}) and (\ref{eq:BackwardPropagation}) with the Lanczos algorithm.\cite{Saad1992_MatrixExponential}
The accuracy of this approximation can be controlled as suggested in Ref.~\citenum{Saad1992_MatrixExponential}. 
For all simulations reported below, a wave function accuracy of 10$^{-10}$ is reached with a 10-dimensional Krylov space.
This third error source is, therefore, negligible compared to the other ones.
A fourth approximation, often overlooked in the literature, is the neglect of the time-dependence of the tangent-space projector that depends on time via the renormalized bases $\ket{a_i^{(l)}(t)}$ and $\ket{a_i^{(r)}(t)}$.
Even though, up to our knowledge, no algorithm for estimating the error underlying this fourth approximation has been proposed, it will vanish in the limit of a converged MPS because the space spanned by the renormalized bases will change smoothly with time.

\subsection{External time-dependent perturbations}

The simulation of molecular spectra requires including the light-matter interaction contribution to the molecular Hamiltonian.
As discussed in Ref.~\citenum{Lubich2014_TimeIntegrationTT}, the formal solution to the differential equation given in Eq.~(\ref{eq:ProjectedTDSchrodinger2}) cannot be written as in Eq.~(\ref{eq:ForwardPropagation}) for time-dependent Hamiltonians $\mathcal{H}(t)$.
However, Eq.~(\ref{eq:ProjectedTDSchrodinger2}) can still be simplified by a Trotter factorization, resulting in the following series of differential equations:

\begin{align}
\frac{\partial \ket{\Psi_\text{MPS}(t)}}{\partial t} 
&= - \mathrm{i} \mathcal{P}_i^{(1)} \mathcal{H}(t) \ket{\Psi_\text{MPS}(t)} 
\label{eq:SeriesOfDifferentialEquation} \\
\frac{\partial \ket{\Psi_\text{MPS}(t)}}{\partial t} 
&= \mathrm{i} \mathcal{P}_i^{(2)} \mathcal{H}(t) \ket{\Psi_\text{MPS}(t)} \, ,
\label{eq:SeriesOfDifferentialEquation2}
\end{align}
where the solution to Eq.~(\ref{eq:SeriesOfDifferentialEquation}) for site $i$ is the initial wave function for Eq.~(\ref{eq:SeriesOfDifferentialEquation2}) for the same site, and the solution of Eq.~(\ref{eq:SeriesOfDifferentialEquation2}) for a given site defines the initial wave function for Eq.~(\ref{eq:SeriesOfDifferentialEquation}) for the following site.
Ref.~\citenum{Lubich2014_TimeIntegrationTT} suggests to solve the local differential equation with a fourth-order Runge-Kutta propagator.
In the present work, we instead rely on two Magnus integrators that are routinely applied to RT-TD-DFT.\cite{Castro2004_Propagators,Rubio2018_Propagators}
The first one is the exponential midpoint rule (EMR2) integrator that solves the Eq.~(\ref{eq:SeriesOfDifferentialEquation}) as (the extension to Eq.~(\ref{eq:SeriesOfDifferentialEquation2}) is straightforward)

\begin{equation}
\ket{\Psi_\text{MPS}(t+\Delta t)}
= e^{-\mathrm{i} \mathcal{P}_i^{(1)} \mathcal{H}(t+\Delta t/2)\Delta t} \ket{\Psi_\text{MPS}(t)} \, ,
\label{eq:CommutatorFree-EMR2}
\end{equation}
and leads to the following equation for the tensor $M_{a_{i-1},a_i}^{\sigma_i}$:

\begin{equation}
\bm{M}^{\sigma_i}(t+\Delta t) = e^{-\mathrm{i}\bm{H}_i^{(1)}(t+\Delta t/2) \Delta t} \bm{M}^{\sigma_i}(t)
\label{eq:EMR2-For-Tensors}
\end{equation}

\noindent Note that we neglect the dependence of $\mathcal{P}_{\ket{\Psi_\text{MPS}(t)}}$ on time and, therefore, $\bm{H}_i^{(1)}(t+\Delta t/2)$ is obtained from the representation of $\mathcal{H}(t+\Delta t/2)$ in the time-independent basis $\ket{a_{i-1}^{(l)} \sigma_i a_i^{(r)}}$.
We will refer to the TD-DMRG algorithm in which the local differential equations are solved with Eq.~(\ref{eq:EMR2-For-Tensors}) as EMR2-TD-DMRG.
The error associated with solving Eqs.~(\ref{eq:SeriesOfDifferentialEquation}) and (\ref{eq:SeriesOfDifferentialEquation2}) with EMR2 scales as $\mathcal{O}(\Delta t^2)$, as for the Trotterization error.
The Trotter error is expected to be very small for large $m$ values, \textit{i.e.} for converged TD-DMRG simulations, and the EMR2 error will then be predominant.
To cure this effect, we also solve the local differential equation with the fourth-order commutator-free (CF4) propagator\cite{Blanes2006_CommutatorFreeMagnus}
that approximates the solution to Eq.~(\ref{eq:SeriesOfDifferentialEquation}) as:

\begin{equation}
\begin{aligned}
\ket{\Psi_\text{MPS}(t+\Delta t)}
= & \exp\left[ -\mathrm{i} \alpha_1 \mathcal{P}_i^{(1)} \mathcal{H}(t+ \Delta t_1)\Delta t
-\mathrm{i} \alpha_2 \mathcal{P}_i^{(1)} \mathcal{H}(t+ \Delta t_2)\Delta t
\right] \times \\
& \exp \left[ -\mathrm{i} \alpha_2 \mathcal{P}_i^{(1)} \mathcal{H}(t+ \Delta t_1)\Delta t
-\mathrm{i} \alpha_1 \mathcal{P}_i^{(1)} \mathcal{H}(t+ \Delta t_2)\Delta t
\right]
\ket{\Psi_\text{MPS}(t)} \, ,
\end{aligned}
\label{eq:CommutatorFree-CF4}
\end{equation}
where the constants $\alpha_1$ and $\alpha_2$ and the time-steps $\Delta t_1$ and $\Delta t_2$ are given in Ref.~\citenum{Blanes2006_CommutatorFreeMagnus}. 
Written in terms of tensors, Eq.~(\ref{eq:CommutatorFree-CF4}) reads:

\begin{equation}
\begin{aligned}
\bm{M}^{\sigma_i}(t+\Delta t) 
= & e^{-\mathrm{i}(\alpha_1\bm{H}_i^{(1)}(t+\Delta t_1)
	+ \alpha_2\bm{H}_i^{(1)}(t+\Delta t_2))\Delta t} \times \\
& e^{-\mathrm{i}(\alpha_2\bm{H}_i^{(1)}(t+\Delta t_1) 
	+ \alpha_1\bm{H}_i^{(1)}(t+\Delta t_2))\Delta t} 
\bm{M}^{\sigma_i}(t)
\end{aligned}
\label{eq:CF4-For-Tensors}
\end{equation}

\noindent We solve Eq.~(\ref{eq:CF4-For-Tensors}) by applying twice the Lanczos algorithm, one per exponential operator. 
Note that $\bm{H}_i^{(1)}(t+\Delta t_1)$ and $\bm{H}_i^{(1)}(t+\Delta t_2)$ do not commute and, therefore, their relative order must be preserved when calculating the Krylov vectors.
We will refer to the resulting TD-DMRG algorithm as CF4-TD-DMRG.
The computational cost of CF4-TD-DMRG is approximately four times higher than that of EMR2-TD-DMRG, because four MPO/MPS contractions must be calculated per Krylov subspace vector. 
However, the CF4 integrator error scales as $\mathcal{O}(\Delta t^4)$ and, therefore, can support larger time steps.
In the following, we will not present results obtained with the fourth-order Runge-Kutta algorithm\cite{Lubich2014_TimeIntegrationTT} since, in all cases, both EMR2-TD-DMRG and CF4-TD-DMRG are largely more stable.

\subsection{Spin-adapted TD-DMRG}

The computational cost of electronic TD-DMRG can be reduced by leveraging the spin symmetries associated with the squared value of the overall electronic spin ($\mathcal{S}^2$) and to its projection on a given axis ($\mathcal{S}_z$).
Both these quantities are conserved during the propagation and, therefore, the wave function can be encoded as a symmetry-adapted MPS.\cite{Vidal2011_DMRG-U1Symm,Troyer2011_PEPS-Symmetry}
For abelian symmetry groups (such as the one associated with the $\mathcal{S}_z$ symmetry), each indices of an MPS tensor ($a_i$) can be assigned univocally to an irreducible representation ($\Gamma_{a_i}$) of the symmetry group.
The non-zero blocks of $M_{a_{i-1},a_i}^{\sigma_i}$ are those for which $\Gamma_{a_{i-1}} \otimes \Gamma_{\sigma_i} = \Gamma_{a_i}$.
This block-diagonal structure can be exploited to enhance the $m$ energy convergence.\cite{Keller2016_SpinAdapted}
Compared to TI-DMRG, the tangent-space TD-DMRG theory introduces an additional step, \textit{i.e.} the zero-site tensor propagation (see Eq.~(\ref{eq:BackwardPropagation})).
Expressed in terms of MPS tensors and boundaries, the contraction between $\bm{H}_i^{(2)}$ and the zero-site tensor $\bm{N}$ reads:

\begin{equation}
\bar{N}_{a_l,\bar{a}_l} = \sum_{b_l,\bar{b}_l} \sum_{a_l',\bar{a}_l'} 
L_{a_l,a_l'}^{b_l} N_{a_l', \bar{a}_l'} R_{\bar{a}_l, \bar{a}_l'}^{\bar{b}_l}
\label{eq:ZeroSiteContraction}
\end{equation}

\noindent It follows from Ref.~\citenum{Vidal2011_DMRG-U1Symm} that, for a symmetry-adapted MPS, the only non-zero blocks of $\bm{N}$ are those for which $\Gamma_{a_i} = \Gamma_{a_{i}'}$.
The boundaries have an analog block structure that can be leveraged to reduce the computational cost associated with Eq.~\ref{eq:ZeroSiteContraction}.

\noindent The design of MPS tensors adapted to non-abelian symmetry groups, such as SU(2) that is associated with the $\mathcal{S}^2$ conservation, is less trivial.\cite{Sharma2015_GeneralNonAbelian,Keller2016_SpinAdapted}
In this case, each index $a_i$ is mapped to a combination of $\mathcal{S}_{a_i}^z$ and $\mathcal{S}_{a_i}^2$ quantum numbers.
Following Ref.~\citenum{Keller2016_SpinAdapted}, we express each tensor as the product of a Clebsch-Gordan coefficient that depends on both $\mathcal{S}_{a_i}^z$ and $\mathcal{S}_{a_i}^2$, and a reduced tensor that is independent on $\mathcal{S}_{a_i}^z$.
The efficiency of an SU(2)-adapted DMRG implementation relies on the possibility of expressing all contractions in terms of reduced tensors without calculating their full counterparts.
$N_{a_l', \bar{a}_l'}$ is a rank-0 tensor\cite{Keller2016_SpinAdapted} and, therefore, Eq.~\ref{eq:ZeroSiteContraction} still holds true for the reduced counterpart of $\bm{L}$, $\bm{N}$, and $\bm{R}$.
Note that the same does not hold true for the contraction of a single-site tensor with the MPO, where an additional scaling factor proportional to the 9j-Wigner symbol must be included.\cite{Keller2016_SpinAdapted}

\subsection{Available TD-DMRG variants}

As we highlighted in the previous sections, multiple TD-DMRG variants are obtained depending on the algorithm used to solve the local differential equation and on the approximation of the time-evolution operator.
For time-independent Hamiltonians, we solve the local differential equation with the Lanczos algorithm and refer to the resulting method simply as TD-DMRG.
For time-dependent Hamiltonians, the local equations can be solved with either the EMR2 or the CF4 integrator. We refer to the resulting algorithms as EMR2-TD-DMRG and CF4-TD-DMRG, respectively.
In all cases, either one or two tensors can be propagated at each time step.
We will denote these two classes of TD-DMRG variants by adding ``(SS)'' and ``(TS)'', respectively, at the end of the algorithm name.

\section{Charge dynamics following ionization of benzene}
\label{sec:ChargeDynamics}

We simulate with TD-DMRG the electronic dynamics after ionization of benzene, which is, together with iodoacetylene and phenylalanine,\cite{Calegari2014,Worner2015_Iodoacetylene} one of the few molecules for which experimental attosecond spectra are available.
Upon ionization, a charge oscillation occurs between the $\pi^0$ orbital and the two other degenerate occupied $\pi$ orbitals (referred to in the following as $\pi^1$, a graphical representation of the orbitals is given in Figure~S1 of the Supporting Information).\cite{Mikosch2017_Benzene-Attosecond} 
This is confirmed by third-order algebraic diagrammatic construction (ADC(3)) calculations\cite{Despre2015} that predict an oscillation period of about 950~as, in good agreement with the experimental data.
A more recent adaptive TD-CI simulation based on a CAS(8,8) predicts an oscillation period of about 700~as,\cite{Evangelista2019_Adaptive-TDCI} and the difference with the ADC(3) results is ascribed to missing dynamical correlation effects.
Here, we show that TD-DMRG can target larger active spaces including up to 26 electrons in 26 orbitals, based on the HF canonical orbitals calculated with the cc-pVDZ basis set.
We first optimize the TD-DMRG parameters (bond dimension $m$, time step $\Delta t$ and integration algorithm) for a model active space including 14 electrons in 14 orbitals.
We then simulate the dynamics on larger active spaces based on the resulting optimal parameters.
In all cases, we rely on the non-spin-adapted formulation of TD-DMRG.

\subsection{Optimization of the TD-DMRG parameters}

We generate the initial state for the propagation by optimizing the ground state for a CAS(14,14) with both TI-DMRG(TS) and the imaginary-time variant of TD-DMRG (iTD-DMRG(TS)) that we introduced in Ref.~\citenum{Baiardi2020_tcDMRG}.
We report the resulting optimized energies in Table~\ref{tab:CAS14_Imag}.
In the TI-DMRG calculations, we perturbed the two-site tensor based on the algorithm presented in Ref.~\citenum{McCulloch2015_Mixing} to enhance the efficiency of the sweep-based optimization and avoid convergence to local minima of the energy functional.

\begin{table}[htbp!]
	\centering
	\begin{tabular}{c|ccc}
		\hline \hline
		&   $m$=125   &   $m$=250    &   $m$=500   \\
		\hline
		TI-DMRG   &  -230.76070 &  -230.79079  &  -230.76082 \\
		iTD-DMRG   &  -230.76070 &  -230.76079  &  -230.76082 \\
		\hline \hline 
	\end{tabular}
	\caption{Ground-state energy (in Hartree atomic units) of benzene calculated with TI-DMRG(TS) and iTD-DMRG(TS), based on CAS(14,14) with varying $m$ values. The iTD-DMRG time-step is 500~as.}
	\label{tab:CAS14_Imag}
\end{table}

\noindent The converged iTD-DMRG(TS) ground-state energy ($m$=500) matches the TI-DMRG(TS) result and, as illustrated in Figure~S2 of the Supporting Information, the imaginary-time propagation converges as fast as TI-DMRG (as we already highlighted in our previous work\cite{Baiardi2020_tcDMRG}).
For this reason, if not otherwise specified, in the following we always optimize the initial MPS wave function with iTD-DMRG(TS).

\noindent We simulate the ionization process with the so-called \quotes{sudden ionization} model and assume that one electron is removed instantaneously from the $\pi^0$ orbital.
The initial state wave function for the propagation ($\ket{\Psi_\text{GS}^{N-1}}$) is expressed in terms of the optimized ground state of the neutral molecule ($\ket{\Psi_\text{GS}^{N}}$) as

\begin{equation}
\ket{\Psi_\text{GS}^{N-1}} 
= \frac{\hat{a}_{\pi^0} \ket{\Psi_\text{GS}^N}}
{\ExV{\Psi_\text{GS}^N}{\hat{a}_{\pi^0}^\dagger \hat{a}_{\pi^0}}{\Psi_\text{GS}^N}} \, ,
\label{eq:SuddenIonization}
\end{equation}
where $\hat{a}_{\pi^0}$ is the annihilation operator associated with the $\pi^0$ orbital.
We obtain the MPS representation of $\ket{\Psi_\text{GS}^{N-1}}$ by encoding $\hat{a}_{\pi^0}$ as MPO and evaluating Eq.~(\ref{eq:SuddenIonization}) as MPO/MPS contraction.\cite{Schollwoeck2011_Review-DMRG}
We then construct the MPO representation of the second-quantization Hamiltonian based on the Hartree-Fock orbitals of the neutral state.
Therefore, we neglect orbital relaxation effects.
Even though this approximation limits the calculation accuracy, we will show in the following that TD-DMRG delivers converged simulations of the valence ionization dynamics of benzene also with this non optimal basis set.
Orbital relaxation effects will certainly be more relevant under strong non-equilibrium conditions, such as for core ionizations.
In these cases, the TD-DMRG efficiency can be enhanced by optimizing the molecular orbital coefficients at each time step, together with the MPS entries.\cite{Sato2013_TD-CAS}
However, a detailed description of the resulting algorithm goes beyond the scopes of the present work.

\begin{figure}[htbp!]
	\centering
	\includegraphics[width=.75\textwidth]{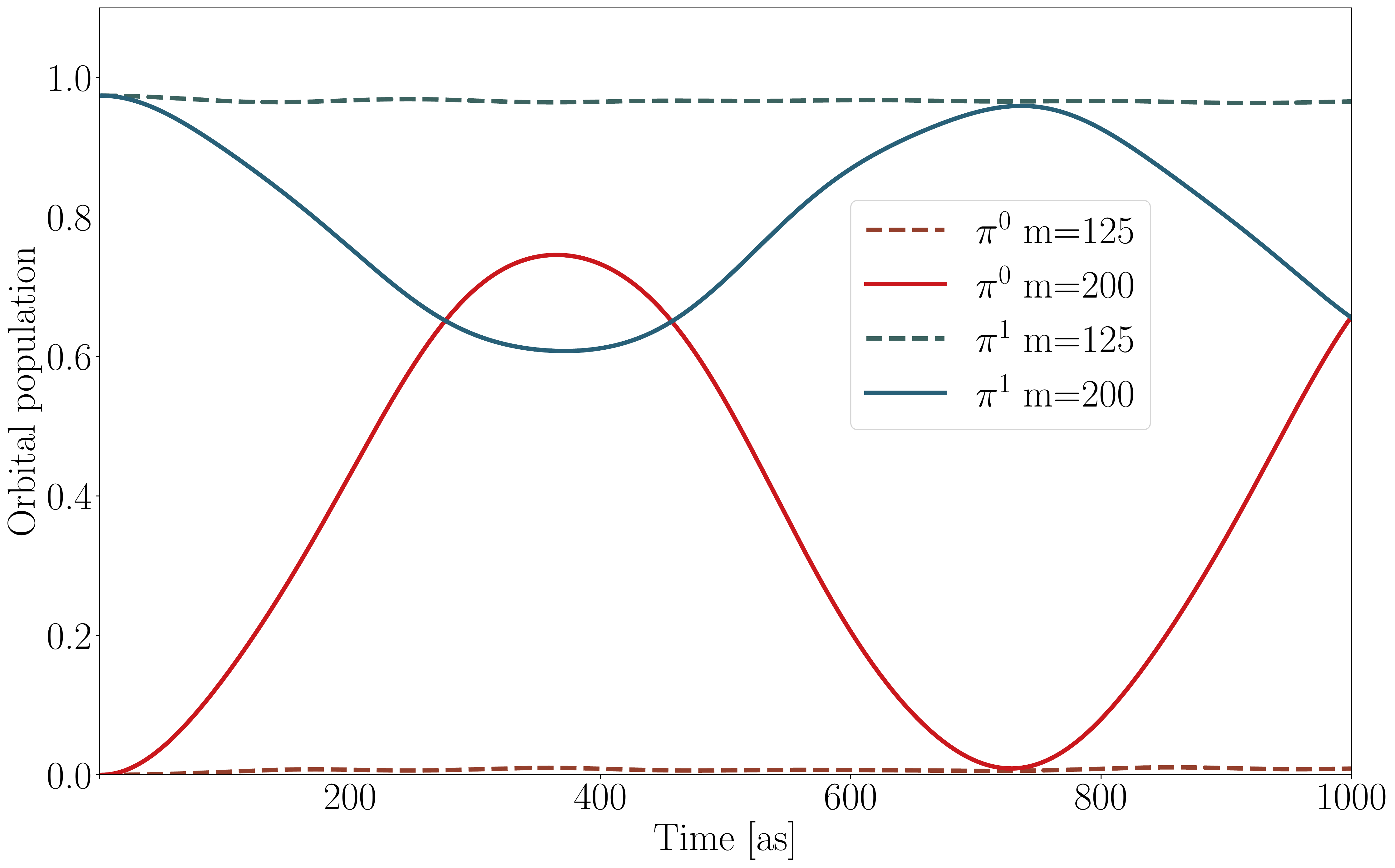}
	\caption{Time evolution of the $\pi^0$ and $\pi^1$ orbital population of benzene calculated with TD-DMRG(TS), $\Delta t$=1~as, and $m$=125 and 200.}
	\label{fig:PopBenzene_CAS14-14}
\end{figure}

\noindent We report in Figure~\ref{fig:PopBenzene_CAS14-14} the time evolution of the $\pi^0$ orbital population simulated with TD-DMRG(TS).
For $m$=125, the hole remains localized on the $\pi^0$ orbital but, starting from $m$=200, we observe the migration of the hole created in the $\pi^0$ spin-orbital to the $\pi^1$ one, in agreement with experimental\cite{Despre2015} and theoretical\cite{Evangelista2019_Adaptive-TDCI} data.
The time-dependent population does not change by further increasing $m$ to 250 and, therefore, it converges with the same $m$ value that delivers a converged ground-state energy.
The maximum value of the $\pi^0$ orbital population is obtained (for $m$=200) after 364~as, which corresponds to an oscillation period of 728~as, in agreement with TD-CI data\cite{Evangelista2019_Adaptive-TDCI} obtained with CAS(8,8).
Moreover, the $\pi^1$ orbital population reaches its minimum at 364~as, therefore confirming that the hole migrates from $\pi^0$ to $\pi^1$.

\begin{figure}[htbp!]
	\includegraphics[width=\textwidth]{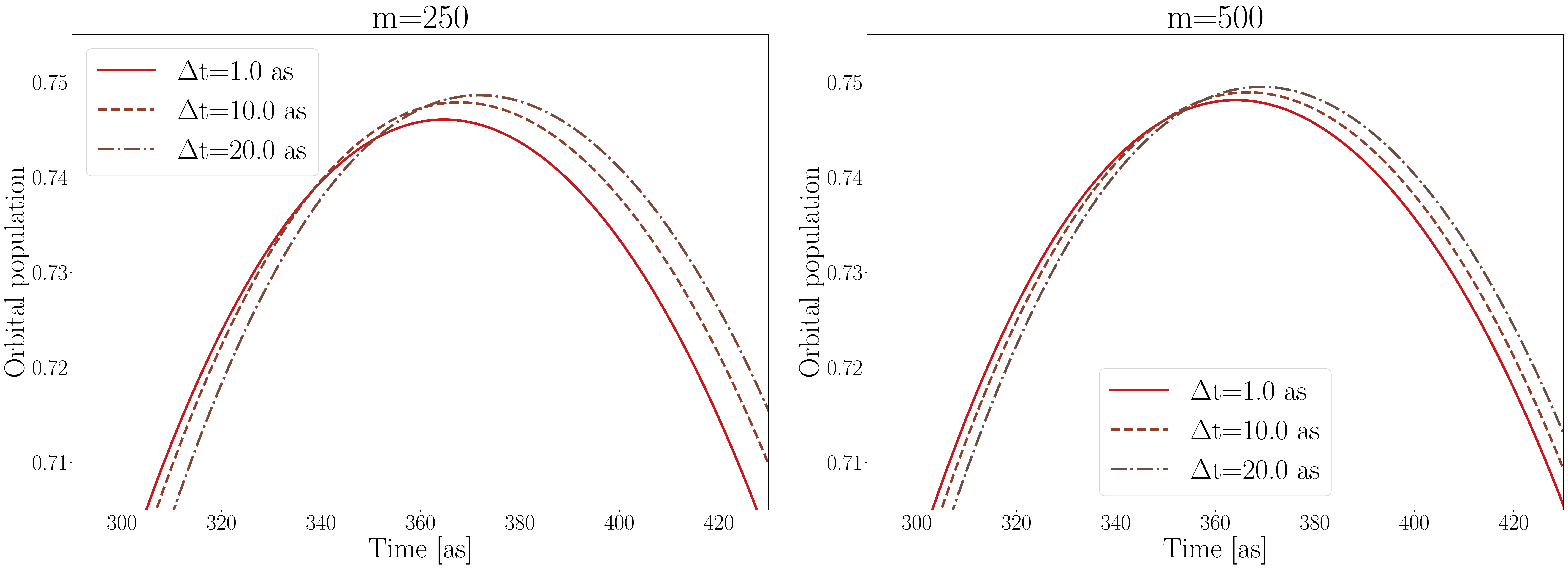}
	\caption{Time evolution of the $\pi^0$ orbital population of benzene following its ionization calculated with TD-DMRG(TS), $m$=250 (left panel) and $m$=500 (right panel), varying $\Delta t$ values, and for a CAS(14,14)}
	\label{fig:CAS14_ConvT}
\end{figure}

\noindent The TD-DMRG efficiency depends crucially on the stability of the integration algorithm with respect to the time step $\Delta t$.
As we show in Figure~S3 of the Supporting Information, the TD-DMRG(TS) population of the $\pi^0$ orbital, calculated with different integration time-steps, is qualitatively reproduced for time steps ranging from 1 to 20~as.
Such values are larger than the steps employed in the reference TD-CI work\cite{Evangelista2019_Adaptive-TDCI}, and comparable to the time-steps employed in other TD-DMRG algorithms.\cite{Frahm2019_TD-DMRG_Ultrafast}

\begin{table}
	\begin{tabular}{c|ccccc}
		\hline \hline
		$\Delta t$  &   1   &   5   &  10   &  15   &  20   \\
		\hline 
		$m$=250     &  365  &  366  &  368  &  370  &  372  \\
		$m$=500     &  365  &  365  &  367  &  368  &  369  \\
		\hline
		\hline
	\end{tabular}
	\caption{Half oscillation period ($t_{1/2}$) of the $\pi^0$ orbital population (in as) calculated with TD-DMRG(TS) with different $m$ and $\Delta t$ values.}
	\label{tab:MaxPop}
\end{table}

\noindent As displayed in Figure~\ref{fig:CAS14_ConvT} and in Table~\ref{tab:MaxPop}, the difference between the half-period oscillation time $t_{1/2}$ of the $\pi^0$ orbital obtained with different $m$ and $\Delta t$ values are minor.
For $m$=250, the difference between the $t_{1/2}$ value obtained with $\Delta t$=1~as and $\Delta t$=20~as is 7~as, and decreases to 4~as with $m$=500.
These values are smaller than the accuracy of time-resolved experiments, and to other effects that are neglected in our model, such as orbital relaxation and vibrational effects.
Nevertheless, this suggests that TD-DMRG supports large $\Delta t$ values, especially for large $m$ values. 
In fact, by increasing the bond dimension $m$, the truncation error decreases and larger $\Delta t$ values can be employed without increasing the Trotter error.

\noindent All results reported above are obtained with TD-DMRG(TS) because no electron dynamics is observed with TD-DMRG(SS), as we illustrate in Figure~S4 of the Supporting Information.
As discussed in Ref.~\citenum{McCulloch2015_Mixing}, TI-DMRG(SS) is prone to local minima convergence, and we observe the same effect here in TD-DMRG.
Let $i_\text{ion}$ be the site corresponding to the ionized orbital. 
By contracting all MPS tensors for sites $i \neq i_\text{ion}$, the MPS can be written as

\begin{equation}
\ket{\Psi_\text{GS}^{N}}
= \sum_{\sigma_{i_\text{ion}}} \sum_{a_{i_\text{ion}}}  \sum_{a_{i_\text{ion}+1}} 
M_{a_{i_\text{ion}} a_{i_\text{ion}+1}}^{\sigma_{i_\text{ion}}} 
\ket{\sigma_\text{ion}} \otimes \ket{a_{i_\text{ion}}^{(l)}}
\otimes \ket{a_{i_\text{ion}+1}^{(r)}} \, .
\label{eq:RightRenormalizedContraction}
\end{equation}

\noindent As discussed in Section~\ref{sec:theory}, both $\ket{a_{i_\text{ion}}^{(l)}}$ and $\ket{a_{i_\text{ion}+1}^{(r)}}$ can be assigned to an irreducible representation of the Hamiltonian symmetry group, \textit{i.e.} to a given number of alpha ($N^\alpha(a_{i_\text{ion}}^{(l)})$ and $N^\alpha(a_{i_\text{ion}+1}^{(r)})$) and beta ($N^\beta(a_{i_\text{ion}}^{(l)})$ and $N^\beta(a_{i_\text{ion}+1}^{(r)})$) electrons.
After applying the annihilation operator (see Eq.~(\ref{eq:SuddenIonization})), $M_{a_{i_\text{ion}} a_{i_\text{ion}+1}}^{\sigma_{i_\text{ion}}} \neq 0$ only if $N^\alpha(\sigma_{i_\text{ion}})=0$.
Hole migration from $\pi^0$ to $\pi^1$ will be observed only if blocks for which $N^{\alpha}(\sigma_{i_\text{ion}})=0$ become non-null along the propagation.
This is, however, not possible because the time evolution does not break the block structure and, therefore, the hole remains localized on the $i_\text{ion}$-th site.
We do not observe this effect with TD-DMRG(TS)\cite{Schollwoeck2011_Review-DMRG,McCulloch2015_Mixing} that propagates simultaneously two adjacent tensors and, therefore, breaks the site symmetry.
For this reason, if not otherwise specified, all results reported in the following are obtained with TD-DMRG(TS).

\subsection{Active space selection}

A migration half-period of 367~as is in reasonable agreement with the reference TD-CI data but is lower than the ADC(3) value of 467~as.\cite{Despre2015}
We simulate the ionization dynamics based on CAS(20,20) and CAS(26,26) to study the impact of the active space size on the ionization dynamics.

\begin{figure}[htbp!]
	\includegraphics[width=\textwidth]{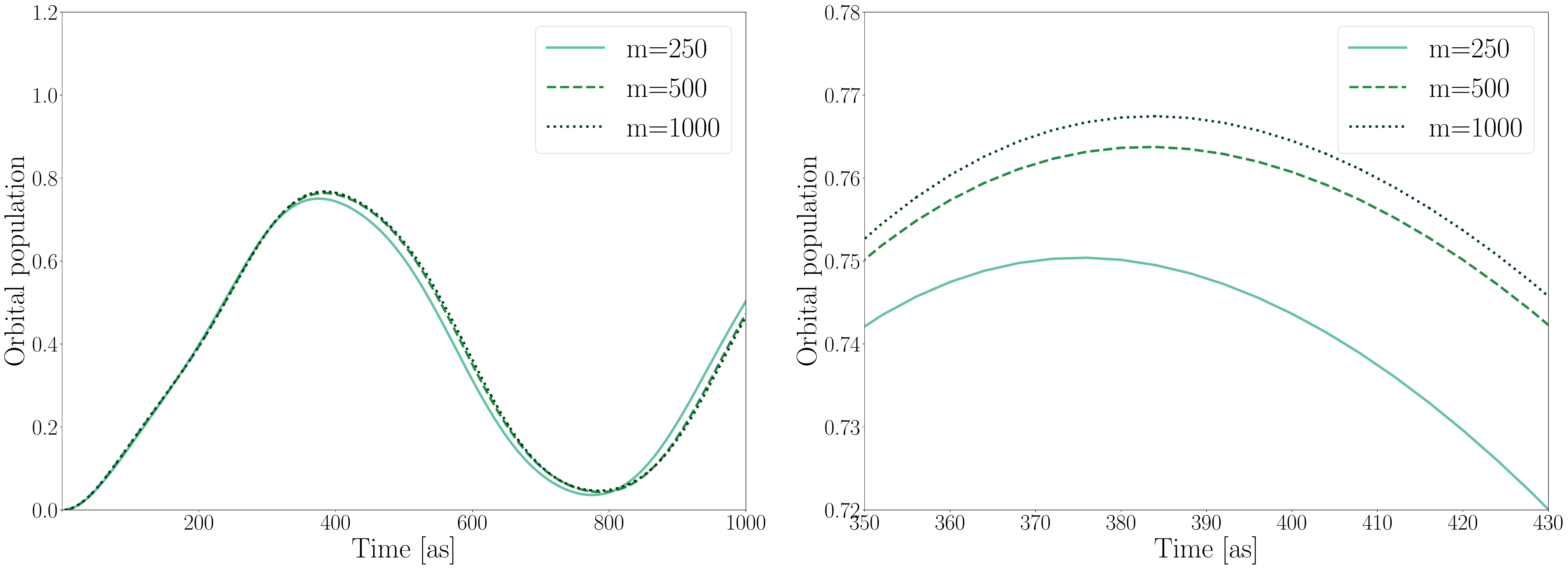}
	\caption{Time evolution of the $\pi^0$ orbital population calculated with TD-DMRG(TS), $m$=250, 500, and 1000, and $\Delta t$=4~as based on CAS(20,20). We report in the right panel an enlarged representation of the full propagation (reported in the left panel) between 350 and 430~as.}
	\label{fig:CAS20_mVarying}
\end{figure}

\noindent We report in Figure~\ref{fig:CAS20_mVarying} the $\pi^0$ population calculated with TD-DMRG(TS), varying $m$ values, $\Delta t$=4~as and based on CAS(20,20) (as we show in Figure~S5 of the Supporting Information, no differences are observed with smaller time steps).
We optimize the initial MPS with iTD-DMRG(TS), and we report the corresponding energy in Table~S1 of the Supporting Information.
As for the energy, also the time-dependent population of the $\pi^0$ orbital is converged with $m$=500, and minor differences are observed with $m$=1000.
The resulting oscillation period is 800~as, in better agreement with the experimental value.

\begin{figure}[htbp!]
	\centering
	\includegraphics[width=.75\textwidth]{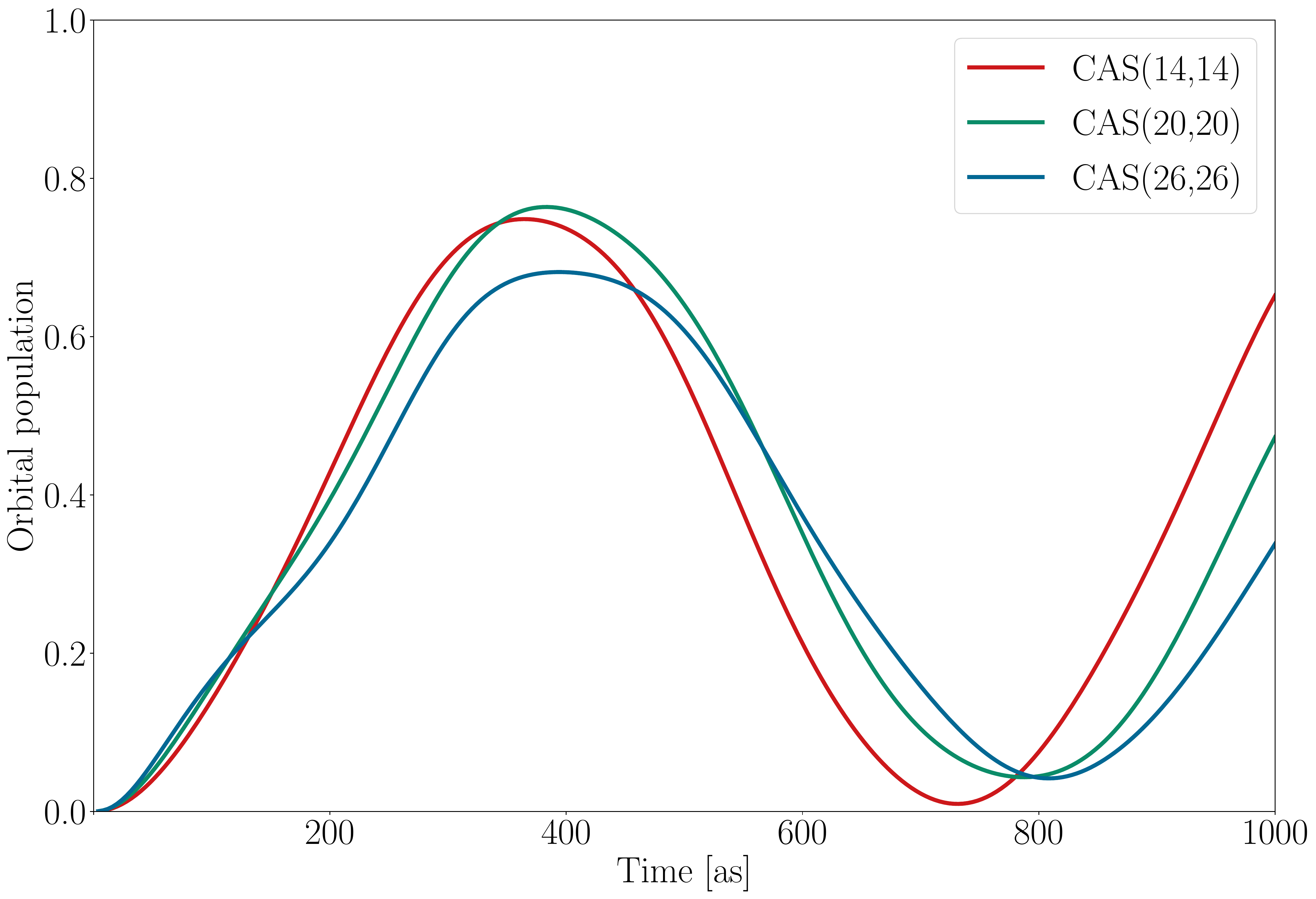}
	\caption{Time evolution of the $\pi^0$ orbital population of benzene following ionization calculated with TD-DMRG(TS), $m$=500, $\Delta t$=4~as and three different active spaces: CAS(14,14) (solid, red line), CAS(20,20) (solid, green line), and CAS(26,26) (solid, blue line).}
	\label{fig:Benzene_AllCAS}
\end{figure}

\noindent We report in Figure~\ref{fig:Benzene_AllCAS} the time evolution of the $\pi^0$ orbital population for CAS(14,14), CAS(20,20), and CAS(26,26).
The maximum $\pi^0$ orbital population consistently decreases by increasing the active space size.
The hole oscillation period for CAS(26,26) is 804~as, which is a significantly better match of the experimental data, of about 900~as, compared to CAS(14,14).
Note that our simplified model of benzene neglects vibrational effects and relies on a double-$\zeta$ basis set.
These limitations must be overcome to improve further the simulation accuracy.

\subsection{Time-dependent entanglement measures}

Quantum information-based metrics\cite{Legeza2003_OrderingOptimization} have been largely exploited to quantify correlation effects in molecular system and to automatize the selection of active spaces in multiconfigurational calculations.\cite{Stein2016_AutomatedSelection,Stein2016_DelicateBalance,Stein2017_AutoCAS-Chemia,Stein2017_MultireferenceQuantification,Stein2019_AutoCAS-Implementation}
Such metrics are based on the single- ($s_i(1)$) and two-orbital ($s_{ij}(2)$) entropy for orbitals $i$ and $j$, defined as

\begin{equation}
s_i(1) = -\sum_{\alpha=1}^4 w_{i,\alpha}^{(1)} \ln w_{i,\alpha}^{(1)} \,
\label{eq:SingleOrbitalEntropy}
\end{equation}
and

\begin{equation}
s_{ij}(2) = -\sum_{\alpha=1}^{16} w_{ij,\alpha}^{(2)} \ln w_{ij,\alpha}^{(2)} \, ,
\label{eq:TwoOrbitalEntropy}
\end{equation}
where $w_{i,\alpha}^{(1)}$ and $w_{i,\alpha}^{(2)}$ are the $\alpha$-th eigenvalues of the one- and two-orbital density matrix, respectively.
Large $s_i(1)$ values identify strongly correlated orbitals in time-independent wave functions.\cite{Stein2016_AutomatedSelection}
Here we will show that the same ideas can be extended to TD-DMRG and track changes in the multireference character of the time-dependent wave function.

\begin{figure}[htbp!]
	\includegraphics[width=.49\textwidth]{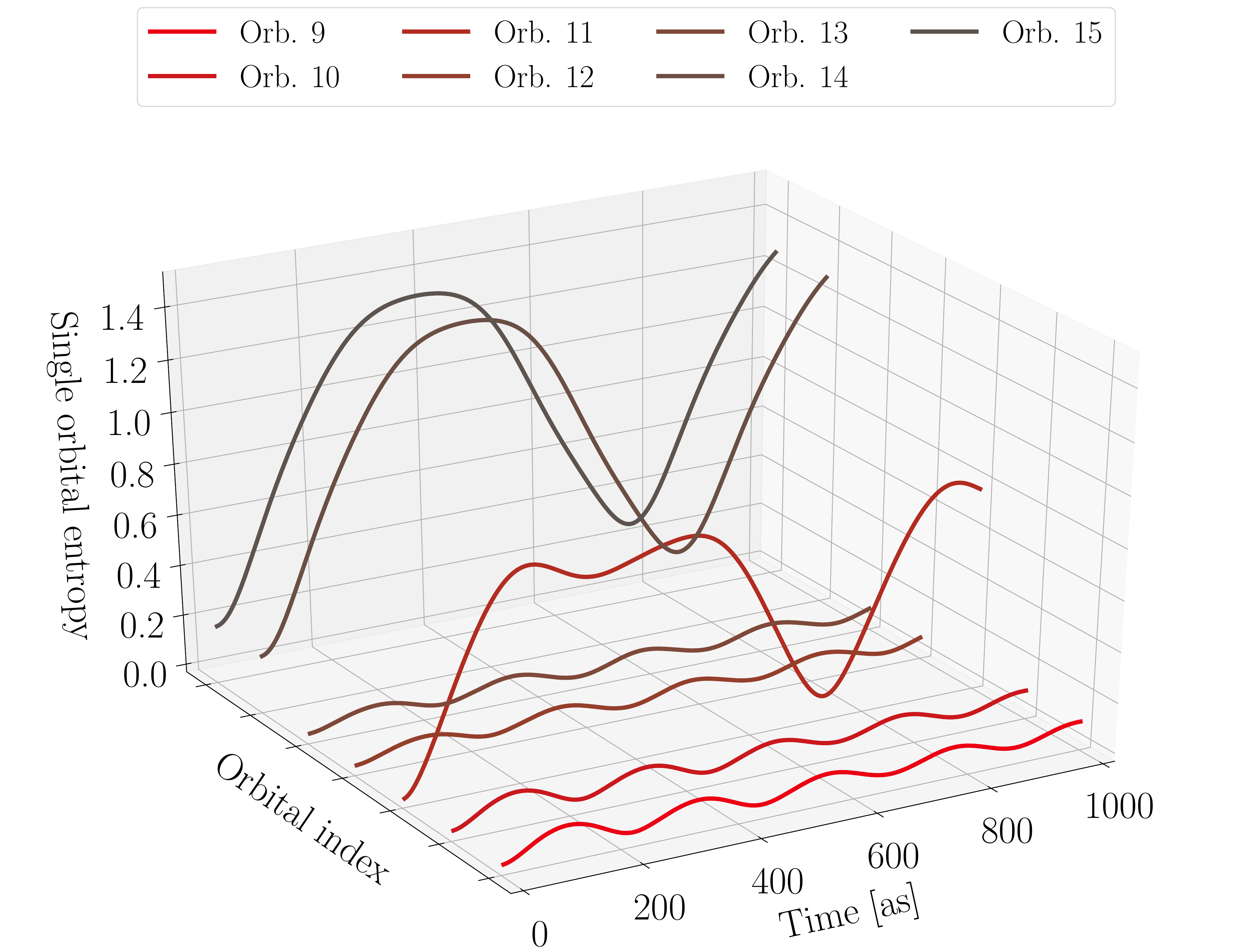}
	\includegraphics[width=.49\textwidth]{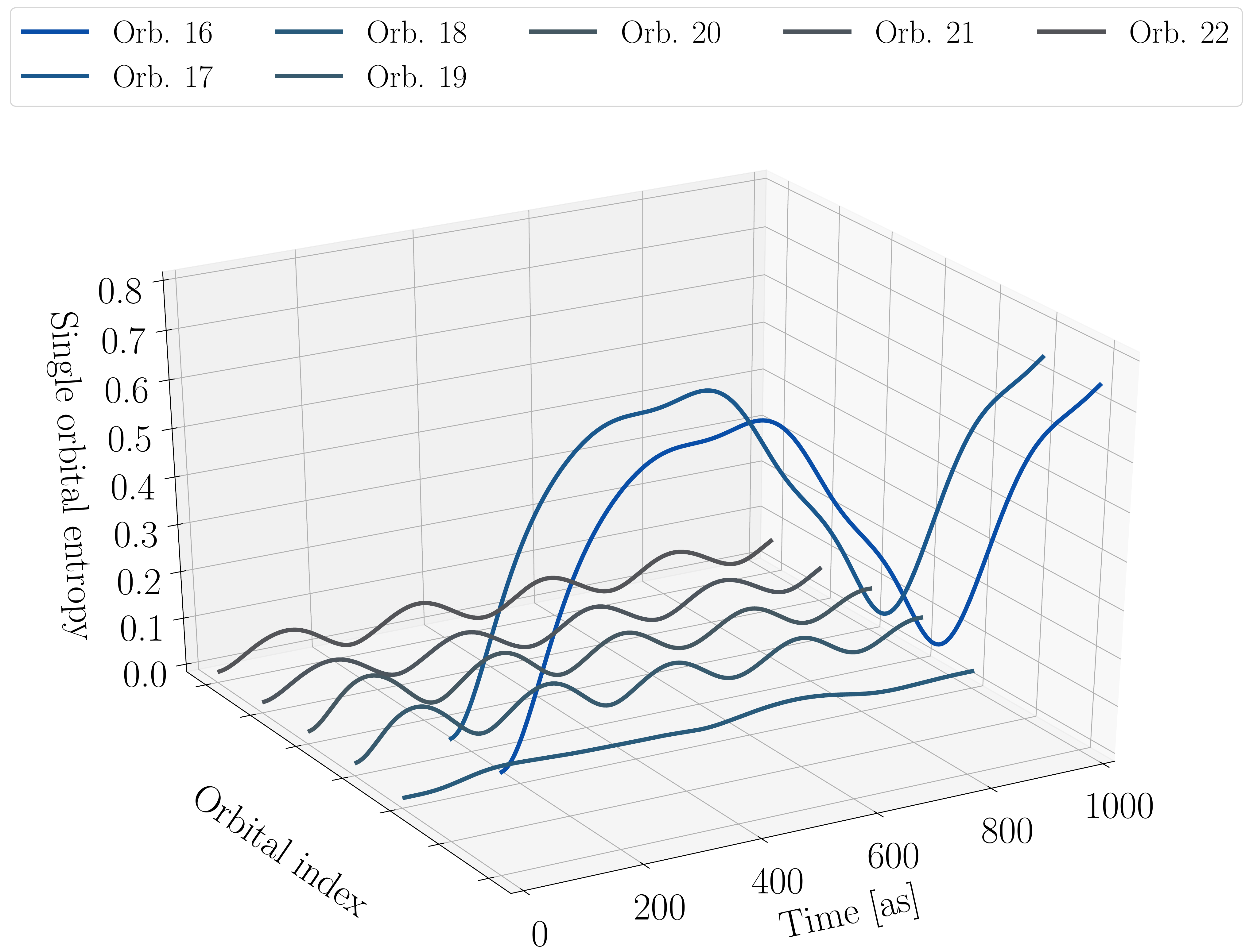}
	\caption{Time evolution of the single-orbital entropy $s^{(1)}$ for the occupied (left panel) and virtual (right panel) orbitals of benzene calculated with TD-DMRG(TS), CAS(14,14), $m$=500, and $\Delta t$=4~as. Orbitals are numbered as in Figure~S1 of the Supporting Information.}
	\label{fig:Entanglement_CAS14-14}
\end{figure}

\noindent We report in Figure~\ref{fig:Entanglement_CAS14-14} the time evolution of $s^{(1)}$ for benzene calculated with TD-DMRG(TS) and based on CAS(14,14).
As expected, the largest $s^{(1)}$ variation is observed for the ionized orbital (labeled as 11 in Figure~\ref{fig:Entanglement_CAS14-14}) and for the $\pi^1$ orbitals that are involved in the hole migration, labeled as 14 and 15.
As shown in the right panel of Figure~\ref{fig:Entanglement_CAS14-14}, the single-orbital entropy is large also for orbitals 16 and 17, \textit{i.e.} the two degenerate lowest-energy virtual $\pi$ orbitals, which are therefore involved in the hole-migration process.
The remaining orbitals have a single-orbital entropy smaller than 0.1 at for all $t$ values.

\begin{figure}[htbp!]
	\includegraphics[width=.49\textwidth]{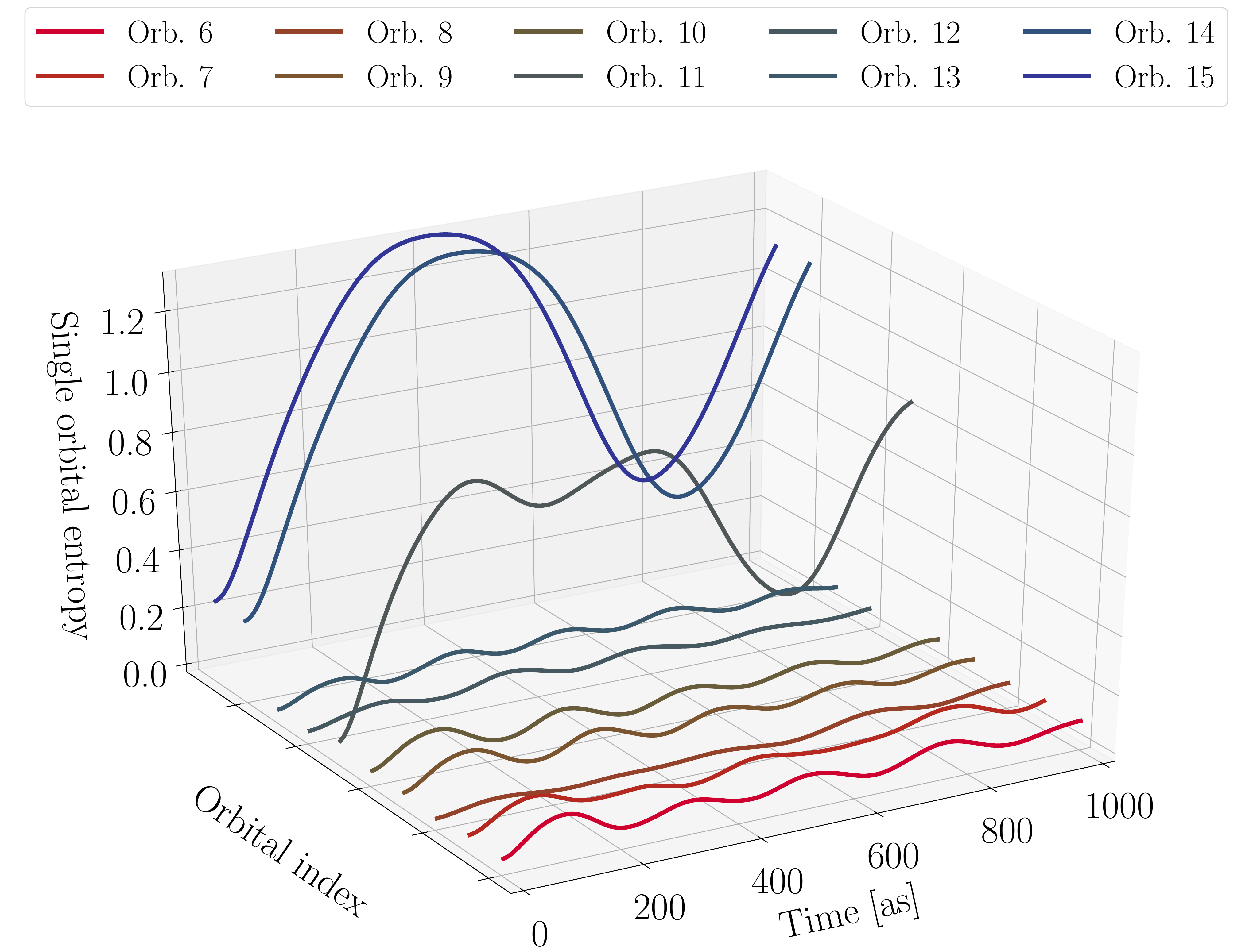}
	\includegraphics[width=.49\textwidth]{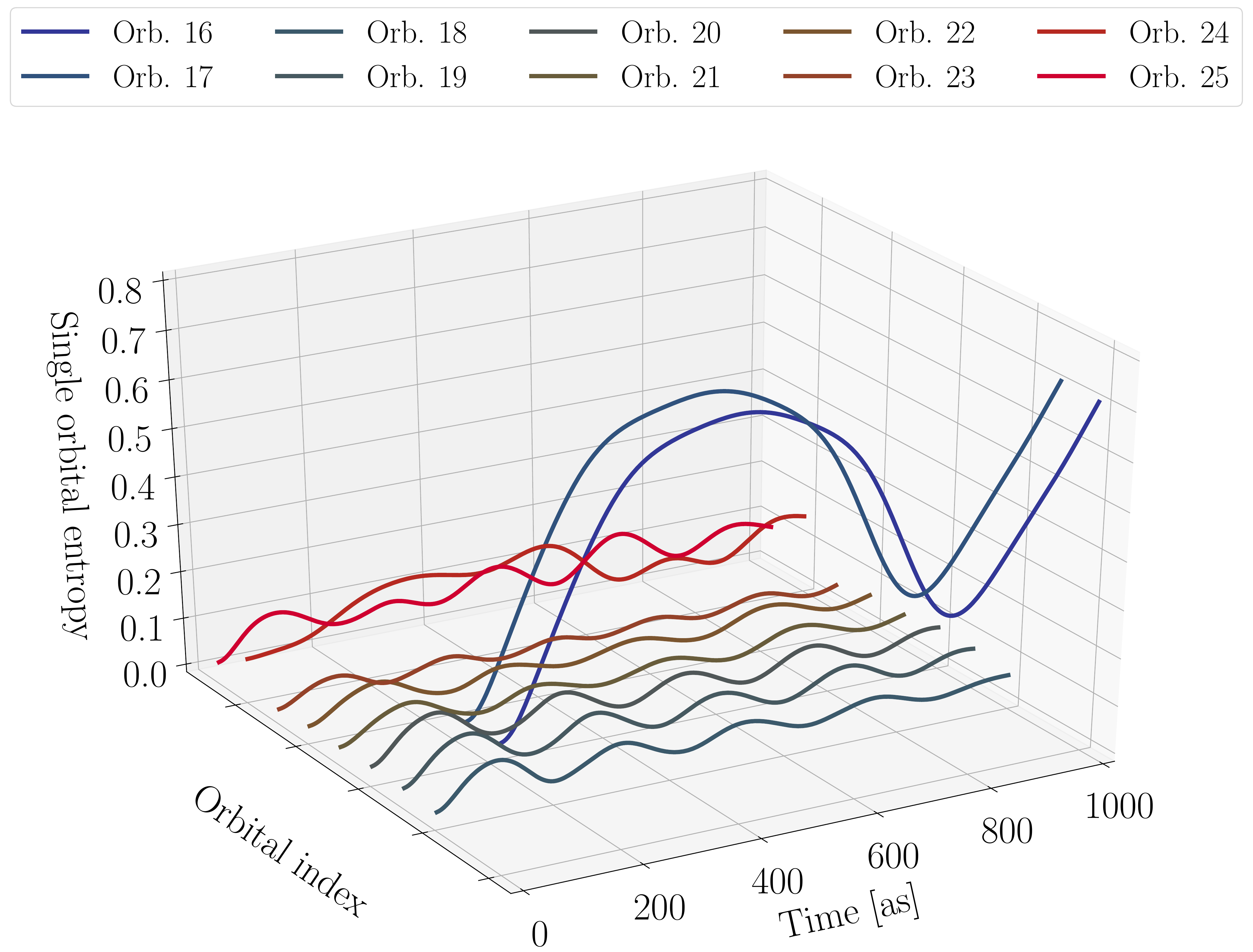}
	\caption{Time evolution of the single-orbital entropy $s^{(1)}$ for the occupied (left panel) and virtual (right panel) orbitals of benzene calculated with TD-DMRG(TS) based on CAS(20,20), $m$=500, and $\Delta t$=4~as. Orbitals are numbered as in Figure~S1 of the Supporting Information.}
	\label{fig:Entanglement_CAS20-20}
\end{figure}

\noindent We report in Figure~\ref{fig:Entanglement_CAS20-20} the same plot obtained with the larger CAS(20,20).
Comparison with Figure~\ref{fig:Entanglement_CAS14-14} shows that the single-orbital entropy of orbital 24, which was not included in the CAS(14,14), is larger than 0.2 between 400 and 600~as.
The selection criteria implemented in the \texttt{AutoCAS} algorithm\cite{Stein2016_AutomatedSelection} would define this orbital as strongly correlated and, therefore, it is expected to give the largest contribution to the increase of the half-oscillation period.
Note also that the maximum $s_{24}(1)$ value is observed between 400 and 600~as, \textit{i.e.} the $t$ values for which the largest difference between CAS(14,14) and CAS(20,20) is observed.
As we show in Figure~S6 of the Supporting Information, the $s_1$ value of the orbitals included in the CAS(26,26) and not in CAS(20,20) is below 0.1 in all cases.
This agrees with the observation that this additional increase of the active space size does not further change the population dynamics (see Figure~4).
This also suggests that changes in the multireference character of the time-dependent wave function can be monitored based on the orbital entropies.
Even though here we analyze the multireference character of the wave function \textit{a posteriori}, it would be possible to apply the \texttt{AutoCAS} algorithm\cite{Stein2019_AutoCAS-Implementation} to automatically define an active space at each time step and construct the MPS wave function only with the resulting active orbitals.
Implementing, in practice, such an algorithm would require deriving the DMRG parallel of the time-dependent complete active space self-consistent field method,\cite{Sato2013_TD-CAS} a task that goes beyond the scopes of the present work.
For this reason here we analyze the time evolution of the orbital entropies only qualitatively.

\section{Electronic absorption spectra from TD-DMRG}
\label{sec:AbsorptionSpectra}

Besides enabling the simulation of non-equilibrium electron dynamics, real-time electronic-structure methods offer an alternative to TI algorithms to calculate absorption spectra.
This approach has been applied to semiempirical methods,\cite{Ghosh2017_TD-Semiempirical,Gagliardi2019_TD-Semiempirical} RT-TD-DFT\cite{Lopata2011_RealTimeTDDFT,Cramer2015_RTTDDFT-Comparison,Repisky2015_RealTime4c,Lopata2016_Absorption-RTTDDFT,Kummel2018_RT-TDDFT} and TD-CC.\cite{Nascimento2016_TDCC}
The TD route to electronic spectra is particularly appealing for X-ray spectra\cite{Lopata2012_RTTDDFT-Core,Ruud2015_XRay-Relativistic-RT,DePrince2017_CoreCC-RealTime,Li2018_XRay-RealTime} that are difficult to target with TI-based methods due to the need of optimizing high-energy excited states.
Here, we apply this idea to TD-DMRG.
The electronic absorption cross section $\sigma(E)$ is expressed, in the time-domain, as\cite{Schuurman2018_XRay}

\begin{equation}
\sigma(E) = \frac{E}{3\pi} \int_{-\infty}^{+\infty} \, \text{d}t
\ExV{\Psi_0}{\hat{\mu} \, e^{-\mathrm{i} \mathcal{H} t} \, \hat{\mu}}{\Psi_0} 
= \frac{E}{3\pi} \int_{-\infty}^{+\infty} \, \text{d}t \; C(t) \, ,
\label{eq:Autocorr}
\end{equation}
where $\ket{\Psi_0}$ is the electronic ground state wave function and $\hat{\mu}$ is the dipole operator expressed, in second quantization, as

\begin{equation}
\hat{\mu} = \sum_{pq=1}^L \ExV{\phi_p}{\mu}{\phi_q} \, a_p^\dagger a_q \, ,
\label{eq:DipoleSQ}
\end{equation}
where $\bm{\phi} = \{ \phi_1 , \ldots , \phi_L \}$ is the reference orbital basis. 
The autocorrelation function $C(t)=\ExV{\Psi_0}{\hat{\mu} \, e^{-\mathrm{i} \mathcal{H} t} \, \hat{\mu} }{\Psi_0}$ can be calculated from TD-DMRG by 1) encoding $\hat{\mu} \ket{\Psi_0}$ as an MPS and 2) propagating the resulting wave function with the time-independent Hamiltonian $\mathcal{H}$ of Eq.~(\ref{eq:SQEleHams}).
The MPO representation of the dipole operator $\hat{\mu}$ reads

\begin{equation}
\hat{\mu} = \sum_{\bm{\sigma}, \bm{\sigma}'} \sum_{b_1,\ldots,b_{L-1}} 
\mu_{1,b_1}^{\sigma_1,\sigma_1'} \mu_{b_1,b_2}^{\sigma_2,\sigma_2'}
\cdots \mu_{b_{L-1},1}^{\sigma_L,\sigma_L'} \ket{\bm{\sigma}} \bra{\bm{\sigma}'} \, .
\label{eq:MPO_Dipole}
\end{equation}

\noindent $\hat{\mu}$ is a one-body operator and, therefore, can be easily encoded as MPO.
Combining Eq.~(\ref{eq:MPO_Dipole}) and Eq.~(\ref{eq:MPS_Definition}), $\hat{\mu} \ket{\Psi_0}$ can be written as

\begin{equation}
\begin{aligned}
\hat{\mu} \ket{\Psi_0} &= \sum_{\bm{\sigma},\bm{\sigma}'} 
\sum_{b_1,\ldots,b_{L-1}} \sum_{a_1',\ldots,a_{L-1}'} 
\mu_{1,b_1}^{\sigma_1,\sigma_1'} \mu_{b_1,b_2}^{\sigma_2,\sigma_2'} 
\cdots \mu_{b_{L-1},1}^{\sigma_L,\sigma_L'} 
M_{1,a_1'}^{\sigma_1'} M_{a_1' a_2'}^{\sigma_2'} 
\cdots M_{a_{L-1}',1}^{\sigma_L'} \ket{\bm{\sigma}'} \\
&= \sum_{\bm{\sigma}} \sum_{a_1,\ldots,a_{L-1}} 
D_{1,a_1}^{\sigma_1} D_{a_1, a_2}^{\sigma_2} 
\cdots D_{a_{L-1},1}^{\sigma_L} \ket{\bm{\sigma}} 
\end{aligned}
\label{eq:MPO_on_MPS}
\end{equation}
where the index $a_i$ takes all possible values of the product $a_i' \times b_i$, and the $\bm{D}^{\sigma_i}$ tensors are defined as

\begin{equation}
D_{a_{i-1},a_i}^{\sigma_i} \equiv D_{\left( a_{i-1}'b_{i-1},a_i' b_i \right)}^{\sigma_i}
= \sum_{\sigma_i'} \mu_{b_{i-1},b_i}^{\sigma_i,\sigma_i'}
M_{a_{i-1}',a_i'}^{\sigma_i'} \, .
\label{eq:BiggerTensor}
\end{equation}

\noindent Eq.~(\ref{eq:BiggerTensor}) highlights that, for site $i$, the bond dimension of the MPS representation of $\hat{\mu} \ket{\Psi_0}$ is $b_i$ times larger than that of $\ket{\Psi_0}$.
To keep the bond dimension fixed, we truncate the MPS representation of $\hat{\mu} \ket{\Psi_0}$ before the propagation starts.

\subsection{Electronic absorption spectra of decacene}

We simulate the absorption spectrum of decacene with TD-DMRG and compare our results to reference TD-CI data\cite{Peng2018_TDCI} based on CAS(10,10), the 6-31G* basis, and natural orbitals obtained from a CI singles calculation.
We begin with the same CAS as in Ref.~\citenum{Peng2018_TDCI}, and then enlarge it and monitor the convergence of the absorption spectrum.

\begin{figure}[htbp!]
	\centering
	\includegraphics[width=.8\textwidth]{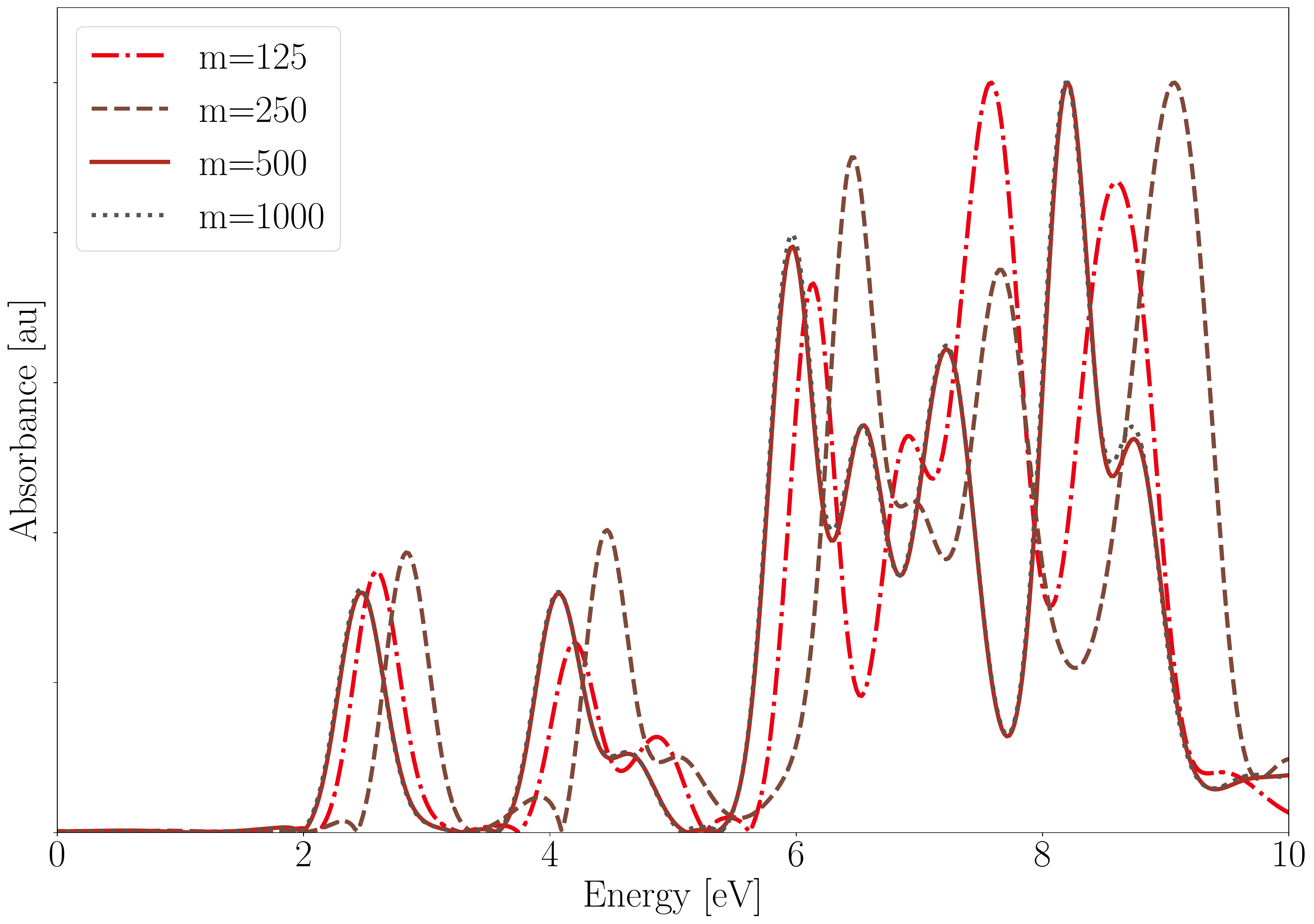}
	\caption{Absorption spectrum of decacene obtained with TD-DMRG(TS), CAS(10,10), $\Delta$t=20~as, and different $m$ values.}
	\label{fig:Decacene_CAS10-10_Spectrum}
\end{figure}

\noindent After optimizing the ground-state wave function with iTD-DMRG(TS) (we report in Table~S2 and in Figure~S7 of the Supporting Information the corresponding energy), we obtain the initial state of the propagation by applying the $y$ component of the dipole operator onto the MPS.\cite{Peng2018_TDCI} 
We calculate the Fourier transform of the autocorrelation function numerically after an overall propagation time of 100~fs, and with a window function of 20~fs.
We simulate the spectra with TD-DMRG(TS) because, as discussed in Section~3.1, TD-DMRG(SS) does not reproduce the correct dynamics for instantaneous perturbations.
We report in Figure~\ref{fig:Decacene_CAS10-10_Spectrum} the absorption spectra calculated for $m$ values ranging between 125 and 1000.
A bond dimension $m$=500 is required to obtain a fully-converged band shape and, therefore, the $m$ convergence of the absorption spectrum is slower than that of the ground state energy.
The initial state for the propagation obtained by applying $\hat{\mu}$ on the ground-state MPS is, in fact, the superposition of multiple excited states and a larger bond dimension is required to represent the resulting wave function.
We show in Figure~S8 of the Supporting Information that the overall band shape obtained with TD-DMRG(TS) based on CAS(10,10), $m$=500, and different integration schemes remains unchanged for time steps that range from 20 to 80~as.
The tangent-space TD-DMRG algorithm is, therefore, more stable than the Runge-Kutta integrator of Ref.~\citenum{Peng2018_TDCI} that diverges for time steps larger than 20~as.

\begin{figure}[htbp!]
	\centering
	\includegraphics[width=.8\textwidth]{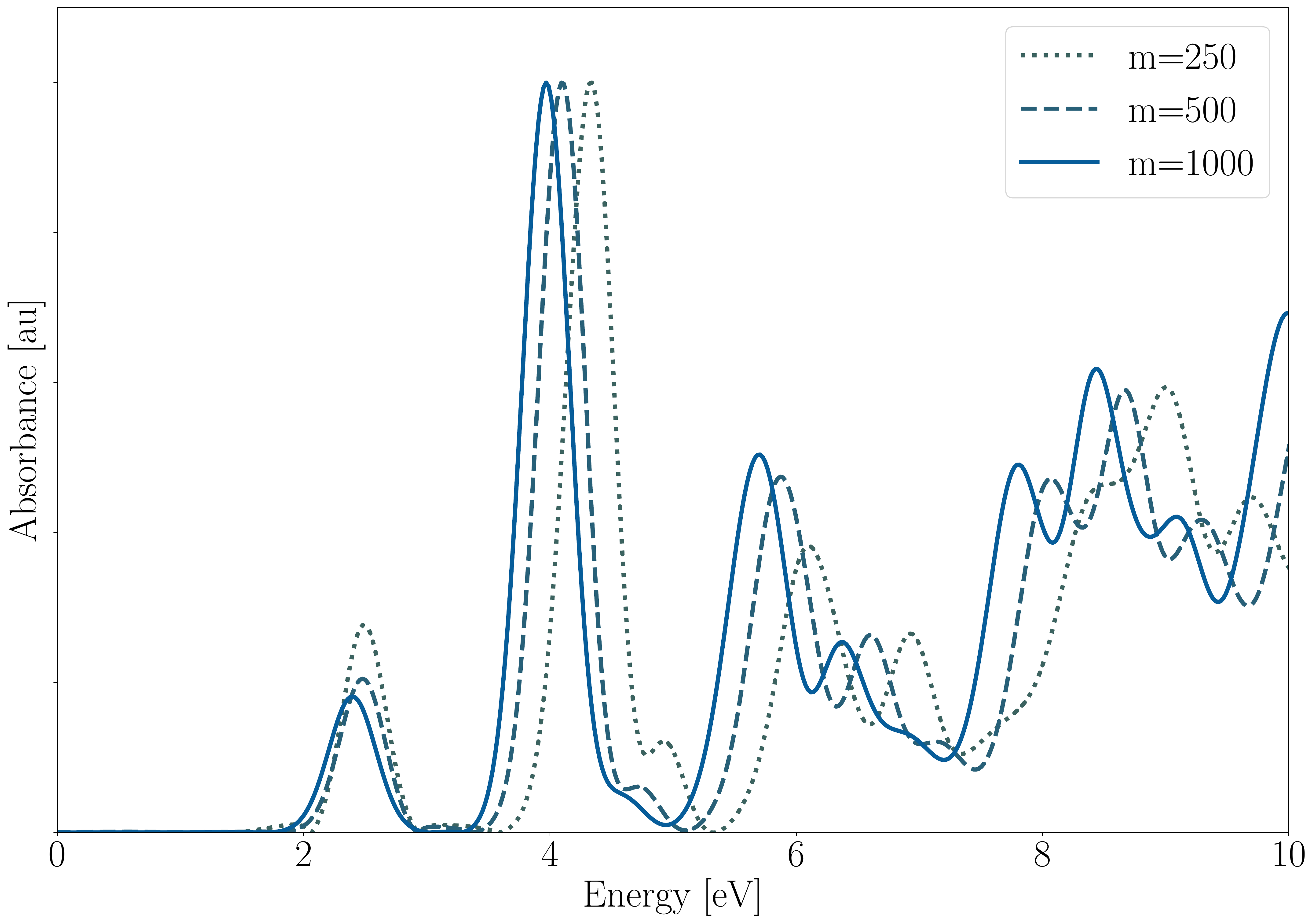}
	\caption{Absorption spectrum of decacene obtained with TD-DMRG(TS) based on CAS(14,14) for different $m$ values. $\Delta t$=20~as in all cases.}
	\label{fig:Decacene_CAS14_Spectrum}
\end{figure}

\noindent Despite being converged, the $m$=1000 spectrum does not match the reference data of Ref.~\citenum{Peng2018_TDCI} that relies on natural orbitals, whereas we construct the MPS wave function from the canonical ones.
We show in Figure~\ref{fig:Decacene_CAS14_Spectrum} that, by increasing the active space size to CAS(14,14), the agreement with the TD-CI spectrum improves remarkably.
As expected, larger $m$ values are required to converge the band shape for a larger CAS, and slight differences are observed between the $m$=500 and $m$=1000 results.
However, the main difference between the $m$=500 and $m$=1000 spectra is an overall shift of the higher-energy bands, while their relative intensity remains nearly unchanged.

\begin{figure}[htbp!]
	\centering
	\includegraphics[width=.8\textwidth]{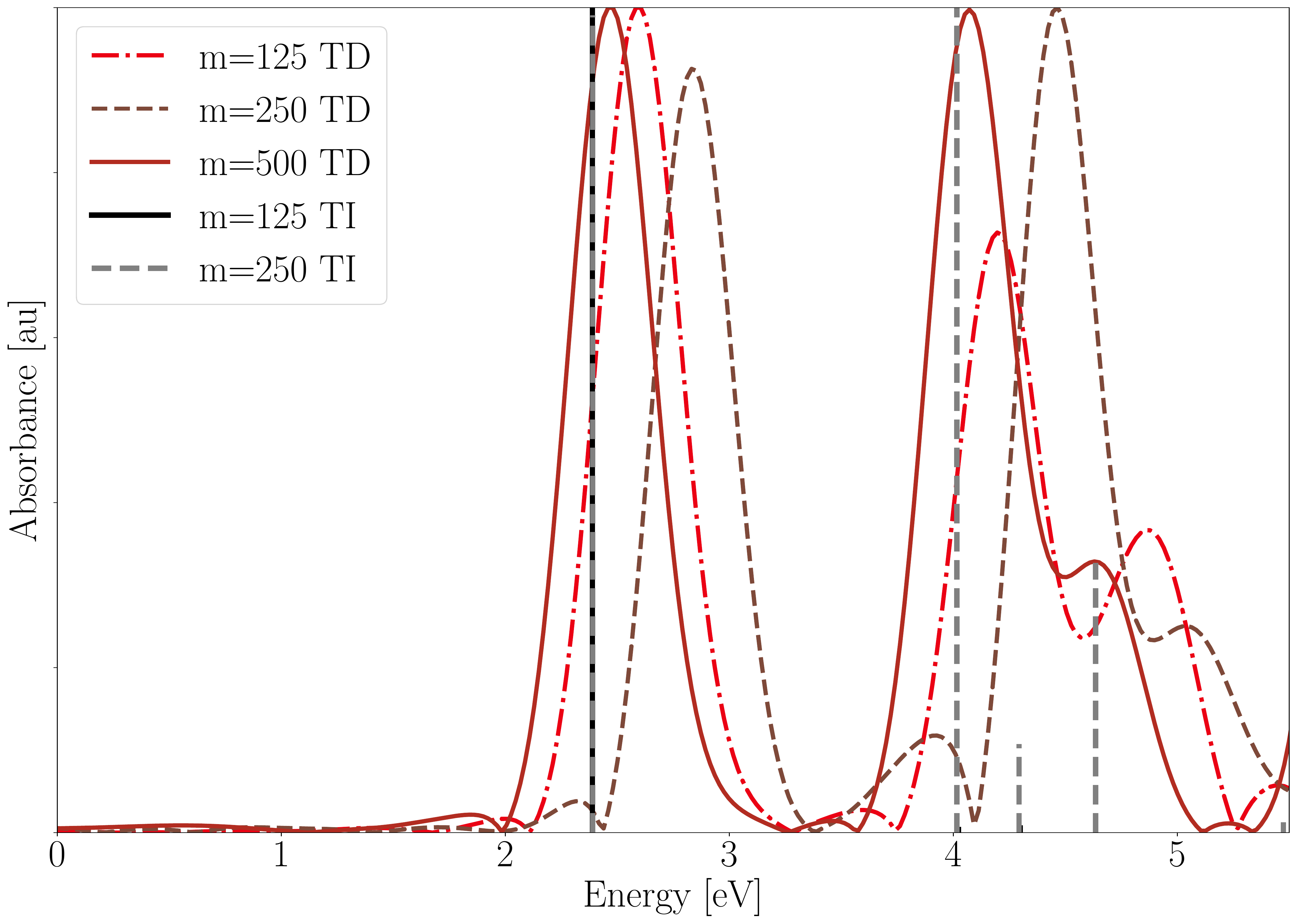}
	\caption{Absorption spectrum of decacene obtained with a TD-DMRG(TS) (red lines) and with TI-DMRG(TS) (black vertical bars) based on CAS(10,10) for different $m$ values and $\Delta t$=20~as.}
	\label{fig:Decacene_TIvsTD_Spectrum}
\end{figure}

\noindent We compare in Figure~\ref{fig:Decacene_TIvsTD_Spectrum} the absorption spectrum of decacene obtained with CAS(10,10) with TD-DMRG(TS) and TI-DMRG(TS).
We report the individual excitation energies and their respective transition dipole moments calculated with TI-DMRG in Table~S3 of the Supporting Information.
The TD-DMRG spectrum is already qualitatively converged with $m$=125, even if all bands are blue-shifted by about 0.5~eV, while the TI-DMRG spectrum obtained with the same bond dimension is qualitatively wrong and dominated by a single, intense band at about 2.5~eV.
Conversely, as we show in Table~S3, TI-DMRG excitation energies are nearly converged with $m$=125, and therefore the convergence of the transition dipole moment with $m$ is slower than for the energy.
Note that both the position and the intensity of all bands in the TI-DMRG spectrum obtained with $m$=250 are correct, while a blue shift is still observed for the TD-DMRG spectrum obtained with the same $m$ value, especially in the high-energy region.
We rationalize the slower convergence of TD-DMRG with $m$ based on the fact that TD-DMRG obtains the spectrum from a single propagation, while TI-DMRG optimizes each state separately.
The time evolving MPS must encode simultaneously all excited states and a larger bond dimension $m$ is required to represent such a complex wave function.
We recall that state-specific TI-DMRG optimizes the excited state MPS with a constrained optimization scheme\cite{Keller2015_MPS-MPO-SQHamiltonian} that is intrinsically sequential and becomes extremely inefficient for high-energy states.
Energy-specific DMRG variants\cite{Dorando2007_TargetingExcitedStates,Devakul2017,Yu2017_ShiftAndInvertMPS,Baiardi2019_HighEnergy-vDMRG} are, in principle, more efficient, but they still suffer from the limitation that each excited state must be optimized separately.
TD-DMRG is an efficient solution to these problems because the absorption spectrum in a given energy range is calculated from a single simulation.

\section{Calculation of dynamical response properties}
\label{sec:ResponseProperties}

\subsection{Time-dependent finite difference method}

The calculation of frequency-dependent molecular properties is a paradigmatic example of a problem that can be targeted by time-independent schemes,\cite{Helgaker2012_ReviewProperties} but that is conveniently solved with real-time electronic-structure methods.\cite{Saalfrank2007_Properties-RealTime,Ding2013_Polarizability-RealTimeDFT,Li2018_RealTime-GUGA_CI,Li2020_RTES-Review}
The time-dependent dipole $\bm{\mu}(t)$ of a molecule interacting with a monochromatic electromagnetic field with frequency $\omega_I$ and maximum electric field $\bm{E}_0$ can be expressed as a superposition of terms oscillating at frequencies that are integer multiples of $\omega_I$.
By including only the first and second harmonics, the $i$-th component of the dipole reads

\begin{equation}
\mu_i(t) = \sum_{j \in \{x,y,z\}} \mu_{ij}(t) E_{0,j} 
+ \sum_{j,k \in \{x,y,z\}} \mu_{ijk}(t) E_{0,j} E_{0,k} \, ,
\label{eq:TD_Dipole}
\end{equation}
with

\begin{equation}
\begin{aligned}
\mu_{ij}(t)  & = \alpha_{ij}(-\omega_I; \omega_I) \cos(\omega_I t) \\
\mu_{ijk}(t) & = \frac{1}{4}  \left( \beta_{ijk}(-2\omega_I; \omega_I, \omega_I) \cos(2\omega_I t) 
+ \beta_{ijk}(0; \omega_I, -\omega_I) \right) \, ,
\end{aligned}
\label{eq:FunctionalForm}
\end{equation}
where $\bm{\alpha}(-\omega_I, \omega_I)$ is the dynamical polarizability tensor, and $\bm{\beta}(-2\omega_I; \omega_I, \omega_I)$ and $\bm{\beta}(0; \omega_I, -\omega_I)$ are the first hyperpolarizability tensors.
We will refer in the following to $\mu_{ij}(t)$ and $\mu_{ijk}(t)$ as first- and second-order dipole response, respectively.
Low-order properties are routinely obtained from time-dependent perturbation theory,\cite{Olsen1992_MCSCF-ResponseTheory,Koch1994_FrequencyDependent,Agren2002_TDHF-DFT,Helgaker2012_ReviewProperties} but this route becomes less efficient for second- or higher-order response properties.
Saalfrank and co-workers\cite{Saalfrank2007_Properties-RealTime} and, later, Li and co-workers\cite{Ding2013_Polarizability-RealTimeDFT,Li2018_RealTime-GUGA_CI} designed a time-dependent finite difference method to calculate molecular properties based on quantum dynamics simulations.
The electronic ground state $\ket{\Psi_0}$ is first propagated under the action of a perturbing oscillating electric field ($\bm{E}_0 \cos(\omega_I t)$) for a given $\bm{E}_0$ value.
The propagation is repeated for field values $-\bm{E}_0$, $2\bm{E}_0$, and $-2\bm{E}_0$, and the first-order dipole response $\mu_{ij}(t)$ is calculated with the following finite-difference expression\cite{Ding2013_Polarizability-RealTimeDFT}

\begin{equation}
\mu_{ij}(t) = \frac{  8 \left( \mu_i(t, E_j) - \mu_i(t, -E_j) \right)
	- 12 \left( \mu_i(t, 2E_j) - \mu_i(t, -2E_j) \right) }{12 E_j} \, .
\label{eq:FiniteDifference}
\end{equation}

\noindent The first-order dipole response $\mu_{ij}(t)$ obtained from Eq.~(\ref{eq:FiniteDifference}) is fitted to the functional form of Eq.~(\ref{eq:FunctionalForm}) to calculate $\bm{\alpha}$.
Eq.~(\ref{eq:FiniteDifference}) can be extended to first and second hyperpolarizabilities and, in the latter case, the propagation must be repeated also for a field of $3\bm{E}_0$ and $-3\bm{E}_0$.\cite{Ding2013_Polarizability-RealTimeDFT}
This scheme has been applied so-far to RT-TD-DFT\cite{Ding2013_Polarizability-RealTimeDFT} and TD-CI,\cite{Li2018_RealTime-GUGA_CI} and we apply it here to the calculation of high-order molecular response properties with DMRG beyond the linear-response approximation.\cite{Dorando2009_AnalyticalResponseFunction,Nakatani2014_LinearResponseDMRG}
We keep only the dipole term in the light-matter interaction, so that the time-dependent Hamiltonian $\mathcal{H}(t)$ reads:

\begin{equation}
\mathcal{H}(t) = \mathcal{H}_\text{ele} - \bm{\mu} \cdot \bm{E}(t) \,
\label{eq:TDPerturbation}
\end{equation}
where $\bm{E}(t)$ is the time-dependent electric field and $\bm{\mu}$ is defined in Eq.~(\ref{eq:DipoleSQ}).
Eq.~(\ref{eq:TDPerturbation}) can be encoded as an MPO with time-dependent $\bm{W}^{\sigma_i,\sigma_i'}$ tensors based on Eq.~\ref{eq:MPO_Dipole}.

\subsection{TD-DMRG polarizabilities of BH}

We calculate the dynamical (hyper)polarizability of BH with the algorithm introduced above and compare our results to FCIQMC\cite{Booth2018_FCIQMC-Properties} and FCI\cite{Olsen1998_FCI-Polarizabilities} reference data.
Note that FCIQMC data are obtained with static response theory\cite{Blunt2015_Krylov-FCIQMC,Booth2018_FCIQMC-Properties} and, therefore, will reproduce the TD-DMRG data only in the $\omega_I \rightarrow 0$ limit.
We optimize the TD-DMRG parameters (time-step $\Delta t$, field $E_0$) on the cc-pVDZ basis, which includes 19 orbitals for BH, and apply the resulting optimal parameter set to the larger aug-cc-pVDZ basis set, which includes 32 orbitals.
We simulate the electron dynamics with TD-DMRG(SS) because the perturbation is not instantaneous, as it was for the previous two applications, and is instead switched on adiabatically.\cite{Ding2013_Polarizability-RealTimeDFT}

\begin{table}[htbp!]
	\begin{tabular}{cc|cc|cc}
		\hline \hline
		& \multirow{2}{*}{$E_0$} & \multicolumn{2}{c|}{$\Delta t$=10~as}   
		& \multicolumn{2}{c}{$\Delta t$=20~as} \\
		&        &  $m$=125   &   $m$=250   &   $m$=125  &    $m$=250 \\
		\hline
		\multirow{3}{*}{EMR2}
		& 0.003  &  21.9736   &  21.9730    &  22.0096   &   21.9730   \\
		& 0.006  &  21.9739   &  21.9732    &  21.9711   &   21.9728   \\
		& 0.010  &  21.9758   &  21.9760    &  21.9730   &   21.9750   \\
		\hline
		\multirow{3}{*}{CF4}
		& 0.003  &  21.9799   &  21.9716    &  21.9928   &   21.9977   \\
		& 0.006  &  21.9764   &  21.9813    &  21.9968   &   21.9974   \\
		& 0.010  &  21.9785   &  21.9856    &  21.9987   &   21.9988   \\
		\hline \hline
	\end{tabular}
	\caption{First dynamical polarizability (in atomic units) $\alpha_{xx}(-\omega_I, \omega_I)$ of BH calculated at $\omega_I$=488~nm with EMR2-TD-DMRG(SS)/cc-pVDZ and CF4-TD-DMRG(SS)/cc-pVDZ for varying integration time-steps $\Delta t$, bond dimension $m$, and electric field $E_0$.}
	\label{tab:BH_Properties_DZ}
\end{table}

\noindent We run both EMR2-TD-DMRG and CF4-TD-DMRG propagations starting from the MPS optimized with iTD-DMRG (we report the corresponding energies in Table~S4).
All results reported in the following are obtained with the spin-adapted TD-DMRG algorithm.
We report in Table~\ref{tab:BH_Properties_DZ} the dynamical polarizability $\alpha_{xx}(-\omega_I, \omega_I)$ of BH (where $x$ is the molecular axis of BH) obtained for different $m$ and $E_0$ values.
Following Ref.~\citenum{Olsen1998_FCI-Polarizabilities}, we set the incident frequency to 0.093368~a.u. (488~nm).
The calculated $\alpha_{xx}(-\omega_I, \omega_I)$ value is stable with respect to changes of all relevant parameters: variations below 10$^{-2}$~a.u. are observed between $E_0$=0.003 and $E_0$=0.01~a.u., in agreement with the optimal $E_0$ value of 0.003 reported in Ref.~\citenum{Li2018_RealTime-GUGA_CI}.
As for the energy, also the first polarizability is converged with $m$ for all $E_0$ and $\Delta t$ values with $m$=125.
Deviations below 10$^{-2}$~a.u. are observed between the $\Delta t$=10~as and $\Delta t$=20~as results, indicating that both EMR2-TD-DMRG and CF4-TD-DMRG are as stable as the Lanczos-based integrator.
We report in Figure~S9 of the Supporting Information the EMR2-TD-DMRG(SS) first-order dipole response $\mu_{xx}$, obtained from Eq.~\ref{eq:FiniteDifference}, with $E_0=0.003$~a.u. and varying time steps. 
$\mu_{xx}$ oscillates at the same frequency of the perturbing field.
Therefore, the time evolution of $\mu_{xx}$ reproduces correctly the analytical function of Eq.~(\ref{eq:FunctionalForm}) and the finite-differentiation error is negligible.

\begin{table}[htbp!]
	\begin{tabular}{cc|cc|cc}
		\hline \hline
		& \multirow{2}{*}{$E_0$} & \multicolumn{2}{c|}{$\Delta t$=10~as} & \multicolumn{2}{c}{$\Delta t$=20~as} \\
		&        &  $m$=125   &   $m$=250   &   $m$=125  &    $m$=250  \\
		\hline
		\multirow{3}{*}{EMR2}
		& 0.003  & -63.5868   &  -63.7395   & -73.2764   &  -63.5011   \\
		& 0.006  & -63.9726   &  -63.9725   & -63.7697   &  -63.7867   \\
		& 0.010  & -63.9091   &  -63.8303   & -63.9482   &  -63.8908   \\
		\hline
		\multirow{3}{*}{CF4}
		& 0.003  & -63.6866   &  -63.7500   & -63.5397   &  -63.5601   \\
		& 0.006  & -63.3128   &  -63.8216   & -63.8533   &  -63.7638   \\
		& 0.010  & -63.8978   &  -64.2867   & -64.1987   &  -64.2090   \\
		\hline \hline
	\end{tabular}
	\caption{Dynamical hyperpolarizability $\beta_{xxx}(-2\omega_I, \omega_I, \omega_I)$ (in au) of BH calculated at $\omega_I$=488~nm with EMR2-TD-DMRG(SS) and CF4-TD-DMRG(SS) based on the cc-pVDZ basis set for varying integration time-steps $\Delta t$, bond dimension $m$, and electric field $E_0$.}
	\label{tab:BH_HyperProperties_DZ}
\end{table}

\begin{figure}[htbp!]
	\centering
	\includegraphics[width=\textwidth]{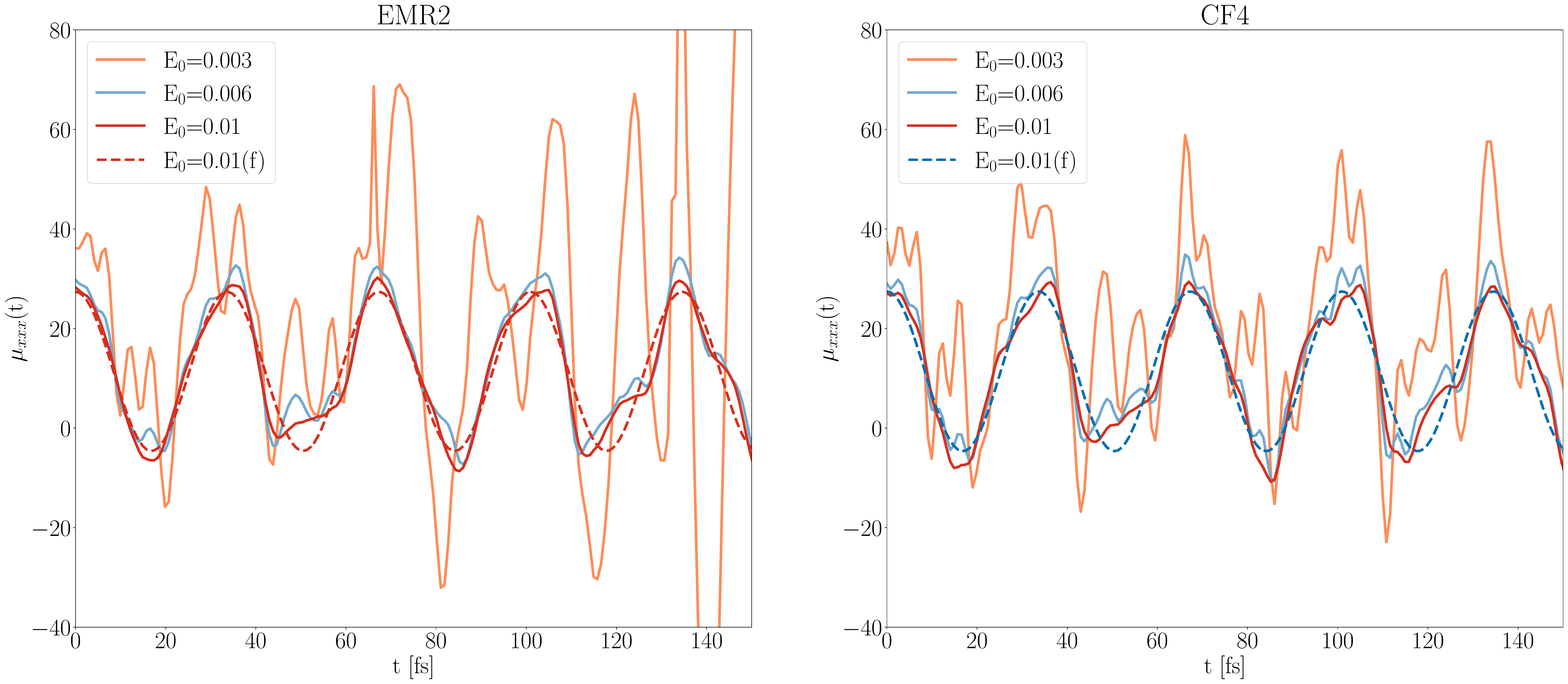}
	\caption{Second-order dipole response $\mu_{xxx}(t)$ of BH calculated with EMR2-TD-DMRG(SS)/cc-pVDZ (left panel) and CF4-TD-DMRG(SS)/cc-pVDZ (right panel), $\Delta t$=20~as, and for different $E_0$ values. Plots labeled as \quotes{(f)} are obtained fitting $\mu_{xxx}$ to Eq.~(\ref{eq:FunctionalForm}).}
	\label{fig:BH_Hyper}
\end{figure}

\noindent As shown in Table~\ref{tab:BH_HyperProperties_DZ}, the finite-difference calculation of hyperpolarizability $\beta_{xxx}(-2\omega_I, \omega_I, \omega_I)$ (denoted as $\beta_{xxx}$ in the following for simplicity) is more sensible to the simulation parameters than that of $\alpha_{xx}$.
The EMR2-TD-DMRG value obtained with $E_0$=0.003 and $\Delta t$=20~as deviates by approximately 10~a.u. from the corresponding $m$=250 value.
The same difference falls below 0.003~a.u. with CF4-TD-DMRG.
For small fields, the error of the EMR2 integration algorithm becomes comparable to the dipole variation, and this renders the finite-difference procedure less accurate.
We highlight this effect in Figure~\ref{fig:BH_Hyper}, where we report the time-evolution of $\mu_{xxx}$.
The finite-difference formula of Eq.~\ref{eq:FiniteDifference} holds if $\mu_{xxx}(t)$ is a sinus-like function.
This is the case for CF4-TD-DMRG (right panel of Figure~\ref{fig:BH_Hyper}), while the EMR2-TD-DMRG propagation deviates from the expected periodic time-evolution, especially in the long-time limit.

\begin{table}[htbp!]
	\begin{tabular}{c|cccc|cc}
		\hline \hline
		& \multicolumn{2}{c}{EMR2-TD-DMRG} & \multicolumn{2}{c|}{CF4-TD-DMRG} 
		& \multirow{2}{*}{FCI\cite{Olsen1998_FCI-Polarizabilities}} 
		& \multirow{2}{*}{FCIQMC\cite{Booth2018_FCIQMC-Properties}} \\
		&  $m$=125   &   $m$=250   &   $m$=125  &    $m$=250  &            &          \\
		\hline
		$\alpha_{xx}$(1064~nm)  &   24.1685  &   24.1359   &  24.1432   &   24.1491   &     23.74  &   20.29  \\
		$\alpha_{xx}$(488~nm)   &   26.1083  &   26.1171   &  26.1821   &   26.1912   &     25.63  &   20.29  \\
		\hline
		$\beta_{xxx}$(1064~nm)  &  -24.0167  &  -21.9264   &  -23.3424  &  -23.0983   &    -31.26  &    -     \\
		$\beta_{xxx}$(488~nm)   &  -77.9075  &  -78.5977   &  -75.1769  &  -74.4685   &   -118.10  &    -     \\
		\hline \hline
	\end{tabular}
	\caption{Polarizability $\alpha_{xx}$ and hyperpolarizability $\beta_{xxx}$ of BH calculated with EMR2-TD-DMRG and CF4-TD-DMRG with the aug-cc-pVDZ basis, $\Delta t$=20~as, $E_0$=0.006, and for two different $\omega_I$ values, 488~nm and 1064~nm. We report reference data obtained with FCI frequency-dependent response theory, and with FCIQMC static response theory.}
	\label{tab:BH_HyperProperties_augDZ}
\end{table}

\noindent The $\alpha_{xx}$ and $\beta_{xxx}$ values reported in Table~\ref{tab:BH_Properties_DZ} and \ref{tab:BH_HyperProperties_DZ} do not match the full CI reference data\cite{Olsen1998_FCI-Polarizabilities} of 25.63~a.u. and -118.10~a.u., respectively, obtained with the larger d-aug-ccpVTZ basis.
We repeat the calculation with the aug-cc-pVDZ basis set, with $\Delta t$=20~as, $E_0$=0.006.
As we show in Table~S4 of the Supporting Information, iTD-DMRG(TS) converges with $m$=125 both for the cc-pVDZ and the aug-cc-pVDZ bases. 
This suggests that increasing the basis set site does not modify the entanglement structure of the MPS and, therefore, that the simulation parameters optimized for the cc-pVDZ basis will be equally accurate based on the aug-cc-pVDZ basis.
We report in Table~\ref{tab:BH_HyperProperties_augDZ} the dynamical response properties calculated both at 488~nm and at 1064~nm to verify if TD-DMRG reproduces correctly the difference in $\alpha_{xx}$ and $\beta_{xxx}$ at these two different frequencies.
As above, the polarizability $\alpha_{xx}$ is converged below 0.01~a.u. with $m$=125, even though TD-DMRG overestimates the FCI data by approximately 0.4~a.u..
Note, however, that the difference between $\alpha_{xx}$(488~nm) and $\alpha_{xx}$(1064~nm) calculated with TD-DMRG is approximately 2~a.u., in agreement with the trend obtained with FCI.
Moreover, the calculated $\alpha_{xx}$(1064~nm) and $\alpha_{xx}$(488~nm) values are closer to the FCI reference data,\cite{Olsen1998_FCI-Polarizabilities} which were obtained by properly including the frequency-dependent contribution to the molecular property, than to the FCIQMC one,\cite{Booth2018_FCIQMC-Properties} obtained with static perturbation theory.
This indicates that TD-DMRG correctly captures the dynamical contribution to $\alpha_{xx}$ and that the discrepancy with FCI data is due to a basis set effect.

\begin{figure}[htbp!]
	\centering
	\includegraphics[width=.75\textwidth]{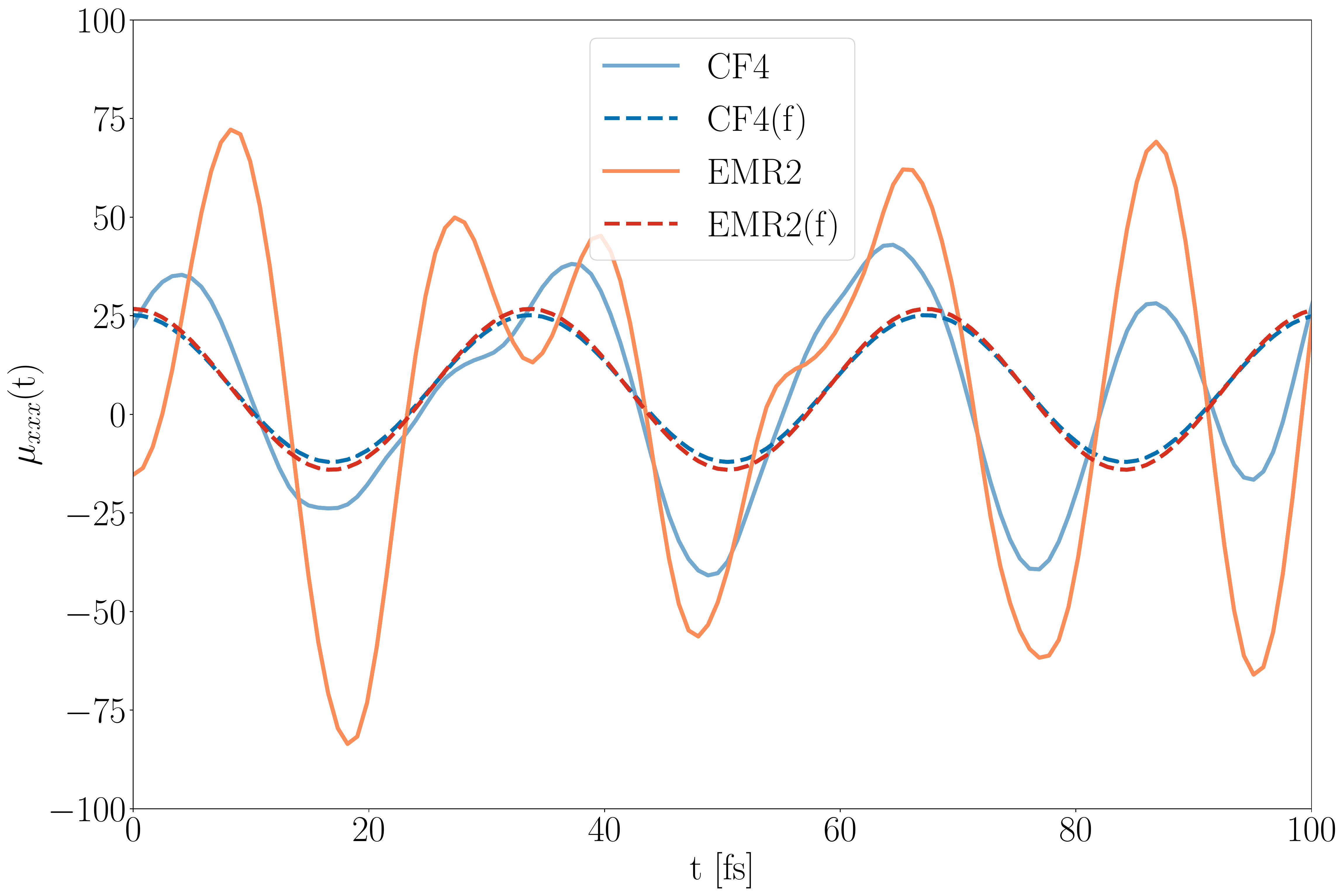}
	\caption{Second-order dipole response $\mu_{xxx}$ calculated EMR2-TD-DMRG (red lines) and CF4-TD-DMRG (blue lines), $m$=250, $\Delta t$=20~as, $E_0$=0.006, $\omega_I$=488~nm for the aug-cc-pVTZ basis. Plots labeled as \quotes{(f)} are obtained fitting $\mu_{xxx}$ to Eq.~(\ref{eq:FunctionalForm}).}
	\label{fig:HyperPolarizability_augBasis}
\end{figure}

\noindent As we show in Table~\ref{tab:BH_HyperProperties_augDZ}, also in this case the calculation of $\beta_{xxx}$ is more sensible to the simulation parameters than for $\alpha_{xx}$.
The EMR2-TD-DMRG $\beta_{xxx}$(1064~nm) value calculated with $m$=125 and $m$=250 differ by more than 3~au, and the same difference reduces to 0.3~a.u. with CF4-TD-DMRG, further confirming that the CF4 integrator is more reliable than EMR2 in the calculation of high-order properties.
Moreover, we report the time-dependent second-order dipole response $\mu_{xxx}(t)$ obtained with EMR2-TD-DMRG and CF4-TD-DMRG in Figure~\ref{fig:HyperPolarizability_augBasis}.
Based on the definition of Eq.~(\ref{eq:FunctionalForm}), $\mu_{xxx}(t)$ is expected to oscillate periodically at frequency $2\omega_I$.
This is the case, at least for the first three oscillations, of the CF4-TD-DMRG propagation, but not for EMR2-TD-DMRG.
Interestingly, the response properties obtained by fitting $\mu_{xxx}(t)$ to Eq.~(\ref{eq:FunctionalForm}) is nearly the same for EMR2-TD-DMRG and CF4-TD-DMRG.
Therefore, the calculated $\beta_{xxx}$(488 nm) value are not as different as Figure~\ref{fig:HyperPolarizability_augBasis} would suggest.
Nevertheless, due to its higher numerical stability, CF4-TD-DMRG should be considered as the reference method for simulations on time-dependent Hamiltonians.

\section{Conclusions}
\label{sec:conclusions}

In the present work, we apply the tangent space-based formulation of the time-dependent Density Matrix Renormalization Group (TD-DMRG) theory\cite{Lubich2014_TimeIntegrationTT,Haegeman2016_MPO-TDDMRG,Baiardi2019_TDDMRG} to simulate the many-body electron dynamics of molecular systems.
The resulting algorithm enables simulating non-equilibrium electron dynamics with a proper inclusion of electronic correlation effects.
TD-DMRG limits the computational costs of time-dependent configuration-interaction methods by encoding the wave function as a matrix product state.
Even though there is no formal guarantee that the electronic wave function can be represented as a compact MPS throughout the whole propagation,\cite{Montangero2006_EntanglementIncrease,Cirac2008_EntanglementScaling,Legeza2019_OrbitalOptimization-TD} we show that this holds true in practice for several applications, including ionization processes and electronic dynamics driven by weak electromagnetic perturbations.
We show that, compared to alternative formulations, tangent-space TD-DMRG is particularly well suited to large-scale quantum dynamics simulations since it can be applied to arbitrary complex Hamiltonian operators and is stable for large propagation time-steps.
We show that the algorithm is an appealing alternative to time-independent DMRG for calculating absorption spectra and high-order response molecular properties.
Moreover, TD-DMRG enables simulating electron dynamics triggered by ultrafast ionizations.
In future works, we will further enhance the efficiency of this pilot TD-DMRG theory by coupling it with algorithms to adapt dynamically the orbital basis, either with a self-consistent field-based optimization\cite{Sato2013_TD-CAS,Madsen2014_TD-RASSCF,Chan2018_TD-CASSCF-Surfaces} or with entanglement-based algorithms.\cite{Legeza2003_OrderingOptimization,Legeza2019_OrbitalOptimization-TD}
These extensions will pave the route towards a systematic application of TD-DMRG to the study of ultrafast attosecond molecular spectroscopy.\cite{Nisoli2017_Review,Martin2019_Attosecond-Review}

\begin{acknowledgement}
This work was supported by ETH Z\"{u}rich through the ETH Fellowship No. FEL-49 18-1.
The author is very grateful to Prof.~Dr.~Markus Reiher for helpful discussions and for his insightful comments about this manuscript.
\end{acknowledgement}

\section*{Description of the Supporting Information}

The Supporting Information contains the following additional data on the simulation of the electron dynamics in benzene: the graphical representation of the molecular orbitals, the analysis of the iTD-DMRG(TS) convergence with the bond dimension $m$ and the active space sizes, the time-dependent orbital population obtained with varying $\Delta t$ and $m$ values, the time-dependent single-orbital entropy of the three highest and three lowest orbitals of benzene for CAS(26,26).
It also contains the following additional data for decacene: the comparison of the iTD-DMRG(TS) and TI-DMRG(TS) energy convergence, the excitation energies and dipole strengths calculated with TI-DMRG and varying $m$ values, the convergence of the absorption spectrum calculated with $\Delta t$ based on CAS(10,10).
Lastly, it reports the following data for BH: ground-state energy calculated with iTD-DMRG(SS) and iTD-DMRG(TS) and varying $m$ values, first-order dipole response calculated with EMR2-TD-DMRG(SS) and different $\Delta t$ values.

\bibliography{library}

\providecommand{\latin}[1]{#1}
\makeatletter
\providecommand{\doi}
  {\begingroup\let\do\@makeother\dospecials
  \catcode`\{=1 \catcode`\}=2 \doi@aux}
\providecommand{\doi@aux}[1]{\endgroup\texttt{#1}}
\makeatother
\providecommand*\mcitethebibliography{\thebibliography}
\csname @ifundefined\endcsname{endmcitethebibliography}
  {\let\endmcitethebibliography\endthebibliography}{}
\begin{mcitethebibliography}{121}
\providecommand*\natexlab[1]{#1}
\providecommand*\mciteSetBstSublistMode[1]{}
\providecommand*\mciteSetBstMaxWidthForm[2]{}
\providecommand*\mciteBstWouldAddEndPuncttrue
  {\def\EndOfBibitem{\unskip.}}
\providecommand*\mciteBstWouldAddEndPunctfalse
  {\let\EndOfBibitem\relax}
\providecommand*\mciteSetBstMidEndSepPunct[3]{}
\providecommand*\mciteSetBstSublistLabelBeginEnd[3]{}
\providecommand*\EndOfBibitem{}
\mciteSetBstSublistMode{f}
\mciteSetBstMaxWidthForm{subitem}{(\alph{mcitesubitemcount})}
\mciteSetBstSublistLabelBeginEnd
  {\mcitemaxwidthsubitemform\space}
  {\relax}
  {\relax}

\bibitem[Corkum and Krausz(2007)Corkum, and
  Krausz]{Corkum2007_AttosecondScience}
Corkum,~P.~B.; Krausz,~F. {Attosecond science}. \emph{Nat. Phys.}
  \textbf{2007}, \emph{3}, 381--387\relax
\mciteBstWouldAddEndPuncttrue
\mciteSetBstMidEndSepPunct{\mcitedefaultmidpunct}
{\mcitedefaultendpunct}{\mcitedefaultseppunct}\relax
\EndOfBibitem
\bibitem[Kling and Vrakking(2008)Kling, and
  Vrakking]{Vrakking2007_AttosecondMolecularDynamics}
Kling,~M.~F.; Vrakking,~M.~J. {Attosecond Electron Dynamics}. \emph{Ann. Rev.
  Phys. Chem.} \textbf{2008}, \emph{59}, 463--492\relax
\mciteBstWouldAddEndPuncttrue
\mciteSetBstMidEndSepPunct{\mcitedefaultmidpunct}
{\mcitedefaultendpunct}{\mcitedefaultseppunct}\relax
\EndOfBibitem
\bibitem[Krausz and Ivanov(2009)Krausz, and
  Ivanov]{Krausz2009_AttosecondPhysicsReview}
Krausz,~F.; Ivanov,~M. {Attosecond physics}. \emph{Rev. Mod. Phys.}
  \textbf{2009}, \emph{81}, 163--234\relax
\mciteBstWouldAddEndPuncttrue
\mciteSetBstMidEndSepPunct{\mcitedefaultmidpunct}
{\mcitedefaultendpunct}{\mcitedefaultseppunct}\relax
\EndOfBibitem
\bibitem[Gallmann \latin{et~al.}(2012)Gallmann, Cirelli, and
  Keller]{Keller2012_AttosecondScience}
Gallmann,~L.; Cirelli,~C.; Keller,~U. {Attosecond Science: Recent Highlights
  and Future Trends}. \emph{Ann. Rev. Phys. Chem..} \textbf{2012}, \emph{63},
  447--469\relax
\mciteBstWouldAddEndPuncttrue
\mciteSetBstMidEndSepPunct{\mcitedefaultmidpunct}
{\mcitedefaultendpunct}{\mcitedefaultseppunct}\relax
\EndOfBibitem
\bibitem[Palacios and Mart{\'{i}}n(2020)Palacios, and
  Mart{\'{i}}n]{Martin2019_Attosecond-Review}
Palacios,~A.; Mart{\'{i}}n,~F. {The quantum chemistry of attosecond molecular
  science}. \emph{Wiley Interdiscip. Rev. Comput. Mol. Sci.} \textbf{2020},
  \emph{1}, e1430\relax
\mciteBstWouldAddEndPuncttrue
\mciteSetBstMidEndSepPunct{\mcitedefaultmidpunct}
{\mcitedefaultendpunct}{\mcitedefaultseppunct}\relax
\EndOfBibitem
\bibitem[Kraus \latin{et~al.}(2015)Kraus, Mignolet, Baykusheva, Rupenyan,
  Horn{\'{y}}, Penka, Grassi, Tolstikhin, Schneider, Jensen, Madsen, Bandrauk,
  Remacle, and W{\"{o}}rner]{Worner2015_Iodoacetylene}
Kraus,~P.~M.; Mignolet,~B.; Baykusheva,~D.; Rupenyan,~A.; Horn{\'{y}},~L.;
  Penka,~E.~F.; Grassi,~G.; Tolstikhin,~O.~I.; Schneider,~J.; Jensen,~F.;
  Madsen,~L.~B.; Bandrauk,~A.~D.; Remacle,~F.; W{\"{o}}rner,~H.~J. {Measurement
  and laser control of attosecond charge migration in ionized iodoacetylene}.
  \emph{Science} \textbf{2015}, \emph{350}, 790--795\relax
\mciteBstWouldAddEndPuncttrue
\mciteSetBstMidEndSepPunct{\mcitedefaultmidpunct}
{\mcitedefaultendpunct}{\mcitedefaultseppunct}\relax
\EndOfBibitem
\bibitem[H{\"{u}}tten \latin{et~al.}(2018)H{\"{u}}tten, Mittermair, Stock,
  Beerwerth, Shirvanyan, Riemensberger, Duensing, Heider, Wagner, Guggenmos,
  Fritzsche, Kabachnik, Kienberger, and
  Bernhardt]{Hutten2019_Kripton-Attosecond}
H{\"{u}}tten,~K.; Mittermair,~M.; Stock,~S.~O.; Beerwerth,~R.; Shirvanyan,~V.;
  Riemensberger,~J.; Duensing,~A.; Heider,~R.; Wagner,~M.~S.; Guggenmos,~A.;
  Fritzsche,~S.; Kabachnik,~N.~M.; Kienberger,~R.; Bernhardt,~B. {Ultrafast
  quantum control of ionization dynamics in krypton}. \emph{Nat. Commun.}
  \textbf{2018}, \emph{9}, 1--5\relax
\mciteBstWouldAddEndPuncttrue
\mciteSetBstMidEndSepPunct{\mcitedefaultmidpunct}
{\mcitedefaultendpunct}{\mcitedefaultseppunct}\relax
\EndOfBibitem
\bibitem[Timmers \latin{et~al.}(2019)Timmers, Zhu, Li, Kobayashi, Sabbar,
  Hollstein, Reduzzi, Mart{\'{i}}nez, Neumark, and
  Leone]{Martinez2019_Br2-Attosecond}
Timmers,~H.; Zhu,~X.; Li,~Z.; Kobayashi,~Y.; Sabbar,~M.; Hollstein,~M.;
  Reduzzi,~M.; Mart{\'{i}}nez,~T.~J.; Neumark,~D.~M.; Leone,~S.~R.
  {Disentangling conical intersection and coherent molecular dynamics in methyl
  bromide with attosecond transient absorption spectroscopy}. \emph{Nat.
  Commun.} \textbf{2019}, \emph{10}, 1--8\relax
\mciteBstWouldAddEndPuncttrue
\mciteSetBstMidEndSepPunct{\mcitedefaultmidpunct}
{\mcitedefaultendpunct}{\mcitedefaultseppunct}\relax
\EndOfBibitem
\bibitem[Li \latin{et~al.}(2020)Li, Govind, Isborn, DePrince, and
  Lopata]{Li2020_RTES-Review}
Li,~X.; Govind,~N.; Isborn,~C.; DePrince,~A.~E.; Lopata,~K. {Real-Time
  Time-Dependent Electronic Structure Theory}. \emph{Chem. Rev.} \textbf{2020},
  \emph{120}, 9951--9993\relax
\mciteBstWouldAddEndPuncttrue
\mciteSetBstMidEndSepPunct{\mcitedefaultmidpunct}
{\mcitedefaultendpunct}{\mcitedefaultseppunct}\relax
\EndOfBibitem
\bibitem[Cheng \latin{et~al.}(2006)Cheng, Evans, and {Van
  Voorhis}]{VanVoorhis2006_PredictorCorrector}
Cheng,~C.-L.~L.; Evans,~J.~S.; {Van Voorhis},~T. {Simulating molecular
  conductance using real-time density functional theory}. \emph{Phys. Rev. B}
  \textbf{2006}, \emph{74}, 155112\relax
\mciteBstWouldAddEndPuncttrue
\mciteSetBstMidEndSepPunct{\mcitedefaultmidpunct}
{\mcitedefaultendpunct}{\mcitedefaultseppunct}\relax
\EndOfBibitem
\bibitem[Lopata and Govind(2011)Lopata, and Govind]{Lopata2011_RealTimeTDDFT}
Lopata,~K.; Govind,~N. {Modeling Fast Electron Dynamics with Real-Time
  Time-Dependent Density Functional Theory: Application to Small Molecules and
  Chromophores}. \emph{J. Chem. Theory Comput.} \textbf{2011}, \emph{7},
  1344--1355\relax
\mciteBstWouldAddEndPuncttrue
\mciteSetBstMidEndSepPunct{\mcitedefaultmidpunct}
{\mcitedefaultendpunct}{\mcitedefaultseppunct}\relax
\EndOfBibitem
\bibitem[Repisky \latin{et~al.}(2015)Repisky, Konecny, Kadek, Komorovsky,
  Malkin, Malkin, and Ruud]{Repisky2015_RealTime4c}
Repisky,~M.; Konecny,~L.; Kadek,~M.; Komorovsky,~S.; Malkin,~O.~L.;
  Malkin,~V.~G.; Ruud,~K. {Excitation Energies from Real-Time Propagation of
  the Four-Component Dirac-Kohn-Sham Equation}. \emph{J. Chem. Theory Comput.}
  \textbf{2015}, \emph{11}, 980--991\relax
\mciteBstWouldAddEndPuncttrue
\mciteSetBstMidEndSepPunct{\mcitedefaultmidpunct}
{\mcitedefaultendpunct}{\mcitedefaultseppunct}\relax
\EndOfBibitem
\bibitem[Provorse and Isborn(2016)Provorse, and
  Isborn]{Isborn2016_ElectronDynamics-TDDFT}
Provorse,~M.~R.; Isborn,~C.~M. {Electron dynamics with real-time time-dependent
  density functional theory}. \emph{Int. J. Quantum Chem.} \textbf{2016},
  \emph{116}, 739--749\relax
\mciteBstWouldAddEndPuncttrue
\mciteSetBstMidEndSepPunct{\mcitedefaultmidpunct}
{\mcitedefaultendpunct}{\mcitedefaultseppunct}\relax
\EndOfBibitem
\bibitem[Goings \latin{et~al.}(2018)Goings, Lestrange, and
  Li]{Li2018_RealTime-Review}
Goings,~J.~J.; Lestrange,~P.~J.; Li,~X. {Real-time time-dependent electronic
  structure theory}. \emph{Wiley Interdiscip. Rev. Comput. Mol. Sci.}
  \textbf{2018}, \emph{8}, e1341\relax
\mciteBstWouldAddEndPuncttrue
\mciteSetBstMidEndSepPunct{\mcitedefaultmidpunct}
{\mcitedefaultendpunct}{\mcitedefaultseppunct}\relax
\EndOfBibitem
\bibitem[Habenicht \latin{et~al.}(2014)Habenicht, Tani, Provorse, and
  Isborn]{Isborn2014_RabiOscillations}
Habenicht,~B.~F.; Tani,~N.~P.; Provorse,~M.~R.; Isborn,~C.~M. {Two-electron
  Rabi oscillations in real-time time-dependent density-functional theory}.
  \emph{J. Chem. Phys.} \textbf{2014}, \emph{141}, 184112\relax
\mciteBstWouldAddEndPuncttrue
\mciteSetBstMidEndSepPunct{\mcitedefaultmidpunct}
{\mcitedefaultendpunct}{\mcitedefaultseppunct}\relax
\EndOfBibitem
\bibitem[Pigg \latin{et~al.}(2012)Pigg, Hagen, Nam, and
  Papenbrock]{Pigg2012_TDCC-ImaginaryTime}
Pigg,~D.~A.; Hagen,~G.; Nam,~H.; Papenbrock,~T. {Time-dependent coupled-cluster
  method for atomic nuclei}. \emph{Phys. Rev. C} \textbf{2012}, \emph{86},
  14308\relax
\mciteBstWouldAddEndPuncttrue
\mciteSetBstMidEndSepPunct{\mcitedefaultmidpunct}
{\mcitedefaultendpunct}{\mcitedefaultseppunct}\relax
\EndOfBibitem
\bibitem[Kvaal(2012)]{Kvaal2012_TDCC}
Kvaal,~S. {Ab initio quantum dynamics using coupled-cluster}. \emph{J. Chem.
  Phys.} \textbf{2012}, \emph{136}, 194109\relax
\mciteBstWouldAddEndPuncttrue
\mciteSetBstMidEndSepPunct{\mcitedefaultmidpunct}
{\mcitedefaultendpunct}{\mcitedefaultseppunct}\relax
\EndOfBibitem
\bibitem[Nascimento and DePrince(2016)Nascimento, and
  DePrince]{Nascimento2016_TDCC}
Nascimento,~D.~R.; DePrince,~A.~E. {Linear Absorption Spectra from Explicitly
  Time-Dependent Equation-of-Motion Coupled-Cluster Theory}. \emph{J. Chem.
  Theory Comput.} \textbf{2016}, \emph{12}, 5834--5840\relax
\mciteBstWouldAddEndPuncttrue
\mciteSetBstMidEndSepPunct{\mcitedefaultmidpunct}
{\mcitedefaultendpunct}{\mcitedefaultseppunct}\relax
\EndOfBibitem
\bibitem[Nascimento and DePrince(2017)Nascimento, and
  DePrince]{DePrince2017_CoreCC-RealTime}
Nascimento,~D.~R.; DePrince,~A.~E. {Simulation of Near-Edge X-ray Absorption
  Fine Structure with Time-Dependent Equation-of-Motion Coupled-Cluster
  Theory}. \emph{J. Phys. Chem. Lett.} \textbf{2017}, \emph{8},
  2951--2957\relax
\mciteBstWouldAddEndPuncttrue
\mciteSetBstMidEndSepPunct{\mcitedefaultmidpunct}
{\mcitedefaultendpunct}{\mcitedefaultseppunct}\relax
\EndOfBibitem
\bibitem[Nascimento and DePrince(2019)Nascimento, and
  DePrince]{DePrince2019_EOM-TD-Spectroscopy}
Nascimento,~D.~R.; DePrince,~A.~E. {A general time-domain formulation of
  equation-of-motion coupled-cluster theory for linear spectroscopy}. \emph{J.
  Chem. Phys.} \textbf{2019}, \emph{151}, 204107\relax
\mciteBstWouldAddEndPuncttrue
\mciteSetBstMidEndSepPunct{\mcitedefaultmidpunct}
{\mcitedefaultendpunct}{\mcitedefaultseppunct}\relax
\EndOfBibitem
\bibitem[Koulias \latin{et~al.}(2019)Koulias, Williams-Young, Nascimento,
  DePrince, and Li]{Li2019_Relativistic-RTCC}
Koulias,~L.~N.; Williams-Young,~D.~B.; Nascimento,~D.~R.; DePrince,~A.~E.;
  Li,~X. {Relativistic Real-Time Time-Dependent Equation-of-Motion
  Coupled-Cluster}. \emph{J. Chem. Theory Comput.} \textbf{2019}, \emph{15},
  6617--6624\relax
\mciteBstWouldAddEndPuncttrue
\mciteSetBstMidEndSepPunct{\mcitedefaultmidpunct}
{\mcitedefaultendpunct}{\mcitedefaultseppunct}\relax
\EndOfBibitem
\bibitem[Kristiansen \latin{et~al.}(2020)Kristiansen, Sch{\o}yen, Kvaal, and
  Pedersen]{BondoPedersen2020_Stability-TDCC}
Kristiansen,~H.~E.; Sch{\o}yen,~{\O}.~S.; Kvaal,~S.; Pedersen,~T.~B. {Numerical
  stability of time-dependent coupled-cluster methods for many-electron
  dynamics in intense laser pulses}. \emph{J. Chem. Phys.} \textbf{2020},
  \emph{152}, 071102\relax
\mciteBstWouldAddEndPuncttrue
\mciteSetBstMidEndSepPunct{\mcitedefaultmidpunct}
{\mcitedefaultendpunct}{\mcitedefaultseppunct}\relax
\EndOfBibitem
\bibitem[Skeidsvoll \latin{et~al.}(2020)Skeidsvoll, Balbi, and
  Koch]{Balbi2020-RT_CC}
Skeidsvoll,~A.~S.; Balbi,~A.; Koch,~H. {Time-dependent coupled-cluster theory
  for ultrafast transient-absorption spectroscopy}. \emph{Phys. Rev. A}
  \textbf{2020}, \emph{102}, 023115\relax
\mciteBstWouldAddEndPuncttrue
\mciteSetBstMidEndSepPunct{\mcitedefaultmidpunct}
{\mcitedefaultendpunct}{\mcitedefaultseppunct}\relax
\EndOfBibitem
\bibitem[Greenman \latin{et~al.}(2010)Greenman, Ho, Pabst, Kamarchik,
  Mazziotti, and Santra]{Mazziotti2010_TDCI}
Greenman,~L.; Ho,~P.~J.; Pabst,~S.; Kamarchik,~E.; Mazziotti,~D.~A.; Santra,~R.
  {Implementation of the time-dependent configuration-interaction singles
  method for atomic strong-field processes}. \emph{Phys. Rev. A} \textbf{2010},
  \emph{82}, 023406\relax
\mciteBstWouldAddEndPuncttrue
\mciteSetBstMidEndSepPunct{\mcitedefaultmidpunct}
{\mcitedefaultendpunct}{\mcitedefaultseppunct}\relax
\EndOfBibitem
\bibitem[Lestrange \latin{et~al.}(2018)Lestrange, Hoffmann, and
  Li]{Li2018_RealTime-GUGA_CI}
Lestrange,~P.~J.; Hoffmann,~M.~R.; Li,~X. {Time-Dependent Configuration
  Interaction Using the Graphical Unitary Group Approach: Nonlinear Electric
  Properties}. \emph{Adv. Quantum Chem.} \textbf{2018}, \emph{76},
  295--313\relax
\mciteBstWouldAddEndPuncttrue
\mciteSetBstMidEndSepPunct{\mcitedefaultmidpunct}
{\mcitedefaultendpunct}{\mcitedefaultseppunct}\relax
\EndOfBibitem
\bibitem[Peng \latin{et~al.}(2018)Peng, Fales, and Levine]{Peng2018_TDCI}
Peng,~W.~T.; Fales,~B.~S.; Levine,~B.~G. {Simulating Electron Dynamics of
  Complex Molecules with Time-Dependent Complete Active Space Configuration
  Interaction}. \emph{J. Chem. Theory Comput.} \textbf{2018}, \emph{14},
  4129--4138\relax
\mciteBstWouldAddEndPuncttrue
\mciteSetBstMidEndSepPunct{\mcitedefaultmidpunct}
{\mcitedefaultendpunct}{\mcitedefaultseppunct}\relax
\EndOfBibitem
\bibitem[Miranda \latin{et~al.}(2011)Miranda, Fisher, Stella, and
  Horsfield]{Miranda2011_TD-CASSCF}
Miranda,~R.~P.; Fisher,~A.~J.; Stella,~L.; Horsfield,~A.~P. {A
  multiconfigurational time-dependent Hartree-Fock method for excited
  electronic states. I. General formalism and application to open-shell
  states}. \emph{J. Chem. Phys.} \textbf{2011}, \emph{134}, 244101\relax
\mciteBstWouldAddEndPuncttrue
\mciteSetBstMidEndSepPunct{\mcitedefaultmidpunct}
{\mcitedefaultendpunct}{\mcitedefaultseppunct}\relax
\EndOfBibitem
\bibitem[Sato and Ishikawa(2013)Sato, and Ishikawa]{Sato2013_TD-CAS}
Sato,~T.; Ishikawa,~K.~L. {Time-dependent complete-active-space
  self-consistent-field method for multielectron dynamics in intense laser
  fields}. \emph{Phys. Rev. A} \textbf{2013}, \emph{88}, 023402\relax
\mciteBstWouldAddEndPuncttrue
\mciteSetBstMidEndSepPunct{\mcitedefaultmidpunct}
{\mcitedefaultendpunct}{\mcitedefaultseppunct}\relax
\EndOfBibitem
\bibitem[Miyagi and Madsen(2014)Miyagi, and Madsen]{Madsen2014_TD-RASSCF}
Miyagi,~H.; Madsen,~L.~B. {Time-dependent restricted-active-space
  self-consistent-field theory for laser-driven many-electron dynamics. II.
  Extended formulation and numerical analysis}. \emph{Phys. Rev. A}
  \textbf{2014}, \emph{89}, 063416\relax
\mciteBstWouldAddEndPuncttrue
\mciteSetBstMidEndSepPunct{\mcitedefaultmidpunct}
{\mcitedefaultendpunct}{\mcitedefaultseppunct}\relax
\EndOfBibitem
\bibitem[Sato and Ishikawa(2015)Sato, and Ishikawa]{Sato2015_TD-RAS}
Sato,~T.; Ishikawa,~K.~L. {Time-dependent multiconfiguration
  self-consistent-field method based on the occupation-restricted
  multiple-active-space model for multielectron dynamics in intense laser
  fields}. \emph{Phys. Rev. A} \textbf{2015}, \emph{91}, 023417\relax
\mciteBstWouldAddEndPuncttrue
\mciteSetBstMidEndSepPunct{\mcitedefaultmidpunct}
{\mcitedefaultendpunct}{\mcitedefaultseppunct}\relax
\EndOfBibitem
\bibitem[Kretchmer and Chan(2018)Kretchmer, and
  Chan]{Chan2018_TD-CASSCF-Surfaces}
Kretchmer,~J.~S.; Chan,~G. K.-L. {The Fate of Atomic Spin in Atomic Scattering
  off Surfaces}. \emph{J. Phys. Chem. Lett.} \textbf{2018}, \emph{9},
  2863--2868\relax
\mciteBstWouldAddEndPuncttrue
\mciteSetBstMidEndSepPunct{\mcitedefaultmidpunct}
{\mcitedefaultendpunct}{\mcitedefaultseppunct}\relax
\EndOfBibitem
\bibitem[Wahyutama \latin{et~al.}(2019)Wahyutama, Sato, and
  Ishikawa]{Sato2019_TDCASSCF-HHG}
Wahyutama,~I.~S.; Sato,~T.; Ishikawa,~K.~L. {Time-dependent multiconfiguration
  self-consistent-field study on resonantly enhanced high-order harmonic
  generation from transition-metal elements}. \emph{Phys. Rev. A}
  \textbf{2019}, \emph{99}, 063420\relax
\mciteBstWouldAddEndPuncttrue
\mciteSetBstMidEndSepPunct{\mcitedefaultmidpunct}
{\mcitedefaultendpunct}{\mcitedefaultseppunct}\relax
\EndOfBibitem
\bibitem[Paeckel \latin{et~al.}(2019)Paeckel, K{\"{o}}hler, Swoboda, Manmana,
  Schollw{\"{o}}ck, and Hubig]{Paeckel2019_Review}
Paeckel,~S.; K{\"{o}}hler,~T.; Swoboda,~A.; Manmana,~S.~R.;
  Schollw{\"{o}}ck,~U.; Hubig,~C. {Time-evolution methods for matrix-product
  states}. \emph{Ann. Phys.} \textbf{2019}, \emph{411}, 167998\relax
\mciteBstWouldAddEndPuncttrue
\mciteSetBstMidEndSepPunct{\mcitedefaultmidpunct}
{\mcitedefaultendpunct}{\mcitedefaultseppunct}\relax
\EndOfBibitem
\bibitem[White(1992)]{White1992_DMRGBasis}
White,~S.~R. {Density matrix formulation for quantum renormalization groups}.
  \emph{Phys. Rev. Lett.} \textbf{1992}, \emph{69}, 2863--2866\relax
\mciteBstWouldAddEndPuncttrue
\mciteSetBstMidEndSepPunct{\mcitedefaultmidpunct}
{\mcitedefaultendpunct}{\mcitedefaultseppunct}\relax
\EndOfBibitem
\bibitem[White(1993)]{White1993_DMRGBasis}
White,~S.~R. {Density-matrix algorithms for quantum renormalization groups}.
  \emph{Phys. Rev. B} \textbf{1993}, \emph{48}, 10345--10356\relax
\mciteBstWouldAddEndPuncttrue
\mciteSetBstMidEndSepPunct{\mcitedefaultmidpunct}
{\mcitedefaultendpunct}{\mcitedefaultseppunct}\relax
\EndOfBibitem
\bibitem[Muth and McCulloch(2007)Muth, and
  McCulloch]{McCulloch2007_FromMPStoDMRG}
Muth,~D.; McCulloch,~I.~P. {From density-matrix renormalization group to matrix
  product states}. \emph{J. Stat. Mech. Theory Exp.} \textbf{2007},
  \emph{2007}, P10014\relax
\mciteBstWouldAddEndPuncttrue
\mciteSetBstMidEndSepPunct{\mcitedefaultmidpunct}
{\mcitedefaultendpunct}{\mcitedefaultseppunct}\relax
\EndOfBibitem
\bibitem[Chan \latin{et~al.}(2008)Chan, Dorando, Ghosh, Hachmann, Neuscamman,
  Wang, and Yanai]{Chan2008_Review}
Chan,~G. K.-L.; Dorando,~J.~J.; Ghosh,~D.; Hachmann,~J.; Neuscamman,~E.;
  Wang,~H.; Yanai,~T. \emph{Frontiers in Quantum Systems in Chemistry and
  Physics}; Springer Netherlands, 2008; pp 49--65\relax
\mciteBstWouldAddEndPuncttrue
\mciteSetBstMidEndSepPunct{\mcitedefaultmidpunct}
{\mcitedefaultendpunct}{\mcitedefaultseppunct}\relax
\EndOfBibitem
\bibitem[Chan and Zgid(2009)Chan, and Zgid]{Zgid2009_Review}
Chan,~G. K.~L.; Zgid,~D. {The Density Matrix Renormalization Group in Quantum
  Chemistry}. \emph{Annual Reports in Computational Chemistry} \textbf{2009},
  \emph{5}, 149--162\relax
\mciteBstWouldAddEndPuncttrue
\mciteSetBstMidEndSepPunct{\mcitedefaultmidpunct}
{\mcitedefaultendpunct}{\mcitedefaultseppunct}\relax
\EndOfBibitem
\bibitem[Chan and Sharma(2011)Chan, and Sharma]{Chan2011_Review}
Chan,~G. K.-L.; Sharma,~S. {The Density Matrix Renormalization Group in Quantum
  Chemistry}. \emph{Annu. Rev. Phys. Chem.} \textbf{2011}, \emph{62},
  465--481\relax
\mciteBstWouldAddEndPuncttrue
\mciteSetBstMidEndSepPunct{\mcitedefaultmidpunct}
{\mcitedefaultendpunct}{\mcitedefaultseppunct}\relax
\EndOfBibitem
\bibitem[Wouters and {Van Neck}(2013)Wouters, and {Van
  Neck}]{Wouters2013_Review}
Wouters,~S.; {Van Neck},~D. {The density matrix renormalization group for ab
  initio quantum chemistry}. \emph{Eur. Phys. J. D} \textbf{2013}, \emph{31},
  395--402\relax
\mciteBstWouldAddEndPuncttrue
\mciteSetBstMidEndSepPunct{\mcitedefaultmidpunct}
{\mcitedefaultendpunct}{\mcitedefaultseppunct}\relax
\EndOfBibitem
\bibitem[Keller and Reiher(2014)Keller, and Reiher]{Keller2014}
Keller,~S.~F.; Reiher,~M. {Determining Factors for the Accuracy of DMRG in
  Chemistry}. \emph{Chimia} \textbf{2014}, \emph{68}, 200--203\relax
\mciteBstWouldAddEndPuncttrue
\mciteSetBstMidEndSepPunct{\mcitedefaultmidpunct}
{\mcitedefaultendpunct}{\mcitedefaultseppunct}\relax
\EndOfBibitem
\bibitem[Kurashige(2014)]{Kurashige2014_Review}
Kurashige,~Y. {Multireference electron correlation methods with density matrix
  renormalisation group reference functions}. \emph{Mol. Phys.} \textbf{2014},
  \emph{112}, 1485--1494\relax
\mciteBstWouldAddEndPuncttrue
\mciteSetBstMidEndSepPunct{\mcitedefaultmidpunct}
{\mcitedefaultendpunct}{\mcitedefaultseppunct}\relax
\EndOfBibitem
\bibitem[Olivares-Amaya \latin{et~al.}(2015)Olivares-Amaya, Hu, Nakatani,
  Sharma, Yang, and Chan]{Olivares2015_DMRGInPractice}
Olivares-Amaya,~R.; Hu,~W.; Nakatani,~N.; Sharma,~S.; Yang,~J.; Chan,~G. K.-L.
  {The ab-initio density matrix renormalization group in practice}. \emph{J.
  Chem. Phys.} \textbf{2015}, \emph{142}, 34102\relax
\mciteBstWouldAddEndPuncttrue
\mciteSetBstMidEndSepPunct{\mcitedefaultmidpunct}
{\mcitedefaultendpunct}{\mcitedefaultseppunct}\relax
\EndOfBibitem
\bibitem[Szalay \latin{et~al.}(2015)Szalay, Pfeffer, Murg, Barcza, Verstraete,
  Schneider, and Legeza]{Szalay2015_Review}
Szalay,~S.; Pfeffer,~M.; Murg,~V.; Barcza,~G.; Verstraete,~F.; Schneider,~R.;
  Legeza,~{\"{O}}. {Tensor product methods and entanglement optimization for ab
  initio quantum chemistry}. \emph{Int. J. Quantum Chem.} \textbf{2015},
  \emph{115}, 1342--1391\relax
\mciteBstWouldAddEndPuncttrue
\mciteSetBstMidEndSepPunct{\mcitedefaultmidpunct}
{\mcitedefaultendpunct}{\mcitedefaultseppunct}\relax
\EndOfBibitem
\bibitem[Yanai \latin{et~al.}(2015)Yanai, Kurashige, Mizukami, Chalupsk{\'{y}},
  Lan, and Saitow]{Yanai2015}
Yanai,~T.; Kurashige,~Y.; Mizukami,~W.; Chalupsk{\'{y}},~J.; Lan,~T.~N.;
  Saitow,~M. {Density matrix renormalization group for ab initio calculations
  and associated dynamic correlation methods: A review of theory and
  applications}. \emph{Int. J. Quantum Chem.} \textbf{2015}, \emph{115},
  283--299\relax
\mciteBstWouldAddEndPuncttrue
\mciteSetBstMidEndSepPunct{\mcitedefaultmidpunct}
{\mcitedefaultendpunct}{\mcitedefaultseppunct}\relax
\EndOfBibitem
\bibitem[Knecht \latin{et~al.}(2016)Knecht, Hedeg{\aa}rd, Keller, Kovyrshin,
  Ma, Muolo, Stein, and Reiher]{Knecht2016_Chimia}
Knecht,~S.; Hedeg{\aa}rd,~E.~D.; Keller,~S.; Kovyrshin,~A.; Ma,~Y.; Muolo,~A.;
  Stein,~C.~J.; Reiher,~M. {New Approaches for ab initio Calculations of
  Molecules with Strong Electron Correlation}. \emph{Chimia} \textbf{2016},
  \emph{70}, 244--251\relax
\mciteBstWouldAddEndPuncttrue
\mciteSetBstMidEndSepPunct{\mcitedefaultmidpunct}
{\mcitedefaultendpunct}{\mcitedefaultseppunct}\relax
\EndOfBibitem
\bibitem[Baiardi and Reiher(2020)Baiardi, and Reiher]{Baiardi2020_Review}
Baiardi,~A.; Reiher,~M. {The density matrix renormalization group in chemistry
  and molecular physics: Recent developments and new challenges}. \emph{J.
  Chem. Phys.} \textbf{2020}, \emph{152}, 040903\relax
\mciteBstWouldAddEndPuncttrue
\mciteSetBstMidEndSepPunct{\mcitedefaultmidpunct}
{\mcitedefaultendpunct}{\mcitedefaultseppunct}\relax
\EndOfBibitem
\bibitem[Frahm and Pfannkuche(2019)Frahm, and
  Pfannkuche]{Frahm2019_TD-DMRG_Ultrafast}
Frahm,~L.-H.; Pfannkuche,~D. {Ultrafast ab-initio Quantum Chemistry Using
  Matrix Product States}. \emph{J. Chem. Theory Comput.} \textbf{2019},
  \emph{15}, 2154--2165\relax
\mciteBstWouldAddEndPuncttrue
\mciteSetBstMidEndSepPunct{\mcitedefaultmidpunct}
{\mcitedefaultendpunct}{\mcitedefaultseppunct}\relax
\EndOfBibitem
\bibitem[Hastings(2007)]{Hastings2007_AreaLaw}
Hastings,~M.~B. {An area law for one-dimensional quantum systems}. \emph{J.
  Stat. Mech. Theory Exp.} \textbf{2007}, \emph{2007}, P08024--P08024\relax
\mciteBstWouldAddEndPuncttrue
\mciteSetBstMidEndSepPunct{\mcitedefaultmidpunct}
{\mcitedefaultendpunct}{\mcitedefaultseppunct}\relax
\EndOfBibitem
\bibitem[Chiara \latin{et~al.}(2006)Chiara, Montangero, Calabrese, and
  Fazio]{Montangero2006_EntanglementIncrease}
Chiara,~G.~D.; Montangero,~S.; Calabrese,~P.; Fazio,~R. {Entanglement entropy
  dynamics of Heisenberg chains}. \emph{J. Stat. Mech. Theory Exp.}
  \textbf{2006}, \emph{2006}, P03001--P03001\relax
\mciteBstWouldAddEndPuncttrue
\mciteSetBstMidEndSepPunct{\mcitedefaultmidpunct}
{\mcitedefaultendpunct}{\mcitedefaultseppunct}\relax
\EndOfBibitem
\bibitem[Schuch \latin{et~al.}(2008)Schuch, Wolf, Vollbrecht, and
  Cirac]{Cirac2008_EntanglementScaling}
Schuch,~N.; Wolf,~M.~M.; Vollbrecht,~K.~G.; Cirac,~J.~I. {On entropy growth and
  the hardness of simulating time evolution}. \emph{New J. Phys.}
  \textbf{2008}, \emph{10}, 33032\relax
\mciteBstWouldAddEndPuncttrue
\mciteSetBstMidEndSepPunct{\mcitedefaultmidpunct}
{\mcitedefaultendpunct}{\mcitedefaultseppunct}\relax
\EndOfBibitem
\bibitem[W{\'{o}}jtowicz \latin{et~al.}(2020)W{\'{o}}jtowicz, Elenewski, Rams,
  and Zwolak]{Zwolak2020_KramersCrossover}
W{\'{o}}jtowicz,~G.; Elenewski,~J.~E.; Rams,~M.~M.; Zwolak,~M. {Open-system
  tensor networks and Kramers' crossover for quantum transport}. \emph{Phys.
  Rev. A} \textbf{2020}, \emph{101}, 050301\relax
\mciteBstWouldAddEndPuncttrue
\mciteSetBstMidEndSepPunct{\mcitedefaultmidpunct}
{\mcitedefaultendpunct}{\mcitedefaultseppunct}\relax
\EndOfBibitem
\bibitem[Rams and Zwolak(2020)Rams, and
  Zwolak]{Zwolak2020_Transport-EntanglementBarrier}
Rams,~M.~M.; Zwolak,~M. {Breaking the Entanglement Barrier: Tensor Network
  Simulation of Quantum Transport}. \emph{Phys. Rev. Lett.} \textbf{2020},
  \emph{124}, 137701\relax
\mciteBstWouldAddEndPuncttrue
\mciteSetBstMidEndSepPunct{\mcitedefaultmidpunct}
{\mcitedefaultendpunct}{\mcitedefaultseppunct}\relax
\EndOfBibitem
\bibitem[Guifre(2004)]{Vidal2004_TEBD}
Guifre,~V. {Efficient simulation of one-dimensional quantum many-body systems}.
  \emph{Phys. Rev. Lett.} \textbf{2004}, \emph{93}, 40501--40502\relax
\mciteBstWouldAddEndPuncttrue
\mciteSetBstMidEndSepPunct{\mcitedefaultmidpunct}
{\mcitedefaultendpunct}{\mcitedefaultseppunct}\relax
\EndOfBibitem
\bibitem[Feiguin and White(2005)Feiguin, and
  White]{Feiguin2005_Adaptive-TDDMRG}
Feiguin,~A.~E.; White,~S.~R. {Time-step targeting methods for real-time
  dynamics using DMRG}. \emph{Phys. Rev. B} \textbf{2005}, \emph{72},
  020404\relax
\mciteBstWouldAddEndPuncttrue
\mciteSetBstMidEndSepPunct{\mcitedefaultmidpunct}
{\mcitedefaultendpunct}{\mcitedefaultseppunct}\relax
\EndOfBibitem
\bibitem[Haegeman \latin{et~al.}(2011)Haegeman, Cirac, Osborne, Pizorn,
  Verschelde, and Verstraete]{Haegeman2011_TDDMRG-MPSMPO}
Haegeman,~J.; Cirac,~J.~I.; Osborne,~T.~J.; Pizorn,~I.; Verschelde,~H.;
  Verstraete,~F. {Time-dependent variational principle for quantum lattices}.
  \emph{Phys. Rev. Lett.} \textbf{2011}, \emph{107}, 070601\relax
\mciteBstWouldAddEndPuncttrue
\mciteSetBstMidEndSepPunct{\mcitedefaultmidpunct}
{\mcitedefaultendpunct}{\mcitedefaultseppunct}\relax
\EndOfBibitem
\bibitem[Zaletel \latin{et~al.}(2015)Zaletel, Mong, Karrasch, Moore, and
  Pollmann]{Zaletel2015}
Zaletel,~M.~P.; Mong,~R. S.~K.; Karrasch,~C.; Moore,~J.~E.; Pollmann,~F.
  {Time-evolving a matrix product state with long-ranged interactions}.
  \emph{Phys. Rev. B} \textbf{2015}, \emph{91}, 165112\relax
\mciteBstWouldAddEndPuncttrue
\mciteSetBstMidEndSepPunct{\mcitedefaultmidpunct}
{\mcitedefaultendpunct}{\mcitedefaultseppunct}\relax
\EndOfBibitem
\bibitem[Ronca \latin{et~al.}(2017)Ronca, Li, Jimenez-Hoyos, and
  Chan]{Ronca2017_TDDMRG-Targeting}
Ronca,~E.; Li,~Z.; Jimenez-Hoyos,~C.~A.; Chan,~G. K.~L. {Time-Step Targeting
  Time-Dependent and Dynamical Density Matrix Renormalization Group Algorithms
  with ab Initio Hamiltonians}. \emph{J. Chem. Theory Comput.} \textbf{2017},
  \emph{13}, 5560--5571\relax
\mciteBstWouldAddEndPuncttrue
\mciteSetBstMidEndSepPunct{\mcitedefaultmidpunct}
{\mcitedefaultendpunct}{\mcitedefaultseppunct}\relax
\EndOfBibitem
\bibitem[Lubich \latin{et~al.}(2015)Lubich, Oseledets, and
  Vandereycken]{Lubich2014_TimeIntegrationTT}
Lubich,~C.; Oseledets,~I.; Vandereycken,~B. {Time integration of tensor
  trains}. \emph{SIAM J. Numer. Anal.} \textbf{2015}, \emph{53}, 917\relax
\mciteBstWouldAddEndPuncttrue
\mciteSetBstMidEndSepPunct{\mcitedefaultmidpunct}
{\mcitedefaultendpunct}{\mcitedefaultseppunct}\relax
\EndOfBibitem
\bibitem[Haegeman \latin{et~al.}(2016)Haegeman, Lubich, Oseledets,
  Vandereycken, and Verstraete]{Haegeman2016_MPO-TDDMRG}
Haegeman,~J.; Lubich,~C.; Oseledets,~I.; Vandereycken,~B.; Verstraete,~F.
  {Unifying time evolution and optimization with matrix product states}.
  \emph{Phys. Rev. B} \textbf{2016}, \emph{94}, 165116\relax
\mciteBstWouldAddEndPuncttrue
\mciteSetBstMidEndSepPunct{\mcitedefaultmidpunct}
{\mcitedefaultendpunct}{\mcitedefaultseppunct}\relax
\EndOfBibitem
\bibitem[Baiardi \latin{et~al.}(2017)Baiardi, Stein, Barone, and
  Reiher]{Baiardi2017_VDMRG}
Baiardi,~A.; Stein,~C.~J.; Barone,~V.; Reiher,~M. {Vibrational Density Matrix
  Renormalization Group}. \emph{J. Chem. Theory Comput.} \textbf{2017},
  \emph{13}, 3764--3777\relax
\mciteBstWouldAddEndPuncttrue
\mciteSetBstMidEndSepPunct{\mcitedefaultmidpunct}
{\mcitedefaultendpunct}{\mcitedefaultseppunct}\relax
\EndOfBibitem
\bibitem[Muolo \latin{et~al.}(2020)Muolo, Baiardi, Feldmann, and
  Reiher]{Muolo2020_NEAP-DMRG}
Muolo,~A.; Baiardi,~A.; Feldmann,~R.; Reiher,~M. {Nuclear-electronic
  all-particle density matrix renormalization group}. \emph{J. Chem. Phys.}
  \textbf{2020}, \emph{152}, 204103\relax
\mciteBstWouldAddEndPuncttrue
\mciteSetBstMidEndSepPunct{\mcitedefaultmidpunct}
{\mcitedefaultendpunct}{\mcitedefaultseppunct}\relax
\EndOfBibitem
\bibitem[Baiardi and Reiher(2019)Baiardi, and Reiher]{Baiardi2019_TDDMRG}
Baiardi,~A.; Reiher,~M. {Large-scale quantum-dynamics with matrix product
  states}. \emph{J. Chem. Theory Comput.} \textbf{2019}, \emph{15},
  3481--3498\relax
\mciteBstWouldAddEndPuncttrue
\mciteSetBstMidEndSepPunct{\mcitedefaultmidpunct}
{\mcitedefaultendpunct}{\mcitedefaultseppunct}\relax
\EndOfBibitem
\bibitem[Keller \latin{et~al.}(2015)Keller, Dolfi, Troyer, and
  Reiher]{Keller2015_MPS-MPO-SQHamiltonian}
Keller,~S.; Dolfi,~M.; Troyer,~M.; Reiher,~M. {An efficient matrix product
  operator representation of the quantum chemical Hamiltonian}. \emph{J. Chem.
  Phys} \textbf{2015}, \emph{143}, 244118\relax
\mciteBstWouldAddEndPuncttrue
\mciteSetBstMidEndSepPunct{\mcitedefaultmidpunct}
{\mcitedefaultendpunct}{\mcitedefaultseppunct}\relax
\EndOfBibitem
\bibitem[Keller and Reiher(2016)Keller, and Reiher]{Keller2016_SpinAdapted}
Keller,~S.; Reiher,~M. {Spin-adapted matrix product states and operators}.
  \emph{J. Chem. Phys.} \textbf{2016}, \emph{144}, 134101\relax
\mciteBstWouldAddEndPuncttrue
\mciteSetBstMidEndSepPunct{\mcitedefaultmidpunct}
{\mcitedefaultendpunct}{\mcitedefaultseppunct}\relax
\EndOfBibitem
\bibitem[Zgid and Nooijen(2008)Zgid, and Nooijen]{Zgid2008_DMRGSpinAdaptation}
Zgid,~D.; Nooijen,~M. {On the spin and symmetry adaptation of the density
  matrix renormalization group method}. \emph{J. Chem. Phys.} \textbf{2008},
  \emph{128}, 014107\relax
\mciteBstWouldAddEndPuncttrue
\mciteSetBstMidEndSepPunct{\mcitedefaultmidpunct}
{\mcitedefaultendpunct}{\mcitedefaultseppunct}\relax
\EndOfBibitem
\bibitem[Wouters \latin{et~al.}(2012)Wouters, Limacher, {Van Neck}, and
  Ayers]{Wouters2012_SpinAdapted}
Wouters,~S.; Limacher,~P.~A.; {Van Neck},~D.; Ayers,~P.~W. {Longitudinal static
  optical properties of hydrogen chains: Finite field extrapolations of matrix
  product state calculations}. \emph{J. Chem. Phys.} \textbf{2012}, \emph{136},
  134110\relax
\mciteBstWouldAddEndPuncttrue
\mciteSetBstMidEndSepPunct{\mcitedefaultmidpunct}
{\mcitedefaultendpunct}{\mcitedefaultseppunct}\relax
\EndOfBibitem
\bibitem[Sharma(2015)]{Sharma2015_GeneralNonAbelian}
Sharma,~S. {A general non-Abelian density matrix renormalization group
  algorithm with application to the C 2 dimer}. \emph{J. Chem. Phys.}
  \textbf{2015}, \emph{142}, 024107\relax
\mciteBstWouldAddEndPuncttrue
\mciteSetBstMidEndSepPunct{\mcitedefaultmidpunct}
{\mcitedefaultendpunct}{\mcitedefaultseppunct}\relax
\EndOfBibitem
\bibitem[Krause \latin{et~al.}(2007)Krause, Klamroth, and
  Saalfrank]{Saalfrank2007_Properties-RealTime}
Krause,~P.; Klamroth,~T.; Saalfrank,~P. {Molecular response properties from
  explicitly time-dependent configuration interaction methods}. \emph{J. Chem.
  Phys.} \textbf{2007}, \emph{127}, 034107\relax
\mciteBstWouldAddEndPuncttrue
\mciteSetBstMidEndSepPunct{\mcitedefaultmidpunct}
{\mcitedefaultendpunct}{\mcitedefaultseppunct}\relax
\EndOfBibitem
\bibitem[Ding \latin{et~al.}(2013)Ding, {Van Kuiken}, Eichinger, and
  Li]{Ding2013_Polarizability-RealTimeDFT}
Ding,~F.; {Van Kuiken},~B.~E.; Eichinger,~B.~E.; Li,~X. {An efficient method
  for calculating dynamical hyperpolarizabilities using real-time
  time-dependent density functional theory}. \emph{J. Chem. Phys.}
  \textbf{2013}, \emph{138}, 64104\relax
\mciteBstWouldAddEndPuncttrue
\mciteSetBstMidEndSepPunct{\mcitedefaultmidpunct}
{\mcitedefaultendpunct}{\mcitedefaultseppunct}\relax
\EndOfBibitem
\bibitem[Oseledets and Dolgov(2012)Oseledets, and Dolgov]{Oseledets2012_ALS}
Oseledets,~I.~V.; Dolgov,~S.~V. {Solution of Linear Systems and Matrix
  Inversion in the TT-Format}. \emph{SIAM J. Sci. Comput.} \textbf{2012},
  \emph{34}, A2718--A2739\relax
\mciteBstWouldAddEndPuncttrue
\mciteSetBstMidEndSepPunct{\mcitedefaultmidpunct}
{\mcitedefaultendpunct}{\mcitedefaultseppunct}\relax
\EndOfBibitem
\bibitem[Schollw{\"{o}}ck(2011)]{Schollwoeck2011_Review-DMRG}
Schollw{\"{o}}ck,~U. {The density-matrix renormalization group in the age of
  matrix product states}. \emph{Ann. Phys.} \textbf{2011}, \emph{326},
  96--192\relax
\mciteBstWouldAddEndPuncttrue
\mciteSetBstMidEndSepPunct{\mcitedefaultmidpunct}
{\mcitedefaultendpunct}{\mcitedefaultseppunct}\relax
\EndOfBibitem
\bibitem[Holtz \latin{et~al.}(2012)Holtz, Rohwedder, and
  Schneider]{Holtz2012_ManifoldTT}
Holtz,~S.; Rohwedder,~T.; Schneider,~R. {On manifolds of tensors of fixed
  TT-rank}. \emph{Numer. Math.} \textbf{2012}, \emph{120}, 701--731\relax
\mciteBstWouldAddEndPuncttrue
\mciteSetBstMidEndSepPunct{\mcitedefaultmidpunct}
{\mcitedefaultendpunct}{\mcitedefaultseppunct}\relax
\EndOfBibitem
\bibitem[Broeckhove \latin{et~al.}(1988)Broeckhove, Lathouwers, Kesteloot, and
  {Van Leuven}]{VanLeuven1988_EquivalenceTDPrinciple}
Broeckhove,~J.; Lathouwers,~L.; Kesteloot,~E.; {Van Leuven},~P. {On the
  equivalence of time-dependent variational principles}. \emph{Chem. Phys.
  Lett.} \textbf{1988}, \emph{149}, 547--550\relax
\mciteBstWouldAddEndPuncttrue
\mciteSetBstMidEndSepPunct{\mcitedefaultmidpunct}
{\mcitedefaultendpunct}{\mcitedefaultseppunct}\relax
\EndOfBibitem
\bibitem[Cramer \latin{et~al.}(2008)Cramer, Dawson, Eisert, and
  Osborne]{Osborne2008_EntanglementBarrier}
Cramer,~M.; Dawson,~C.~M.; Eisert,~J.; Osborne,~T.~J. {Exact relaxation in a
  class of nonequilibrium quantum lattice systems}. \emph{Phys. Rev. Lett.}
  \textbf{2008}, \emph{100}, 030602\relax
\mciteBstWouldAddEndPuncttrue
\mciteSetBstMidEndSepPunct{\mcitedefaultmidpunct}
{\mcitedefaultendpunct}{\mcitedefaultseppunct}\relax
\EndOfBibitem
\bibitem[Goto and Danshita(2019)Goto, and Danshita]{Goto2019_LongTime-TDDMRG}
Goto,~S.; Danshita,~I. {Performance of the time-dependent variational principle
  for matrix product states in the long-time evolution of a pure state}.
  \emph{Phys. Rev. B} \textbf{2019}, \emph{99}, 054307\relax
\mciteBstWouldAddEndPuncttrue
\mciteSetBstMidEndSepPunct{\mcitedefaultmidpunct}
{\mcitedefaultendpunct}{\mcitedefaultseppunct}\relax
\EndOfBibitem
\bibitem[Krumnow \latin{et~al.}(2019)Krumnow, Eisert, and
  Legeza]{Legeza2019_OrbitalOptimization-TD}
Krumnow,~C.; Eisert,~J.; Legeza,~{\"{O}}. {Towards overcoming the entanglement
  barrier when simulating long-time evolution}. \emph{ArXiv e-prints}
  \textbf{2019}, 1904.11999\relax
\mciteBstWouldAddEndPuncttrue
\mciteSetBstMidEndSepPunct{\mcitedefaultmidpunct}
{\mcitedefaultendpunct}{\mcitedefaultseppunct}\relax
\EndOfBibitem
\bibitem[Saad(1992)]{Saad1992_MatrixExponential}
Saad,~Y. {Analysis of Some Krylov Subspace Approximations to the Matrix
  Exponential Operator}. \emph{SIAM J. Numer. Anal.} \textbf{1992}, \emph{29},
  209--228\relax
\mciteBstWouldAddEndPuncttrue
\mciteSetBstMidEndSepPunct{\mcitedefaultmidpunct}
{\mcitedefaultendpunct}{\mcitedefaultseppunct}\relax
\EndOfBibitem
\bibitem[Hochbruck and Lubich(1997)Hochbruck, and
  Lubich]{Lubich1997_MatrixExponential}
Hochbruck,~M.; Lubich,~C. {On Krylov Subspace Approximations to the Matrix
  Exponential Operator}. \emph{SIAM J. Numer. Anal.} \textbf{1997}, \emph{34},
  1911--1925\relax
\mciteBstWouldAddEndPuncttrue
\mciteSetBstMidEndSepPunct{\mcitedefaultmidpunct}
{\mcitedefaultendpunct}{\mcitedefaultseppunct}\relax
\EndOfBibitem
\bibitem[{Van Den Eshof} and Hochbruck(2006){Van Den Eshof}, and
  Hochbruck]{VanDerEshof2006_MatrixExponentialPreconditioner}
{Van Den Eshof},~J.; Hochbruck,~M. {Preconditioning Lanczos Approximations to
  the Matrix Exponential}. \emph{SIAM J. Sci. Comput.} \textbf{2006},
  \emph{27}, 1438--1457\relax
\mciteBstWouldAddEndPuncttrue
\mciteSetBstMidEndSepPunct{\mcitedefaultmidpunct}
{\mcitedefaultendpunct}{\mcitedefaultseppunct}\relax
\EndOfBibitem
\bibitem[Castro \latin{et~al.}(2004)Castro, Marques, and
  Rubio]{Castro2004_Propagators}
Castro,~A.; Marques,~M. A.~L.; Rubio,~A. {Propagators for the time-dependent
  Kohn–Sham equations}. \emph{J. Chem. Phys.} \textbf{2004}, \emph{121},
  3425--3433\relax
\mciteBstWouldAddEndPuncttrue
\mciteSetBstMidEndSepPunct{\mcitedefaultmidpunct}
{\mcitedefaultendpunct}{\mcitedefaultseppunct}\relax
\EndOfBibitem
\bibitem[{G{\'{o}}mez Pueyo} \latin{et~al.}(2018){G{\'{o}}mez Pueyo}, Marques,
  Rubio, and Castro]{Rubio2018_Propagators}
{G{\'{o}}mez Pueyo},~A.; Marques,~M. A.~L.; Rubio,~A.; Castro,~A. {Propagators
  for the Time-Dependent Kohn-Sham Equations: Multistep, Runge-Kutta,
  Exponential Runge-Kutta, and Commutator Free Magnus Methods}. \emph{J. Chem.
  Theory Comput.} \textbf{2018}, \emph{14}, 3040--3052\relax
\mciteBstWouldAddEndPuncttrue
\mciteSetBstMidEndSepPunct{\mcitedefaultmidpunct}
{\mcitedefaultendpunct}{\mcitedefaultseppunct}\relax
\EndOfBibitem
\bibitem[Blanes and Moan(2006)Blanes, and
  Moan]{Blanes2006_CommutatorFreeMagnus}
Blanes,~S.; Moan,~P.~C. {Fourth- and sixth-order commutator-free Magnus
  integrators for linear and non-linear dynamical systems}. \emph{Appl. Num.
  Math.} \textbf{2006}, \emph{56}, 1519--1537\relax
\mciteBstWouldAddEndPuncttrue
\mciteSetBstMidEndSepPunct{\mcitedefaultmidpunct}
{\mcitedefaultendpunct}{\mcitedefaultseppunct}\relax
\EndOfBibitem
\bibitem[Singh \latin{et~al.}(2011)Singh, Pfeifer, and
  Vidal]{Vidal2011_DMRG-U1Symm}
Singh,~S.; Pfeifer,~R. N.~C.; Vidal,~G. {Tensor network states and algorithms
  in the presence of a global U(1) symmetry}. \emph{Phys. Rev. B}
  \textbf{2011}, \emph{83}, 115125\relax
\mciteBstWouldAddEndPuncttrue
\mciteSetBstMidEndSepPunct{\mcitedefaultmidpunct}
{\mcitedefaultendpunct}{\mcitedefaultseppunct}\relax
\EndOfBibitem
\bibitem[Bauer \latin{et~al.}(2011)Bauer, Corboz, Or{\'{u}}s, and
  Troyer]{Troyer2011_PEPS-Symmetry}
Bauer,~B.; Corboz,~P.; Or{\'{u}}s,~R.; Troyer,~M. {Implementing global Abelian
  symmetries in projected entangled-pair state algorithms}. \emph{Phys. Rev. B}
  \textbf{2011}, \emph{83}, 125106\relax
\mciteBstWouldAddEndPuncttrue
\mciteSetBstMidEndSepPunct{\mcitedefaultmidpunct}
{\mcitedefaultendpunct}{\mcitedefaultseppunct}\relax
\EndOfBibitem
\bibitem[Calegari \latin{et~al.}(2014)Calegari, Ayuso, Trabattoni, Anumula,
  Belshaw, Camillis, Frassetto, Anumula, Frassetto, Poletto, Palacios, Decleva,
  Greenwood, and Nisoli]{Calegari2014}
Calegari,~F.; Ayuso,~D.; Trabattoni,~A.; Anumula,~S.; Belshaw,~L.;
  Camillis,~S.~D.; Frassetto,~F.; Anumula,~S.; Frassetto,~F.; Poletto,~L.;
  Palacios,~A.; Decleva,~P.; Greenwood,~J.~B.; Nisoli,~M. {Ultrafast electron
  dynamics in amino acids induced by attosecond pulses}. \emph{Science}
  \textbf{2014}, \emph{346}, 336\relax
\mciteBstWouldAddEndPuncttrue
\mciteSetBstMidEndSepPunct{\mcitedefaultmidpunct}
{\mcitedefaultendpunct}{\mcitedefaultseppunct}\relax
\EndOfBibitem
\bibitem[Galbraith \latin{et~al.}(2017)Galbraith, Scheit, Golubev, Reitsma,
  Zhavoronkov, Despr{\'{e}}, L{\'{e}}pine, Kuleff, Vrakking, Kornilov,
  K{\"{o}}ppel, and Mikosch]{Mikosch2017_Benzene-Attosecond}
Galbraith,~M. C.~E.; Scheit,~S.; Golubev,~N.~V.; Reitsma,~G.; Zhavoronkov,~N.;
  Despr{\'{e}},~V.; L{\'{e}}pine,~F.; Kuleff,~A.~I.; Vrakking,~M. J.~J.;
  Kornilov,~O.; K{\"{o}}ppel,~H.; Mikosch,~J. {Few-femtosecond passage of
  conical intersections in the benzene cation}. \emph{Nat. Commun.}
  \textbf{2017}, \emph{8}, 1018\relax
\mciteBstWouldAddEndPuncttrue
\mciteSetBstMidEndSepPunct{\mcitedefaultmidpunct}
{\mcitedefaultendpunct}{\mcitedefaultseppunct}\relax
\EndOfBibitem
\bibitem[Despr{\'{e}} \latin{et~al.}(2015)Despr{\'{e}}, Marciniak, Loriot,
  Galbraith, Rouz{\'{e}}e, Vrakking, L{\'{e}}pine, Kuleff, and Le]{Despre2015}
Despr{\'{e}},~V.; Marciniak,~A.; Loriot,~V.; Galbraith,~M. C.~E.;
  Rouz{\'{e}}e,~A.; Vrakking,~M. J.~J.; L{\'{e}}pine,~F.; Kuleff,~A.~I.; Le,~F.
  {Attosecond hole migration in benzene molecules surviving nuclear motion}.
  \emph{J. Phys. Chem. Lett.} \textbf{2015}, \emph{6}, 426--431\relax
\mciteBstWouldAddEndPuncttrue
\mciteSetBstMidEndSepPunct{\mcitedefaultmidpunct}
{\mcitedefaultendpunct}{\mcitedefaultseppunct}\relax
\EndOfBibitem
\bibitem[Schriber and Evangelista(2019)Schriber, and
  Evangelista]{Evangelista2019_Adaptive-TDCI}
Schriber,~J.~B.; Evangelista,~F.~A. {Time dependent adaptive configuration
  interaction applied to attosecond charge migration}. \emph{J. Chem. Phys.}
  \textbf{2019}, \emph{151}, 171102\relax
\mciteBstWouldAddEndPuncttrue
\mciteSetBstMidEndSepPunct{\mcitedefaultmidpunct}
{\mcitedefaultendpunct}{\mcitedefaultseppunct}\relax
\EndOfBibitem
\bibitem[Baiardi and Reiher(2020)Baiardi, and Reiher]{Baiardi2020_tcDMRG}
Baiardi,~A.; Reiher,~M. {Transcorrelated density matrix renormalization group}.
  \emph{J. Chem. Phys.} \textbf{2020}, \emph{153}, 164115\relax
\mciteBstWouldAddEndPuncttrue
\mciteSetBstMidEndSepPunct{\mcitedefaultmidpunct}
{\mcitedefaultendpunct}{\mcitedefaultseppunct}\relax
\EndOfBibitem
\bibitem[Hubig \latin{et~al.}(2015)Hubig, McCulloch, Schollw{\"{o}}ck, and
  Wolf]{McCulloch2015_Mixing}
Hubig,~C.; McCulloch,~I.~P.; Schollw{\"{o}}ck,~U.; Wolf,~F.~A. {Strictly
  single-site DMRG algorithm with subspace expansion}. \emph{Phys. Rev. B}
  \textbf{2015}, \emph{91}, 155115\relax
\mciteBstWouldAddEndPuncttrue
\mciteSetBstMidEndSepPunct{\mcitedefaultmidpunct}
{\mcitedefaultendpunct}{\mcitedefaultseppunct}\relax
\EndOfBibitem
\bibitem[Legeza and S{\'{o}}lyom(2003)Legeza, and
  S{\'{o}}lyom]{Legeza2003_OrderingOptimization}
Legeza,~{\"{O}}.; S{\'{o}}lyom,~J. {Optimizing the density-matrix
  renormalization group method using quantum information entropy}. \emph{Phys.
  Rev. B} \textbf{2003}, \emph{68}, 195116\relax
\mciteBstWouldAddEndPuncttrue
\mciteSetBstMidEndSepPunct{\mcitedefaultmidpunct}
{\mcitedefaultendpunct}{\mcitedefaultseppunct}\relax
\EndOfBibitem
\bibitem[Stein and Reiher(2016)Stein, and Reiher]{Stein2016_AutomatedSelection}
Stein,~C.~J.; Reiher,~M. {Automated Selection of Active Orbital Spaces}.
  \emph{J. Chem. Theory Comput.} \textbf{2016}, \emph{12}, 1760--1771\relax
\mciteBstWouldAddEndPuncttrue
\mciteSetBstMidEndSepPunct{\mcitedefaultmidpunct}
{\mcitedefaultendpunct}{\mcitedefaultseppunct}\relax
\EndOfBibitem
\bibitem[Stein \latin{et~al.}(2016)Stein, von Burg, and
  Reiher]{Stein2016_DelicateBalance}
Stein,~C.~J.; von Burg,~V.; Reiher,~M. {The Delicate Balance of Static and
  Dynamic Electron Correlation}. \emph{J. Chem. Theory Comput.} \textbf{2016},
  \emph{12}, 3764--3773\relax
\mciteBstWouldAddEndPuncttrue
\mciteSetBstMidEndSepPunct{\mcitedefaultmidpunct}
{\mcitedefaultendpunct}{\mcitedefaultseppunct}\relax
\EndOfBibitem
\bibitem[Stein and Reiher(2017)Stein, and Reiher]{Stein2017_AutoCAS-Chemia}
Stein,~C.~J.; Reiher,~M. {Automated Identification of Relevant Frontier
  Orbitals for Chemical Compounds and Processes}. \emph{Chimia} \textbf{2017},
  \emph{71}, 170--176\relax
\mciteBstWouldAddEndPuncttrue
\mciteSetBstMidEndSepPunct{\mcitedefaultmidpunct}
{\mcitedefaultendpunct}{\mcitedefaultseppunct}\relax
\EndOfBibitem
\bibitem[Stein and Reiher(2017)Stein, and
  Reiher]{Stein2017_MultireferenceQuantification}
Stein,~C.~J.; Reiher,~M. {Measuring multi-configurational character by orbital
  entanglement}. \emph{Mol. Phys.} \textbf{2017}, \emph{115}, 2110--2119\relax
\mciteBstWouldAddEndPuncttrue
\mciteSetBstMidEndSepPunct{\mcitedefaultmidpunct}
{\mcitedefaultendpunct}{\mcitedefaultseppunct}\relax
\EndOfBibitem
\bibitem[Stein and Reiher(2019)Stein, and
  Reiher]{Stein2019_AutoCAS-Implementation}
Stein,~C.~J.; Reiher,~M. {autoCAS: A Program for Fully Automated
  Multiconfigurational Calculations}. \emph{J. Comput. Chem.} \textbf{2019},
  \emph{40}, 2216\relax
\mciteBstWouldAddEndPuncttrue
\mciteSetBstMidEndSepPunct{\mcitedefaultmidpunct}
{\mcitedefaultendpunct}{\mcitedefaultseppunct}\relax
\EndOfBibitem
\bibitem[Ghosh \latin{et~al.}(2017)Ghosh, Andersen, Gagliardi, Cramer, and
  Govind]{Ghosh2017_TD-Semiempirical}
Ghosh,~S.; Andersen,~A.; Gagliardi,~L.; Cramer,~C.~J.; Govind,~N. {Modeling
  Optical Spectra of Large Organic Systems Using Real-Time Propagation of
  Semiempirical Effective Hamiltonians}. \emph{J. Chem. Theory Comput.}
  \textbf{2017}, \emph{13}, 4410--4420\relax
\mciteBstWouldAddEndPuncttrue
\mciteSetBstMidEndSepPunct{\mcitedefaultmidpunct}
{\mcitedefaultendpunct}{\mcitedefaultseppunct}\relax
\EndOfBibitem
\bibitem[Ghosh \latin{et~al.}(2019)Ghosh, Asher, Gagliardi, Cramer, and
  Govind]{Gagliardi2019_TD-Semiempirical}
Ghosh,~S.; Asher,~J.~C.; Gagliardi,~L.; Cramer,~C.~J.; Govind,~N. {A
  semiempirical effective Hamiltonian based approach for analyzing excited
  state wave functions and computing excited state absorption spectra using
  real-time dynamics}. \emph{J. Chem. Phys.} \textbf{2019}, \emph{150},
  104103\relax
\mciteBstWouldAddEndPuncttrue
\mciteSetBstMidEndSepPunct{\mcitedefaultmidpunct}
{\mcitedefaultendpunct}{\mcitedefaultseppunct}\relax
\EndOfBibitem
\bibitem[Tussupbayev \latin{et~al.}(2015)Tussupbayev, Govind, Lopata, and
  Cramer]{Cramer2015_RTTDDFT-Comparison}
Tussupbayev,~S.; Govind,~N.; Lopata,~K.; Cramer,~C.~J. {Comparison of real-time
  and linear-response time-dependent density functional theories for molecular
  chromophores ranging from sparse to high densities of states}. \emph{J. Chem.
  Theory Comput.} \textbf{2015}, \emph{11}, 1102--1109\relax
\mciteBstWouldAddEndPuncttrue
\mciteSetBstMidEndSepPunct{\mcitedefaultmidpunct}
{\mcitedefaultendpunct}{\mcitedefaultseppunct}\relax
\EndOfBibitem
\bibitem[Bruner \latin{et~al.}(2016)Bruner, LaMaster, and
  Lopata]{Lopata2016_Absorption-RTTDDFT}
Bruner,~A.; LaMaster,~D.; Lopata,~K. {Accelerated Broadband Spectra Using
  Transition Dipole Decomposition and Pad{\'{e}} Approximants}. \emph{J. Chem.
  Theory Comput.} \textbf{2016}, \emph{12}, 3741--3750\relax
\mciteBstWouldAddEndPuncttrue
\mciteSetBstMidEndSepPunct{\mcitedefaultmidpunct}
{\mcitedefaultendpunct}{\mcitedefaultseppunct}\relax
\EndOfBibitem
\bibitem[Schelter and K{\"{u}}mmel(2018)Schelter, and
  K{\"{u}}mmel]{Kummel2018_RT-TDDFT}
Schelter,~I.; K{\"{u}}mmel,~S. {Accurate Evaluation of Real-Time Density
  Functional Theory Providing Access to Challenging Electron Dynamics}.
  \emph{J. Chem. Theory Comput.} \textbf{2018}, \emph{14}, 1910--1927\relax
\mciteBstWouldAddEndPuncttrue
\mciteSetBstMidEndSepPunct{\mcitedefaultmidpunct}
{\mcitedefaultendpunct}{\mcitedefaultseppunct}\relax
\EndOfBibitem
\bibitem[Lopata \latin{et~al.}(2012)Lopata, {Van Kuiken}, Khalil, and
  Govind]{Lopata2012_RTTDDFT-Core}
Lopata,~K.; {Van Kuiken},~B.~E.; Khalil,~M.; Govind,~N. {Linear-Response and
  Real-Time Time-Dependent Density Functional Theory Studies of Core-Level
  Near-Edge X-Ray Absorption}. \emph{J. Chem. Theory Comput.} \textbf{2012},
  \emph{8}, 3284--3292\relax
\mciteBstWouldAddEndPuncttrue
\mciteSetBstMidEndSepPunct{\mcitedefaultmidpunct}
{\mcitedefaultendpunct}{\mcitedefaultseppunct}\relax
\EndOfBibitem
\bibitem[Kadek \latin{et~al.}(2015)Kadek, Konecny, Gao, Repisky, and
  Ruud]{Ruud2015_XRay-Relativistic-RT}
Kadek,~M.; Konecny,~L.; Gao,~B.; Repisky,~M.; Ruud,~K. {X-ray absorption
  resonances near L 2,3 -edges from real-time propagation of the
  Dirac–Kohn–Sham density matrix}. \emph{Phys. Chem. Chem. Phys.}
  \textbf{2015}, \emph{17}, 22566--22570\relax
\mciteBstWouldAddEndPuncttrue
\mciteSetBstMidEndSepPunct{\mcitedefaultmidpunct}
{\mcitedefaultendpunct}{\mcitedefaultseppunct}\relax
\EndOfBibitem
\bibitem[Kasper \latin{et~al.}(2018)Kasper, Lestrange, Stetina, and
  Li]{Li2018_XRay-RealTime}
Kasper,~J.~M.; Lestrange,~P.~J.; Stetina,~T.~F.; Li,~X. {Modeling L 2,3 -Edge
  X-ray Absorption Spectroscopy with Real-Time Exact Two-Component Relativistic
  Time-Dependent Density Functional Theory}. \emph{J. Chem. Theory Comput.}
  \textbf{2018}, \emph{14}, 1998--2006\relax
\mciteBstWouldAddEndPuncttrue
\mciteSetBstMidEndSepPunct{\mcitedefaultmidpunct}
{\mcitedefaultendpunct}{\mcitedefaultseppunct}\relax
\EndOfBibitem
\bibitem[Neville and Schuurman(2018)Neville, and Schuurman]{Schuurman2018_XRay}
Neville,~S.~P.; Schuurman,~M.~S. {A general approach for the calculation and
  characterization of x-ray absorption spectra}. \emph{J. Chem. Phys.}
  \textbf{2018}, \emph{149}, 154111\relax
\mciteBstWouldAddEndPuncttrue
\mciteSetBstMidEndSepPunct{\mcitedefaultmidpunct}
{\mcitedefaultendpunct}{\mcitedefaultseppunct}\relax
\EndOfBibitem
\bibitem[Dorando \latin{et~al.}(2007)Dorando, Hachmann, and
  Chan]{Dorando2007_TargetingExcitedStates}
Dorando,~J.~J.; Hachmann,~J.; Chan,~G. K.-L.~L. {Targeted excited state
  algorithms}. \emph{J. Chem. Phys.} \textbf{2007}, \emph{127}, 84109\relax
\mciteBstWouldAddEndPuncttrue
\mciteSetBstMidEndSepPunct{\mcitedefaultmidpunct}
{\mcitedefaultendpunct}{\mcitedefaultseppunct}\relax
\EndOfBibitem
\bibitem[Devakul \latin{et~al.}(2017)Devakul, Khemani, Pollmann, Huse, and
  Sondhi]{Devakul2017}
Devakul,~T.; Khemani,~V.; Pollmann,~F.; Huse,~D.~A.; Sondhi,~S.~L. {Obtaining
  highly excited eigenstates of the localized XX chain via DMRG-X}.
  \emph{Philos. Trans. R. Soc. A} \textbf{2017}, \emph{375}, 20160431\relax
\mciteBstWouldAddEndPuncttrue
\mciteSetBstMidEndSepPunct{\mcitedefaultmidpunct}
{\mcitedefaultendpunct}{\mcitedefaultseppunct}\relax
\EndOfBibitem
\bibitem[Yu \latin{et~al.}(2017)Yu, Pekker, and
  Clark]{Yu2017_ShiftAndInvertMPS}
Yu,~X.; Pekker,~D.; Clark,~B.~K. {Finding Matrix Product State Representations
  of Highly Excited Eigenstates of Many-Body Localized Hamiltonians}.
  \emph{Phys. Rev. Lett.} \textbf{2017}, \emph{118}, 17201\relax
\mciteBstWouldAddEndPuncttrue
\mciteSetBstMidEndSepPunct{\mcitedefaultmidpunct}
{\mcitedefaultendpunct}{\mcitedefaultseppunct}\relax
\EndOfBibitem
\bibitem[Baiardi \latin{et~al.}(2019)Baiardi, Stein, Barone, and
  Reiher]{Baiardi2019_HighEnergy-vDMRG}
Baiardi,~A.; Stein,~C.~J.; Barone,~V.; Reiher,~M. {Optimization of highly
  excited matrix product states with an application to vibrational
  spectroscopy}. \emph{J. Chem. Phys.} \textbf{2019}, \emph{150}, 094113\relax
\mciteBstWouldAddEndPuncttrue
\mciteSetBstMidEndSepPunct{\mcitedefaultmidpunct}
{\mcitedefaultendpunct}{\mcitedefaultseppunct}\relax
\EndOfBibitem
\bibitem[Helgaker \latin{et~al.}(2012)Helgaker, Coriani, J{\o}rgensen,
  Kristensen, Olsen, and Ruud]{Helgaker2012_ReviewProperties}
Helgaker,~T.; Coriani,~S.; J{\o}rgensen,~P.; Kristensen,~K.; Olsen,~J.;
  Ruud,~K. {Recent Advances in Wave Function-Based Methods of
  Molecular-Property Calculations}. \emph{Chem. Rev.} \textbf{2012},
  \emph{112}, 543--631\relax
\mciteBstWouldAddEndPuncttrue
\mciteSetBstMidEndSepPunct{\mcitedefaultmidpunct}
{\mcitedefaultendpunct}{\mcitedefaultseppunct}\relax
\EndOfBibitem
\bibitem[Hettema \latin{et~al.}(1992)Hettema, Jensen, J{\o}rgensen, and
  Olsen]{Olsen1992_MCSCF-ResponseTheory}
Hettema,~H.; Jensen,~H. J.~A.; J{\o}rgensen,~P.; Olsen,~J. {Quadratic response
  functions for a multiconfigurational self-consistent field wave function}.
  \emph{J. Chem. Phys.} \textbf{1992}, \emph{97}, 1174--1190\relax
\mciteBstWouldAddEndPuncttrue
\mciteSetBstMidEndSepPunct{\mcitedefaultmidpunct}
{\mcitedefaultendpunct}{\mcitedefaultseppunct}\relax
\EndOfBibitem
\bibitem[Kobayashi \latin{et~al.}(1994)Kobayashi, Koch, and
  J{\o}rgensen]{Koch1994_FrequencyDependent}
Kobayashi,~R.; Koch,~H.; J{\o}rgensen,~P. {Calculation of frequency-dependent
  polarizabilities using coupled-cluster response theory}. \emph{Chem. Phys.
  Lett.} \textbf{1994}, \emph{219}, 30--35\relax
\mciteBstWouldAddEndPuncttrue
\mciteSetBstMidEndSepPunct{\mcitedefaultmidpunct}
{\mcitedefaultendpunct}{\mcitedefaultseppunct}\relax
\EndOfBibitem
\bibitem[Sa{\l}ek \latin{et~al.}(2002)Sa{\l}ek, Vahtras, Helgaker, and
  {\AA}gren]{Agren2002_TDHF-DFT}
Sa{\l}ek,~P.; Vahtras,~O.; Helgaker,~T.; {\AA}gren,~H. {Density-functional
  theory of linear and nonlinear time-dependent molecular properties}. \emph{J.
  Chem. Phys.} \textbf{2002}, \emph{117}, 9630--9645\relax
\mciteBstWouldAddEndPuncttrue
\mciteSetBstMidEndSepPunct{\mcitedefaultmidpunct}
{\mcitedefaultendpunct}{\mcitedefaultseppunct}\relax
\EndOfBibitem
\bibitem[Dorando \latin{et~al.}(2009)Dorando, Hachmann, and
  Chan]{Dorando2009_AnalyticalResponseFunction}
Dorando,~J.~J.; Hachmann,~J.; Chan,~G. K.-l. {Analytic response theory for the
  density matrix renormalization group}. \emph{J. Chem. Phys.} \textbf{2009},
  \emph{130}, 184111\relax
\mciteBstWouldAddEndPuncttrue
\mciteSetBstMidEndSepPunct{\mcitedefaultmidpunct}
{\mcitedefaultendpunct}{\mcitedefaultseppunct}\relax
\EndOfBibitem
\bibitem[Nakatani \latin{et~al.}(2014)Nakatani, Wouters, Neck, and
  Chan]{Nakatani2014_LinearResponseDMRG}
Nakatani,~N.; Wouters,~S.; Neck,~D.~V.; Chan,~G. K.-L. {Linear response theory
  for the density matrix renormalization group: Efficient algorithms for
  strongly correlated excited states.} \emph{J. Chem. Phys.} \textbf{2014},
  \emph{140}, 24108\relax
\mciteBstWouldAddEndPuncttrue
\mciteSetBstMidEndSepPunct{\mcitedefaultmidpunct}
{\mcitedefaultendpunct}{\mcitedefaultseppunct}\relax
\EndOfBibitem
\bibitem[Samanta \latin{et~al.}(2018)Samanta, Blunt, and
  Booth]{Booth2018_FCIQMC-Properties}
Samanta,~P.~K.; Blunt,~N.~S.; Booth,~G.~H. {Response Formalism within Full
  Configuration Interaction Quantum Monte Carlo: Static Properties and
  Electrical Response}. \emph{J. Chem. Theory Comput.} \textbf{2018},
  \emph{14}, 3532--3546\relax
\mciteBstWouldAddEndPuncttrue
\mciteSetBstMidEndSepPunct{\mcitedefaultmidpunct}
{\mcitedefaultendpunct}{\mcitedefaultseppunct}\relax
\EndOfBibitem
\bibitem[Larsen \latin{et~al.}(1998)Larsen, H{\"{a}}ttig, Olsen, and
  J{\o}rgensen]{Olsen1998_FCI-Polarizabilities}
Larsen,~H.; H{\"{a}}ttig,~C.; Olsen,~J.; J{\o}rgensen,~P. {A basis set study of
  coupled cluster and full configuration interaction calculations of molecular
  electric properties for BH}. \emph{Chem. Phys. Lett.} \textbf{1998},
  \emph{291}, 536--546\relax
\mciteBstWouldAddEndPuncttrue
\mciteSetBstMidEndSepPunct{\mcitedefaultmidpunct}
{\mcitedefaultendpunct}{\mcitedefaultseppunct}\relax
\EndOfBibitem
\bibitem[Blunt \latin{et~al.}(2015)Blunt, Alavi, and
  Booth]{Blunt2015_Krylov-FCIQMC}
Blunt,~N.~S.; Alavi,~A.; Booth,~G.~H. {Krylov-Projected Quantum Monte Carlo
  Method}. \emph{Phys. Rev. Lett.} \textbf{2015}, \emph{115}, 050603\relax
\mciteBstWouldAddEndPuncttrue
\mciteSetBstMidEndSepPunct{\mcitedefaultmidpunct}
{\mcitedefaultendpunct}{\mcitedefaultseppunct}\relax
\EndOfBibitem
\bibitem[Nisoli \latin{et~al.}(2017)Nisoli, Decleva, Calegari, Palacios, and
  Mart{\'{i}}n]{Nisoli2017_Review}
Nisoli,~M.; Decleva,~P.; Calegari,~F.; Palacios,~A.; Mart{\'{i}}n,~F.
  {Attosecond Electron Dynamics in Molecules}. \emph{Chem. Rev.} \textbf{2017},
  \emph{117}, 10760--10825\relax
\mciteBstWouldAddEndPuncttrue
\mciteSetBstMidEndSepPunct{\mcitedefaultmidpunct}
{\mcitedefaultendpunct}{\mcitedefaultseppunct}\relax
\EndOfBibitem
\end{mcitethebibliography}

\end{document}